\documentclass[%
 aip,
 amsmath,amssymb,
reprint,%
floatfix,
]{revtex4-2}

\usepackage{graphicx}
\usepackage{dcolumn}
\usepackage{bm}

\usepackage[utf8]{inputenc}
\usepackage[T1]{fontenc}
\usepackage{mathptmx}
\usepackage{etoolbox}
\usepackage{xcolor}
\usepackage{amsmath}
\usepackage[]{algorithm2e}
\usepackage{multirow}
\usepackage[precision=2, unit=mm]{lengthconvert}
\usepackage{tikz}
\usepackage{comment}
\usepackage{xr}
\usepackage{prettyref}
\usepackage{tabularray}
\usepackage{hyperref}

\hypersetup{
    colorlinks=true,
    linkcolor=blue,
    urlcolor=blue,
    citecolor=blue
}
\newrefformat{Seq}{Eq.~S\ref{#1}}

\makeatletter
\def\@customfooter@odd{\hfil\footnotesize The following article has been submitted to/accepted by \textit{Physics of Fluids}. After it is published, it will be found at \href{https://doi.org/10.1063/5.0274946}{doi.org/10.1063/5.0274946}.\hfil\llap{\thepage}}
\def\@customfooter@even{\rlap{\thepage}\hfil\footnotesize The following article has been submitted to/accepted by \textit{Physics of Fluids}. After it is published, it will be found at \href{https://doi.org/10.1063/5.0274946}{doi.org/10.1063/5.0274946}.\hfil}

\def\ps@plain{%
  \let\@oddhead\@empty
  \let\@evenhead\@empty
  \def\@oddfoot{\@customfooter@odd}%
  \def\@evenfoot{\@customfooter@even}%
}

\renewcommand{\@oddfoot}{\@customfooter@odd}
\renewcommand{\@evenfoot}{\@customfooter@even}

\def\ps@titlepage{%
  \let\@oddhead\@empty
  \let\@evenhead\@empty
  \def\@oddfoot{\@customfooter@odd}%
  \def\@evenfoot{\@customfooter@even}%
}
\makeatother

\newrefformat{Seq}{Eq.~S\ref{#1}}

\usetikzlibrary{patterns}

\newcommand{\norm}[1]{\left\lVert#1\right\rVert}

\makeatletter
\def\@email#1#2{%
 \endgroup
 \patchcmd{\titleblock@produce}
  {\frontmatter@RRAPformat}
  {\frontmatter@RRAPformat{\produce@RRAP{*#1\href{mailto:#2}{#2}}}\frontmatter@RRAPformat}
  {}{}
}%
\makeatother

\begin{document}

\preprint{AIP/123-QED}

\title[]{Deep Learning-Based Prediction of High Explosive Induced Fluid Dynamics}
\author{Francis G. VanGessel}
\altaffiliation[Formerly at ]{Naval Surface Warfare Center, Indian Head Division}
\affiliation{ 
Department of Mechanical Engineering, University of Maryland, College Park, Maryland 20742, USA
}%
\email{fvangess@umd.edu}
\author{Mitul Pandya}%
\affiliation{ 
Naval Surface Warfare Center, Indian Head Division, Indian Head, Maryland 20640, USA
}%

\date{\today}

\begin{abstract}
Underwater explosions produce complex fluid phenomena relevant to diverse applications including maritime engineering, medical therapeutics, and inertial confinement fusion. These systems exhibit multiphase flows, chemical kinetics, and highly compressible dynamics that challenge traditional computational approaches. Current hydrodynamic solvers, while accurate, are computationally expensive and non-differentiable, limiting their use in design optimization and real-time applications. Here we show that deep neural networks can predict underwater explosion-induced fluid dynamics 4,025 times faster than traditional solvers while maintaining mean absolute percent errors below 0.005\% across all fluid state variables. Our approach maps from explosive material thermodynamic parameters to the temporal evolution of shock fronts and material interfaces, enabling rapid prediction of system behavior for a broad range of ideal explosive materials. Feature importance analysis reveals that exponential decay parameters of the explosive equation of state are the primary drivers of system dynamics, uncovering a previously unknown relationship between thermodynamic compatibility and energy transfer efficiency at material interfaces. Furthermore, we demonstrate an inverse design framework that leverages the differentiability of our neural surrogate to perform parameter discovery, recovering unknown explosive material properties to within 1\% accuracy through gradient-based optimization. This combination of rapid inference, physical insight, and inverse design capabilities provides a route to engineering controlled fluid behavior in underwater explosive systems through material design. We anticipate our approach could enable new applications in defense systems, underwater manufacturing, and medical procedures where precise control of shock waves and bubble dynamics is essential.
\end{abstract}

\maketitle


\section{\label{sec:introduction}Introduction}

Underwater high-explosive (HE) materials produce a fluid dynamic response characterized by numerous fluid and thermal phenomena. In particular, these HE-water systems exhibit multiphase flow, discontinuous solution fields at the shock and material interface locations, detonation fronts coupled with regions of rapid chemical kinetics, and highly compressible flows which sample large regimes of thermodynamic state space \cite{cole1948underwater}. \textcolor{black}{These fluid phenomena extend beyond underwater explosions, manifesting in diverse scientific and engineering domains. Pressure-driven bubble dynamics influence propeller performance and erosion in maritime applications \cite{denner2024kirkwood, yusvika2020achievements}. Chemical kinetics play a crucial role in supersonic and hypersonic aerodynamic flows \cite{bertin2006critical}, while thermodynamically-driven processes govern energy transfer in inertial confinement fusion \cite{gaffney2018review}. Similarly, shock wave propagation enables the operation of rotating detonation engines \cite{mendible2021data}, while understanding shock and bubble interactions is essential for damage mitigation of maritime structures \cite{tong2022fluid}. The controlled application of cavitation, another related fluid phenomenon, has enabled advances in medical therapeutics \cite{haskell2023monitoring}.}

Rapid, accurate, and differentiable neural surrogates of an HE-water system could enable near real-time prediction capabilities \cite{brunton2022data}, investigation of the HE material property space (e.g. thermophysical properties)\cite{chen2019aerodynamic} and its influence on resulting fluid dynamics, and inverse design for engineered system behaviors. However, these objectives are not achievable through currently available hydrodynamic simulators that are computationally costly and non-differentiable. In this article, we investigate the use of scientific machine learning to overcome traditional computing challenges and address technology needs through the development of differentiable neural surrogates for predicting HE-produced fluid dynamics phenomenon. \textcolor{black}{Such a capability could address similar simulation challenges faced across multiple domains, offering potential advancements in fields ranging from hypersonic aerodynamics to medical cavitation therapeutics that share these complex fluid physics characteristics.}

Among the numerous fluid structures present in the HE-water system, the shock and material interface locations hold particular engineering relevance. Namely, the shock location corresponds to the point of maximum pressure and thus dictates the shock-induced damage imparted to nearby structures. Furthermore, it is at the multiphase HE-water interface where mechanical work is done by the high-pressure gaseous bubble on it's surroundings. These bubble dynamics have the potential to damage nearby maritime structures \cite{chen2018model} or even disintegrate kidney and gall stones at smaller scales \cite{brennen2015cavitation}. However, the discontinuous nature of the shock and the multiphase aspect of the material interface makes these fluid structures particularly challenging to model. Specifically, lack of smoothness restricts applicable numerical algorithms while detonation and multiphase phenomena prevent specifying explicit first-principles governing equations for the bubble dynamics \cite{greif2019decay, toro2013riemann}. Additionally, existing HE materials exhibit a wide variety of detonation and thermodynamic behaviors, thus leading to large variability in the resulting shock and interface dynamics. This work focuses on overcoming these numerous computational challenges and developing rapid and differentiable surrogate predictor of the dynamic fluid states at the shock and material interface for the most prominent class of HE materials.

Computation of the fluid response due to HE detonation has traditionally been performed using hydrodynamics simulations (often termed {\it hydrocodes}) which solve the Navier-Stokes equations in the high-pressure regime assuming adiabatic and inviscid flow \cite{cao2017simulation, kira1999underwater}, under these assumptions the governing equations are referred to as the {\it Euler equations}. Traditional hydrodynamic solvers are hampered by significant computational expense and lack of derivative information between simulation inputs and outputs. Lengthy simulation times are a result of the second-order accuracy of the numerical algorithms used in the vast majority of hydrodynamic solvers \cite{toro2013riemann}. The spatial order of accuracy in turn necessitates highly refined spatiotemporal discretization. Furthermore, hydrocode numerical methods may rely on assumptions of solution field discontinuity, numerical regularization, and non-linear flux terms, all of which render application of adjoint methods problematic \cite{toro2013riemann, leveque2002finite, fikl2016comprehensive}. Physics-based reduced order models (ROMs) partially address the issue of calculation cost, but require application of a number of simplifying thermofluidic assumptions that lower prediction accuracy\cite{cole1948underwater, kirkwood1942progress, zhang2021engineering, zhang2021improved, denner2024kirkwood}

In addition to the computational challenges associated with hydrocode modeling, HE materials exhibit a wide array of thermodynamic properties. These varied thermodynamic properties arise from differences in composition and mass ratios of the HE ingredients and can significantly alter the types of fluid dynamics produced by different HE systems. This range of thermodynamic properties dictate the HE detonation properties that provide the initial conditions to the fluid response. Alleviating computational costs would enable exploration of the HE thermodynamic property space, revealing insights into the complex interplay of shocks, multiphase interfaces, and thermodynamic effects present in these fluid systems. The link between thermodynamic properties, over which one can exercise a certain degree of control, and the fluid dynamic response invites the use of inverse design techniques for engineered system response. However, lack of differentiability of hydrocodes precludes their use within gradient-based inverse design paradigms. In summary, rapid-inference differentiable prediction methods are needed to address computational and design challenges inherent in traditional hydrodynamic computing methods.

Scientific machine learning (SML) techniques offer a possible route to predicting the dynamic fluid response via a computationally efficient and differentiable surrogate model. The SML discipline has experienced rapid growth over the past several years and we highlight only a subset of those research efforts with particular relevance to the work presented here. Physics-informed neural networks (PINNs) are widely used partial differential equation (PDE) and ordinary differential equation solvers which obtain an approximate solution field of a physical system by defining a loss term which minimizes the residuals of the governing PDEs and ODEs while also incorporating appropriate initial and/or boundary condition residuals \cite{raissi2019physics}. PINNs have been successfully applied to solutions of the Euler equations for scenarios exhibiting discontinuous shock states \cite{mao2020physics}. PINN approaches, in their standard format, are restricted to a solution of a single instance of a PDE and do not generalize to, e.g., different boundary or initial conditions without retraining. An alternative data-driven technique, principal orthogonal decomposition (POD) and the closely related dynamic mode decomposition (DMD), attempt to approximate system dynamics through projection onto a linear subspace. Subsequently, the dynamics of the linearized approximation can be evolved more efficiently than the full system dynamics, making this approach applicable to flow control engineering\cite{taira2017modal}. Attempts have been made to adapt POD techniques to advection-dominated systems with discontinuous solutions fields, e.g. shocks \cite{mendible2020dimensionality}. However, the slow decay of the Kolmogorov N-width in these systems poses fundamental challenges to linear subspace projection-based approaches. Operator networks are a class of deep neural network architectures developed to approximate the solution operator of a general class of PDEs \cite{lu2019deeponet, kovachki2023neural}. By learning the operator for a class of PDEs, operator networks address the specificity present in PINNs. Once trained, operator networks enable the mapping between functional representations parameterizing a dynamical system, such as time varying forcing functions, porosity fields, or inflow boundary conditions, and the resulting solution fields \cite{li2024physics}. Operator networks have been applied to predicting the flow state and chemical kinetics behind shocks arising in hypersonics applications \cite{mao2021deepm}. Operator networks offer a compelling data-driven approach for learning the mapping between the thermodynamic representation of an HE material and the resulting material interface and shock front dynamics produced by detonation of the HE material.

Applying SML approaches to HE-water systems presents numerous challenges due to the unique fluid dynamic structures present. These structures include, strong discontinuities at the shock fronts and material interfaces, multiphase flow where each phase is characterized by significantly different thermofluid properties, and varying detonation properties of different HE materials. Numerous operator learning and PINN techniques have been published which partially address a subset of these challenges including works that handle shock discontinuities \cite{mao2020physics, patel2022thermodynamically, jagtap2022physics, jagtap2020conservative}, material interface effects \cite{qiu2022physics, wang2021deep, buhendwa2021inferring}, and varying thermofluid properties \cite{mao2021deepm, cheng2024physics, nguyen2023parc}. Furthermore, physics-aware SML techniques have demonstrated significant potential to modeling deflagration to detonation transition at the mesoscale while accounting for varying thermochemical behaviors. However, no existing approaches have been shown to adequately address the full range of fluid behaviors present in underwater HE systems. Therefore, we have developed a data-driven approach that holistically accommodates shocks, multiphase effects, and varying thermochemical and detonation properties of the HE system. In particular, we focus only on the spacetime locations of the material interface and primary shock, neglecting the intervening fields. In summary, we present a neural-based approach that maps from the thermodynamic parameters of the HE system, to the location and fluid state variable quantities at the material interface and shock front. 

The remainder of this paper is structured as follows. In Sec. \ref{sec:methods} we detail our methodology including the system definition, data generation process, data preprocessing, ML model training and model selection procedure. In Sec. \ref{sec:results} we present the trained ML model, analyzing prediction accuracy of the fluid state variables. Section \ref{sec:feature_importance} presents a feature importance study of the trained ML model, identifying and interpreting which thermodynamic features have the largest influence on the fluid dynamic response. In Sec. \ref{sec:inv_design} we investigate the inverse design capabilities afforded by a differentiable surrogate model and demonstrate an approach for performing parameter discovery to identify unknown system dynamics. Finally, Sec. \ref{sec:conclusions} summarizes the findings of this study.

\section{\label{sec:methods}Data Generation and Machine Learning Methodology}

\subsection{\label{sec:system_def}System Definition and Fluid Simulation}

\begin{figure*}[ht!]
\includegraphics[trim={0 2cm 0 2cm},clip,width=0.5\textwidth]{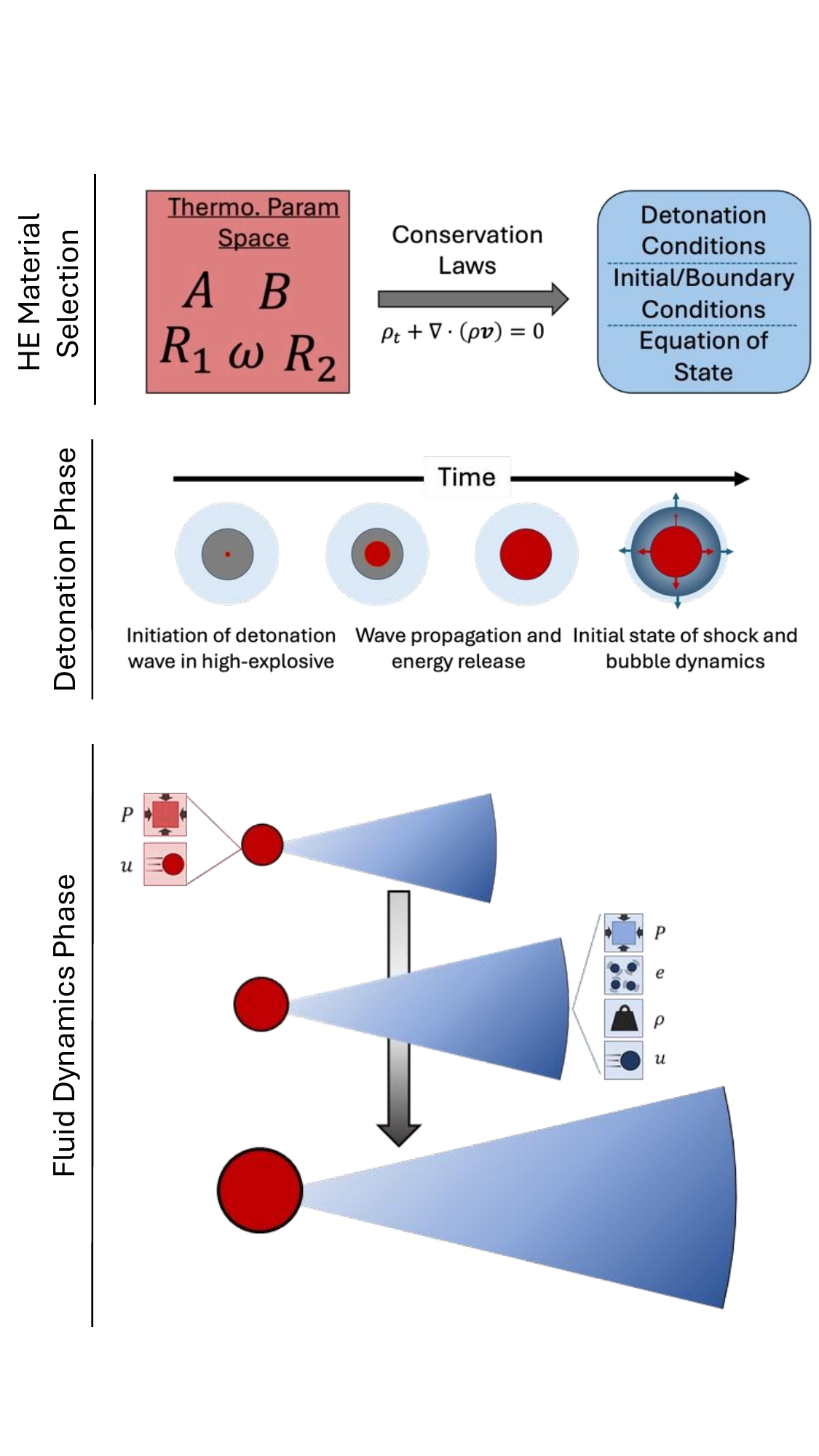}
\caption{\label{fig:undex_cartoon} \textcolor{black}{Illustration of the stages of underwater high-explosive phenomena, demonstrating how equation of state parameters define the detonation state and initial conditions of the subsequent fluid dynamics phase. Three time snapshots of the fluid dynamics phase track the material interface and shock front state variables.}}
\end{figure*}

The dynamics of the HE-water system are calculated through a one-dimensional spherically symmetric simulation of multiphase Euler flow. The geometric configuration of the system is assumed fixed with a ten pound sphere of HE at a constant material density of 1.65 g/cc. Furthermore, the HE material is suspended at the fixed depth of 15.24 meters with the surrounding quiescent water pressure initialized with the appropriate hydrostatic pressure. A summary of the underwater explosion process is visualized in Fig. \ref{fig:undex_cartoon}, depicting the simulation process flow from HE material selection, HE detonation phase, and finally the fluid dynamic response. 

The ground-truth fluid dynamics data in this work is generated using DYSMAS, a finite volume-based hydrodynamics solver developed specifically for simulating underwater explosions. DYSMAS has been extensively validated for predicting underwater shock and bubble dynamics \cite{wardlaw1998coupled, wardlaw2000fluid}. Throughout this study, all performance metrics, including accuracy measurements and computational speed-up comparisons, are benchmarked against the DYSMAS hydrocode simulator.

\textcolor{black}{The product gas resulting from an underwater explosion exhibits pressure values exceeding $5.0\times 10^{11}$ Ba (over $7\times 10^5$ psi) \cite{cole1948underwater}. At these extreme pressures the gaseous bubble and surrounding water is subject to such large forces and strain rates that the effects of fluid viscosity and heat transfer are negligible. In this extreme environment the fluid dynamics obey the Euler equations which are obtained from the Navier-Stokes equations under the assumption of adiabatic and inviscid flow. The Euler equations have been widely used to simulate shock and bubble dynamics, successfully reproducing underwater explosion experimental results \cite{liu2018continuous, wardlaw1998coupled, wardlaw1998spherical, wardlaw2000fluid}. The Euler PDE system is written as}
\begin{subequations}
        \begin{alignat}{4}
        \frac{\partial}{\partial t} \rho_i + \frac{\partial }{\partial r} \left (\rho_i v_i \right)=& \ -\frac{2\rho_i v_i}{r} \\
        \frac{\partial}{\partial t} \left(\rho_i v_i \right) + \frac{\partial }{\partial r} \left (\rho_i v^2_i +p_i \right)=& \ -\frac{2\rho_i v^2_i}{r} \\
        \frac{\partial E_i}{\partial t} + \frac{\partial }{\partial r} \left[v_i \left( E_i + p_i \right) \right]=& \ -\frac{2}{r} \left[v_i\left(E_i + p_i\right)\right] \\
        E_i \ = \ \rho_i \left( \frac{1}{2} v_i^2 + e_i \right) \ \ &; \ \ i \in {w, g} \nonumber
    \end{alignat}
\label{eq:euler_eqs}
\end{subequations}

In eq. \ref{eq:euler_eqs}, $p_i, \rho_i, v_i, \text{and } e_i$ represent the fluid state variables pressure, mass density, velocity, and internal energy respectively. The index $i\in {w, g}$ indexes the materials present in the system, namely water ($w$) and HE gas ($g$). The hydrocode simulator uses an Eulerian scheme. For this multiphase scenario there will exist, at each timestep, a single mixed cell containing both water and HE material. Within this multiphase cell, both phases are enforced to be in pressure and velocity equilibrium. Note that the governing equations require an equation of state (EOS), unique for each material, as a closure to the system of equations. An equation of state provides the relationship between the thermodynamic state variables and can be expressed in the functional form $p=p(\rho, e)$.

Throughout this study we assume the water material properties are constant and governed by a fixed polynomial EOS (see appendix sec. \ref{app:syn_data_gen} for further details). However, the HE gas EOS is allowed to vary to represent the spectrum of thermodynamic properties exhibited by different HE materials. In particular, we represent the explosive equation of state using the Jones-Wilkins-Lee (JWL) EOS\cite{lee1968adiabatic} form, allowing the EOS parameters to vary. The JWL is among the most well-established EOS models for representing explosive thermodynamic properties of ideal HE materials \cite{osti_6530310} , it is commonly expressed as:
\begin{equation}
    p_g(\rho, e) = A\left[1- \frac{\omega \rho}{R_1 \rho_0}\right] \mathrm{e}^{-\frac{R_1 \rho_0 }{\rho}} + B\left[1- \frac{\omega \rho}{R_2 \rho_0}\right] \mathrm{e}^{-\frac{R_2 \rho_0 }{\rho}} + \omega \rho e \label{eq:jwl}
\end{equation}

Here $\rho_0$ is the (assumed fixed) reference density of the explosive material. The thermodynamic parameters, $A, B, R_1, R_2, \text{and } \omega$ are representative of a specific HE material. Hundreds of known explosives have an associated JWL equation of state which can be used to compute the detonation and/or fluid response characteristics for that specific HE material \cite{baker2010theory, osti_6530310}.  Given the validity of the JWL EOS to represent a large class of real world HE materials, in this study we assume that that any HE material, real or notional, can represented through an appropriate choice of the JWL EOS parameters, $\Theta = \{A, B, R_1, R_2, \omega\}$. \textcolor{black}{JWL parameters exhibit significant variability in the literature, with the same HE material often having different calibrated values due to manufacturing variations and experimental methodologies \cite{osti_6530310, giam2020numerical}. Rather than attempting to cover all possible parameter combinations, we selected a domain that is broadly representative of parameter variations exhibited by known ideal explosives. In particular, our domain is inspired by TNT JWL values developed for underwater studies \cite{wardlaw1998spherical}. Furthermore, the bounds defined in Table \ref{tab:jwl_params} capture the majority of mean values reported in \cite{osti_6530310}, with the range width approximately corresponding to one standard deviation of the published parameters. This approach ensures our parameter space is physically realistic while remaining computationally tractable.}

\begin{table*}
\renewcommand{\arraystretch}{1.5}
\caption{\label{tab:jwl_params} JWL EOS Parameter Bounds for Energetic System}
\begin{ruledtabular}
\begin{tabular}{l*{7}{r}}
   & $e_0$ [erg] & $\rho_0 $ [g/cc] & $A$ [Ba] & $B$ [Ba] &  $R_1$ [-] &$R_2$ [-] &$\omega$ [-]\\ \hline
Lower Bounds& 5.33$\times10^{10}$ & 1.65 & 4.5$\times10^{12}$ & 7.5$\times10^{10}$ &  4.0 & 0.90 & 0.25 \\ 
Upper Bounds & 5.33$\times10^{10}$ & 1.65 & 6.0$\times10^{12}$ & 9.5$\times10^{10}$ &  5.5 & 1.15 & 0.35 \\
\end{tabular}
\end{ruledtabular}
\end{table*}

\textcolor{black}{The stages of the underwater explosion phenomenon are depicted in Fig. 1.} The detonation phase of the simulation applies the standard \textit{Chapman-Jouguet} detonation model. Specifically, a Chapman-Jouguet detonation wave is initiated at the explosive center and propagates radially outward, releasing stored chemical energy instantaneously at the shock front as it sweeps through the material to the HE boundary surface \cite{jones1998numerical} (see Fig. \ref{fig:undex_cartoon}). Within the Chapman-Jouguet model, the state of the detonation front is uniquely determined by combining the conservation laws of mass, momentum, and energy, the equation of state of the reacted HE material, $p_g(\rho, e)$, and enforcing equality between the shock velocity and the sum of material and sound velocity. Upon reaching the HE-water interface the detonation phase ends and the fluid dynamic phase, including the material interface and shock front dynamics, begins. Thus, once a set of HE thermodynamic parameters are chosen, the system's HE gas EOS is defined (eq. \ref{eq:jwl}) and the initial conditions of the multiphase fluid dynamic system are uniquely determined through the \textit{Chapman-Jouguet} detonation model.

\begin{figure*}[ht!]
\includegraphics[width=0.95\textwidth]{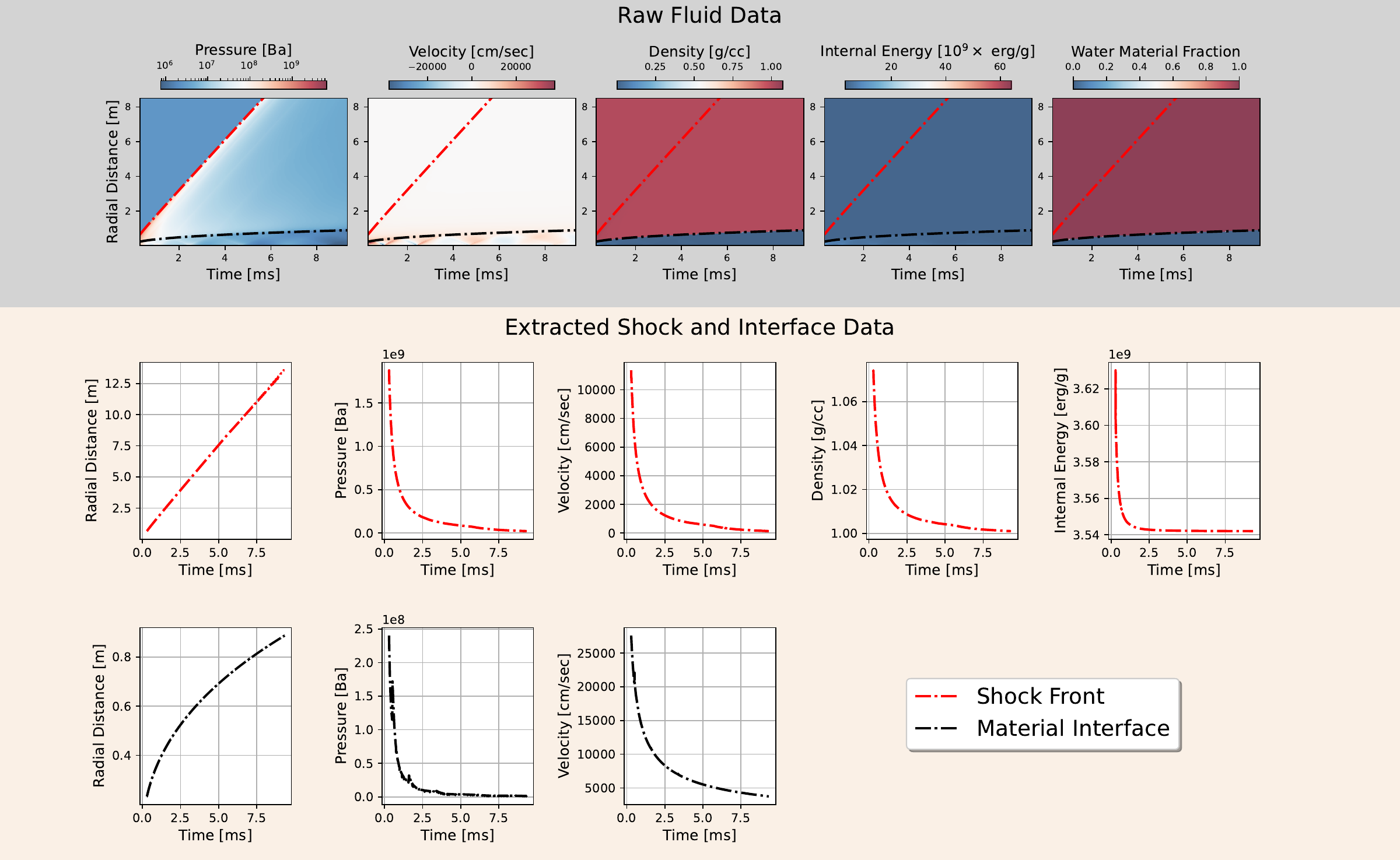}
\caption{\label{fig:sim_field}\textcolor{black}{(Top) Field plots of fluid state variables from a simulation, with dotted lines marking the material interface and primary shock. (Bottom) Extracted fluid state variables at the material interface and shock front.}}
\end{figure*}

The fluid dynamics phase of the simulation produces spatiotemporal data of the multiphase system. The shock front and material interface fluid variable data are subsequently extracted from the full field data. The shock front and material interface locations at each time step correspond to the location of the computational cell with the the maximum water pressure and multiphase material respectively. Thus, we isolate and extract the state variable time histories corresponding to the shock front and bubble interface. In Fig. \ref{fig:sim_field}, we show the  field data, as well as the isolated and extracted shock front and material interface data for a randomly sampled JWL parameter set. A complete description of the simulation inputs, mesh discretization, equation of state forms, detonation model, and numerical algorithm is provided in the supplementary materials sec. \ref{app:syn_data_gen}. 

Inspection of the temporal and spatial variation of the field data reveals several salient features to all HE materials considered in this study. Namely, there is a primary shock front propagating radially outward trailed by a more slowly expanding gas-water (i.e. {\it bubble}) material interface. The density within the gaseous bubble is orders of magnitude less than in the water. The pressure and material velocity fields exhibit reverberating shock and rarefaction waves within the gaseous bubble composed of HE gas. These waves ultimately influence the fluid state at the material interface. Furthermore, interaction of the reflected shock waves with the gaseous bubble generate subsequent shocks within the water that eventually overtake the direct shock. Recall that while the focus of this work is on underwater HE systems, shock-interface dynamics present in these systems also appear in a wider array of multi-material high pressure fluid systems. 

The extracted dynamics represents the targets for our supervised learning task. Specifically, for each input feature matrix $\bm{X}^i = \bm{t}^i \otimes \Theta^i \in \mathbb{R}^{N_t\times5}$, composed of the tensor product of the simulation time vector and HE JWL parameter vector, the corresponding target matrices are $\bm{Y}_s^i=[\bm{x}^i_s, \bm{p}^i_s, \bm{u}^i_s, \bm{\rho}^i_s, \bm{e}_m^i] \in \mathbb{R}^{N_t \times 5}$ and $\bm{Y}_{mi}^i=[\bm{x}^i_{mi}, \bm{p}^i_{mi}, \bm{u}^i_{mi}] \in \mathbb{R}^{N_t \times 3}$. The shock and material interface data are denoted by the subscripts $s$ and $mi$ respectively. Note that there are fewer target fluid variables for the material interface data as the internal energy and density are multi-valued at the material interface and are therefore not included in the prediction task. The time vectors of all simulations are uniformly clipped to maintain a constant dimensionality of $N_t=4349$. With the system definition, simulation procedure, and data extraction detailed, we next describe the data generation process. 

\subsection{\label{sec:data_gen}Data Generation}

\begin{figure*}[ht!]
\includegraphics[width=0.6\textwidth]{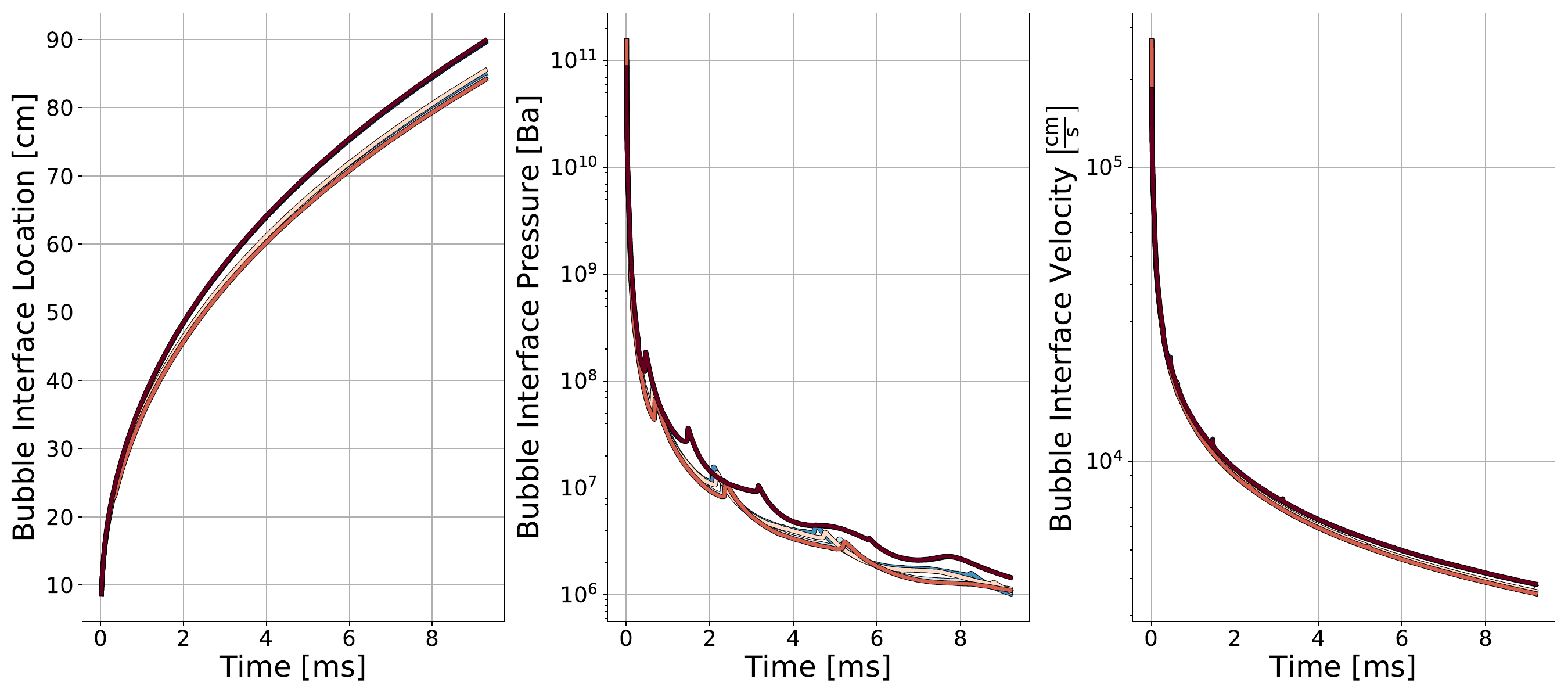}\hfill
\\
\includegraphics[width=\textwidth]{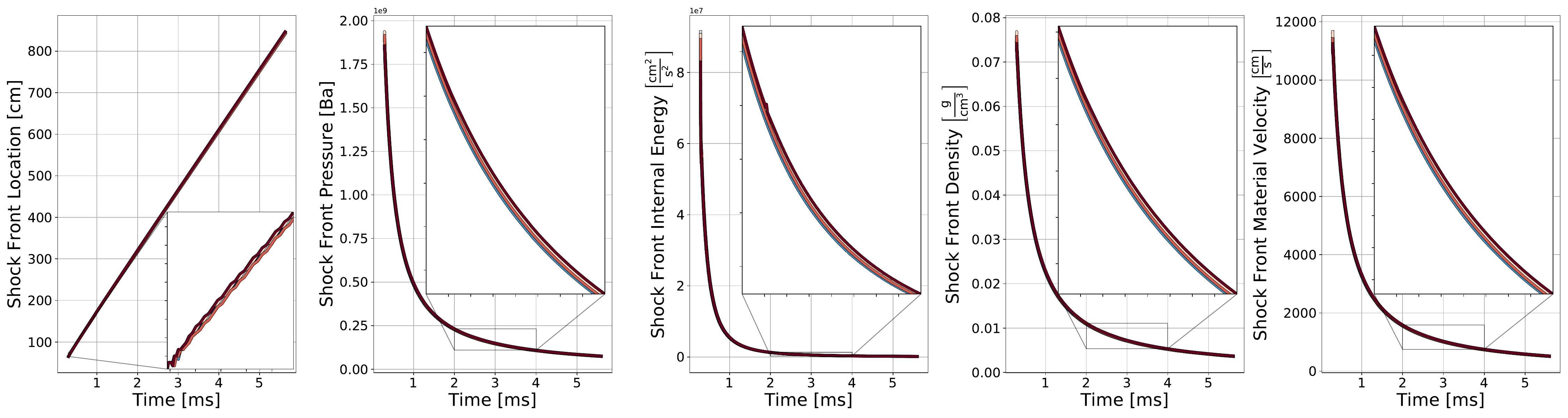}
\caption{\label{fig:fronts} \textcolor{black}{The state variable dynamics at the material interface (top) and shock (bottom) front for five randomly sampled HE materials. Portions of the temporal domain for the shock front have been magnified to distinguish visually similar dynamics.}}
\end{figure*}

A dataset of shock and material interface dynamics is generated by sampling, and subsequently simulating, $N_{JWL}=1,000$ JWL parameter sets. Latin hypercube sampling is used to uniformly sample 1000 points within the JWL EOS parameter space with bounds indicated in table \ref{tab:jwl_params}. Recall, that we sample from a thermodynamic parameter space representative of the majority of ideal HE materials\cite{osti_6530310}. For each JWL parameter set, $\Theta^i$, the dynamics are simulated and extracted as detailed in sec. \ref{sec:system_def}. The entire supervised learning dataset is composed of the input feature and target dynamics datasets; $\{ \bm{X}^i = \bm{t}^i \otimes \Theta^i : i=1,\dots,N_{JWL} \}$, $\{ \bm{Y}_s^i : i=1,\dots,N_{JWL} \}$, , $\{ \bm{Y}_{mi}^i : i=1,\dots,N_{JWL} \}$. Extracted shock and bubble front dynamics for five of the sampled JWL parameter sets are shown in Fig. \ref{fig:fronts}. Note that several of the state variables at the shock and bubble fronts exhibit extremely strong temporal decay properties, e.g. note super-exponential decay present exhibited by the material interface pressure. This decay property, combined with similarity in the dynamics between differing HE materials, presents a challenging learning task. 

\subsection{\label{subsec:ml_methods}Data Processing, Model Training, and Model Selection}

\begin{figure}[ht!]
\includegraphics[width=0.45\textwidth]{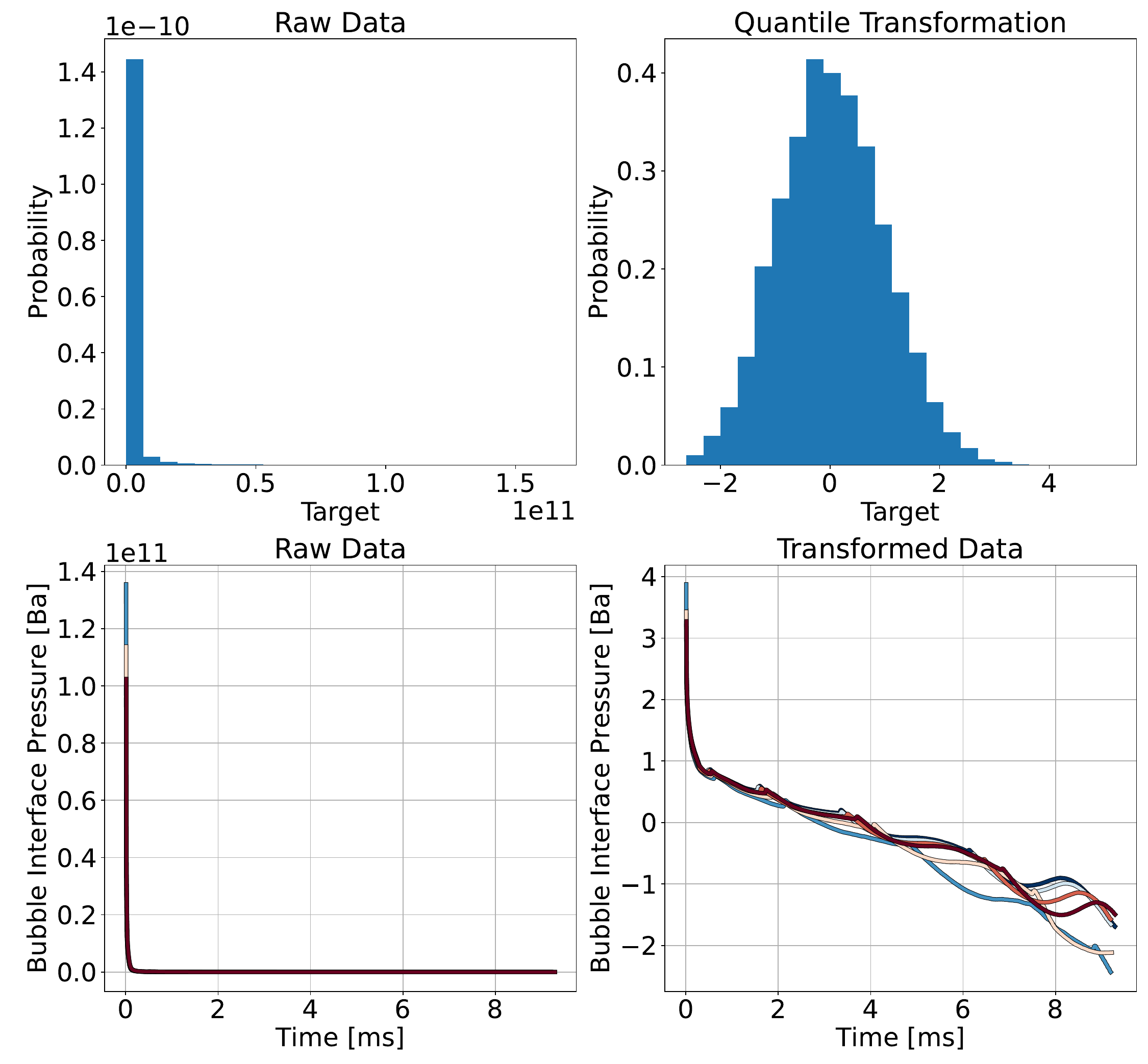}\hfill
\caption{\label{fig:transformer} \textcolor{black}{(Top Row) Distribution of the material interface pressure values before and after the Quantile Transformer is applied. (Bottom Row) Effect of the transformation for five randomly selected energetic materials.}}
\end{figure}

The extracted dynamics data is processed prior to ML model training. The input feature data are individually min-max scaled such that they range from zero to one. The target variables were found to require a more complex scaling procedure to achieve high-accuracy predictions. Specifically, the Quantile Transformer from the SciKit-Learn library \cite{pedregosa2011scikit} was used, transforming each target variable independently and mapping into a normal distribution. We hypothesize that the need to adopt a non-linear scaling technique arises from two sources. First, the relative similarity in the system dynamics, especially evident for the shock front variables (see Fig. \ref{fig:fronts}), across large portions of the thermodynamic parameter space. Second, many fluid variable dynamics exhibit a super-exponential decay (e.g. see the material interface pressure ploted in Fig. \ref{fig:transformer}), requiring strong scaling. An example of the mapping induced by the Quantile Transformer on the material interface pressure variable is given in Fig. \ref{fig:transformer}. The supervised learning datasets are further partitioned into train, validate, and test sets using an 80:10:10 split. Note that this data splitting approach is applied {\it only} to the thermodynamic parameter feature sets characterizing the HE material space. Thus validation and testing results indicate how well the model generalizes to HE materials unseen during training. Under this splitting approach, $N_{train}$=800 JWL parameter set system dynamics are used for model training while two unique sets of dynamics, each containing to 100 unseen JWL parameters sets, are reserved for validation ($N_{val}$) and testing ($N_{test}$).

\begin{figure}[ht!]
\includegraphics[trim={4cm 0cm 2cm 0cm},clip,width=0.45\textwidth]{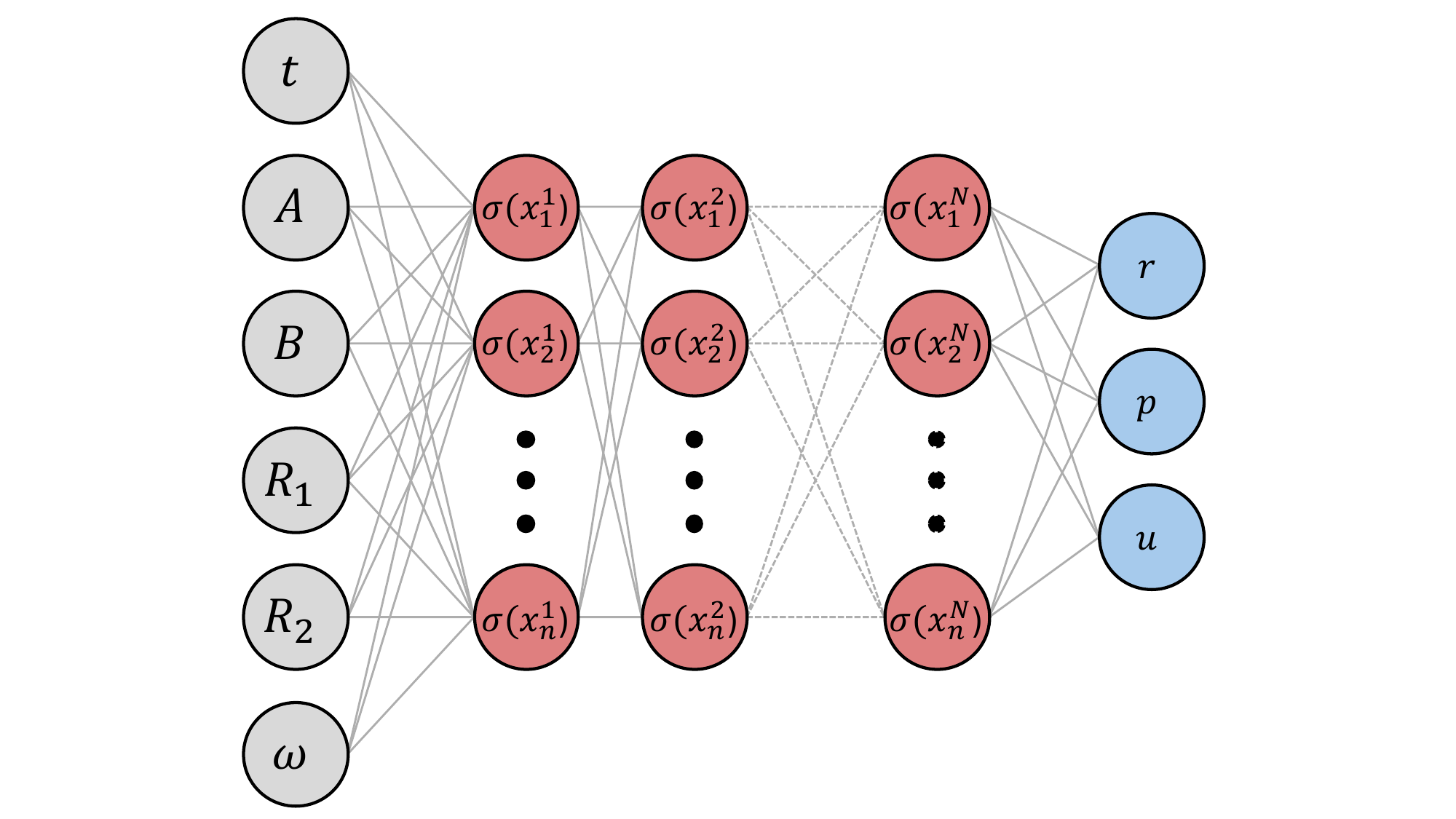}\hfill
\caption{\label{fig:interface_dnn}DNN architecture for the material interface dynamics surrogate.}
\end{figure}

A fully-connected deep neural network (DNN) is used to approximate the operator governing the shock and material interface dynamics. While a number of alternative statistical approaches were explored, including Gaussian process regression and kernel ridge regression (see supplementary materials Sec. \ref{app:ml_comparision} for ML model comparison), the neural network architecture was selected for further development based on its baseline accuracy, scaling capabilities to larger datasets, and differentiable capabilities amenable to inverse design-type problems. Two multi-output DNN models were trained, one for prediction of the material interface dynamics and another for the shock front dynamics. The shock and bubble DNN models are trained to minimize a composite RMSE loss quantifying prediction accuracy of the shock and material interface state variables. The loss for the material interface and shock front is given in eqs. \ref{subeq:1} and \ref{subeq:2} respectively.

\begin{subequations}
\begin{equation}
\mathcal{L}_{mat. int.}(\theta_{mi}) = \sum_{i=1}^{N_{train}} \sum_{j=1}^{N_{var}=3}  \norm{\mathit{NN}_{mi}^{j} (\bm{X}^i; \theta_{mi}) - \bm{Y}_{mi}^{ij}}\label{subeq:1}
\end{equation}
\begin{equation}
\mathcal{L}_{shock}(\theta_s) = \sum_{i=1}^{N_{train}} \sum_{j=1}^{N_{var}=5}  \norm{\mathit{NN}_{s}^{j} (\bm{X}^i; \theta_s) - \bm{Y}_{s}^{ij}}\label{subeq:2}
\end{equation}
\end{subequations}
Here $\mathit{NN}_{mi}^{j} (\bm{X}^i; \theta_{mi})$ and $\mathit{NN}_{s}^{j} (\bm{X}^i; \theta_s)$ represents the DNN prediction for the $j^{th}$ fluid state variables at the material interface and shock front respectively, for the JWL parameter set and time vector encapsulated in $\bm{X}^i = \bm{t}^i \otimes \Theta^i$. Furthermore, $\theta_{mi}$ and $\theta_s$ are the trainable material interface and shock DNN parameters respectively. The architecture of the material interface DNN model is shown in \ref{fig:interface_dnn}. Note that the inputs to the model are a fusing of temporal, $\{t\}$, and thermodynamic, $\{A, B, R_1, R_2, \omega\}$, features. Furthermore, by construction time is an input feature and therefore our model is not autoregressive in nature allowing predictions at any point in time without the need for model rollout.

The shock and material interface DNNs are constructed and trained using the PyTorch machine learning library \cite{paszke2019pytorch}. Full details of the the model architecture, activation functions, and model selection criteria metrics are given in the supplementary materials Sec. \ref{app:nn_training}. The model selection and performance evaluation is performed using the validation and test data sets respectively. Specifically, we select the model at the epoch which maximizes the number of predictions that fall within $\pm .25\%$ of the reference data (see supplementary materials Sec. \ref{app:nn_training} for equation details). The selected model is subsequently evaluated on the test set across a range of well-established error metrics, including root mean-squared error (RMSE), coefficient of determination ($R^2$), and mean absolute percent error (MAPE). Model accuracy results are given in Sec. \ref{sec:results}.

\subsection{\label{subsec:feature_sensitivity}ML Model Feature Sensitivity}

Feature importance is a machine learning technique which provides model interpretability through quantifying the degree to which each input feature influences the model predictions. For our fluid system, feature importance provides fundamental physical insights, namely which thermodynamic input features are the most significant drivers of the resulting system fluid dynamics. We utilize the Feature Ablation (FA) analysis technique (provided through the Captum library \cite{kokhlikyan2020captum}) to assess feature importance. Feature ablation involves perturbing a single input feature dimension while keeping all other dimensions fixed. The resulting shift in model outputs provides information on the strength of the individual feature influence, with more influential features causing a larger magnitude shift in the outputs. Furthermore, a positive (negative) feature ablation value indicates that increasing a feature results in an increase (or decrease for negative values) in the model output. We make one key alteration to the standard FA analysis, which is usually applied to all input feature dimensions. In contrast, we perturb the feature dimensions corresponding to the thermodynamic parameters only, while always keeping the time dimension fixed. Removal of time in the feature importance analysis is motivated from an engineering design perspective. While time is a key driver of underwater energetic system dynamics, it is not a variable over which one can control or tune. In contrast, one can alter the thermodynamic properties and associated the JWL parameter values by controlling HE material composition. Therefore it is in this system design context, that we isolate and study the relationship between system thermodynamics and hydrodynamics through feature importance.

The functional expression for calculating FA values for any output of the shock or bubble dynamics surrogate, $\mathit{NN}^{j}(\mathbf{t}, \Theta; \theta)$, is given below in Eq. \ref{eq:fa}. Note that by excluding time from our feature importance analysis renders the feature ablation value, $AV(t)$, a function of time. 

\begin{equation}
    \label{eq:fa}
    \begin{split}
        AV_i^j(\mathbf{t}) =& \norm{\sum_{\Theta \in \mathcal{T}} (\mathit{NN}^{j}(\mathbf{t}, \Theta; \theta)-\mathit{NN}^{j}(\mathbf{t}, \Theta_{-i}; \theta))}_1 \\
        i=&{A, B, R_1, R_2, \omega} \ \ ; \ \ j=1, \dots, N_{vars}
    \end{split}
\end{equation}

Here $\Theta$ is the unperturbed JWL input parameter set and $\Theta_{-i}$ is the modified input feature with the $i^{th}$ JWL parameter set to a baseline value of 0 (the minimum value under minmax scaling). The trained material interface and shock DNN models, jointly represented here by $\mathit{NN}^{j}$, are used. The summation is taken over a dense uniform sampling (denoted by set $\mathcal{T}$) of 526,700 points within the HE thermodynamic parameter space. We calculate the FA values at 100 uniformly spaced points in the time domain of interest. An ablation value is calculated for all shock and bubble DNN outputs and the results presented and analyzed in sec. \ref{sec:feature_importance}.

\subsection{\label{subsec:param_identification}Gradient-Based Parameter Identification}

The trained DNN surrogates are fully differentiable with respect to their input features, distinguishing these surrogates from traditional hydrodynamics solvers which are not differentiable. Thus our neural surrogates are particularly useful within an inverse design paradigm, utilizing gradient information to iteratively update inputs until a desired output is reached. In particular, our focus is on the canonical problem of parameter discovery for unknown dynamics. Parameter discovery involves dynamics information (i.e. time series) for which the HE material generating those dynamics is unknown. The objective is to then identify which HE material produced the observed fluid dynamic response. Parameter discovery can be formulated as a gradient-based minimization of the residual between DNN model outputs and reference dynamics data. 

Parameter discovery, or material model {\it calibration}, is a common phase during the development and characterization of new materials, including HE materials \cite{osti_6530310}. The most common approach to the calibration challenge involves iteratively updating the HE material parameters until hydrocode calculations match experimental data, using gradient-free optimization techniques such as genetic algorithms, Bayesian techniques, and simplex algorithms \cite{li2024automatic, walters2018bayesian, mortensen2017optimizing}. In contrast, differentiable surrogates will enable parameter discovery via gradient descent approaches which reach optima in far less forward solves and scale better to high dimensional optimization problems.

The approach used to perform parameter discovery is detailed in algorithm \ref{alg:param_discovery}. The parameter discovery framework is tested by taking a randomly sampled HE material (not present in the training data set), generating the dynamics using the hydrodynamics simulator, presenting only the dynamics to the inverse design platform (while withholding the associated JWL parameters) for parameter discovery, and calculating the parameter discovery accuracy. The robustness of the method is evaluated by performing this testing process using 100 randomly sampled HE materials generated by Latin hypercube sampling. \textcolor{black}{Our parameter discovery procedure implements robust safeguards against non-convergence through an adaptive restart strategy: when gradient updates push parameters outside the valid domain, we resample parameters uniformly within bounds, reduce the learning rate by 50\%, and restart optimization with these more conservative conditions. This approach creates a sequence of progressively more conservative optimizations that empirically converged to valid parameter sets in all 100 test cases.} The results and analysis of the inverse design methodology are presented in sec. \ref{sec:inv_design}.

\begin{algorithm}[ht]
 \KwData{A time series of bubble or shock dynamics data $\{(t_i, \hat{y}_i)\}_{i=1}^{N_t}$ produced by a ground-truth, but unknown, JWL parameter set $\Theta_{gt}$. A randomly sampled initial guess for the parameter values $\Theta_{n=0} \sim \mathcal{U}^5$(0,1), sampled from the normalized domain.\\
 Set the number of iterations ($N_{iter}$), and learning rate ($\gamma$). Initialize iteration counter $n=0$}
 \KwResult{Recover numerical approximation of $\Theta_{gt}$\;}
  \While{$n < N_{iter}$}{
  Calculate temporally averaged loss of a differentiable surrogate model $\mathit{NN}(\mathbf{t}, \Theta; \theta)$:\\ $\mathcal{L}_{param}(\Theta_n) =  \frac{1}{N_t} \sum_i^{N_t} \left( \mathit{NN}(t_i, \Theta; \theta) - \hat{y_i} \right)^2$ \\[.25cm]
  Use Automatic Differentiation to calculate the temporally averaged loss gradient:\\ $\nabla_{\Theta} \mathcal{L}_{param}(\Theta)$ \\[.25cm]
  Perform gradient-based update of parameter values:\\ $\Theta_{n+1} = \Theta_{n} - \gamma*\nabla_{\Theta} \mathcal{L}_{param}(\Theta_{n})$\\[.25cm]
  Iterate counter: $n = n + 1$
  Constrain update within domain bounds:\\
 \If{$\Theta_{n+1} \notin [0,1]^5$}{
 Restart iteration count: $n = 0$ \\
  Reinitialize the parameter values within the domain by resampling the parameter values:  $\Theta_{n=0} \sim \mathcal{U}^5$(0,1) \\ [.25cm]
 Reduce the learning rate: $\gamma$ = $\gamma$*0.5 \\[.25cm]
 }

 }
 \caption{Algorithm for performing gradient-based parameter identification.}
 \label{alg:param_discovery}
\end{algorithm}

\section{\label{sec:results}ML Model Performance}

The fluid variable prediction accuracy metrics (see Sec.\ref{subsec:ml_methods}) of the trained shock and material interface models are presented in Table \ref{tab:model_acc}. The accuracy values remain approximately equal across train, validate, and testing splits indicating our models are not overfit. In addition to the RMSE, which was directly minimized during training, $R^2$ and MAPE metric are also reported to provide additional intuition of model accuracy. Both of these metrics indicate that our DNN models produce highly accurate predictions of HE-induced shock and material interface dynamics. Namely, at three significant digits the $R^2$ values are identical to 1, while the MAPE varies between $\sim .7 - 5\times 10^{-3}$\%. \textcolor{black}{Deep neural networks are susceptible to convergence toward local optima due to the non-convex nature of their loss landscapes. To mitigate this challenge, we implemented several optimization strategies: (1) we utilized the Adam optimizer with a learning rate scheduler that reduces the rate upon plateauing of the optimization loss, (2) we employed Xavier initialization to establish appropriate initial weight distributions \cite{glorot2010understanding}, and (3) we conducted multiple training runs with different random initializations. These repeated training procedures yielded consistent performance metrics. While convergence to the global optimum cannot be guaranteed for neural networks, the consistency of our results across multiple initializations, combined with the exceptionally low error metrics achieved (MAPE < $5×10^{-3}$\%), provides strong evidence that our surrogate models accurately and robustly capture the underlying fluid dynamics regardless of initialization conditions.}

\begin{figure*}[ht!]
\includegraphics[width=.3\textwidth]{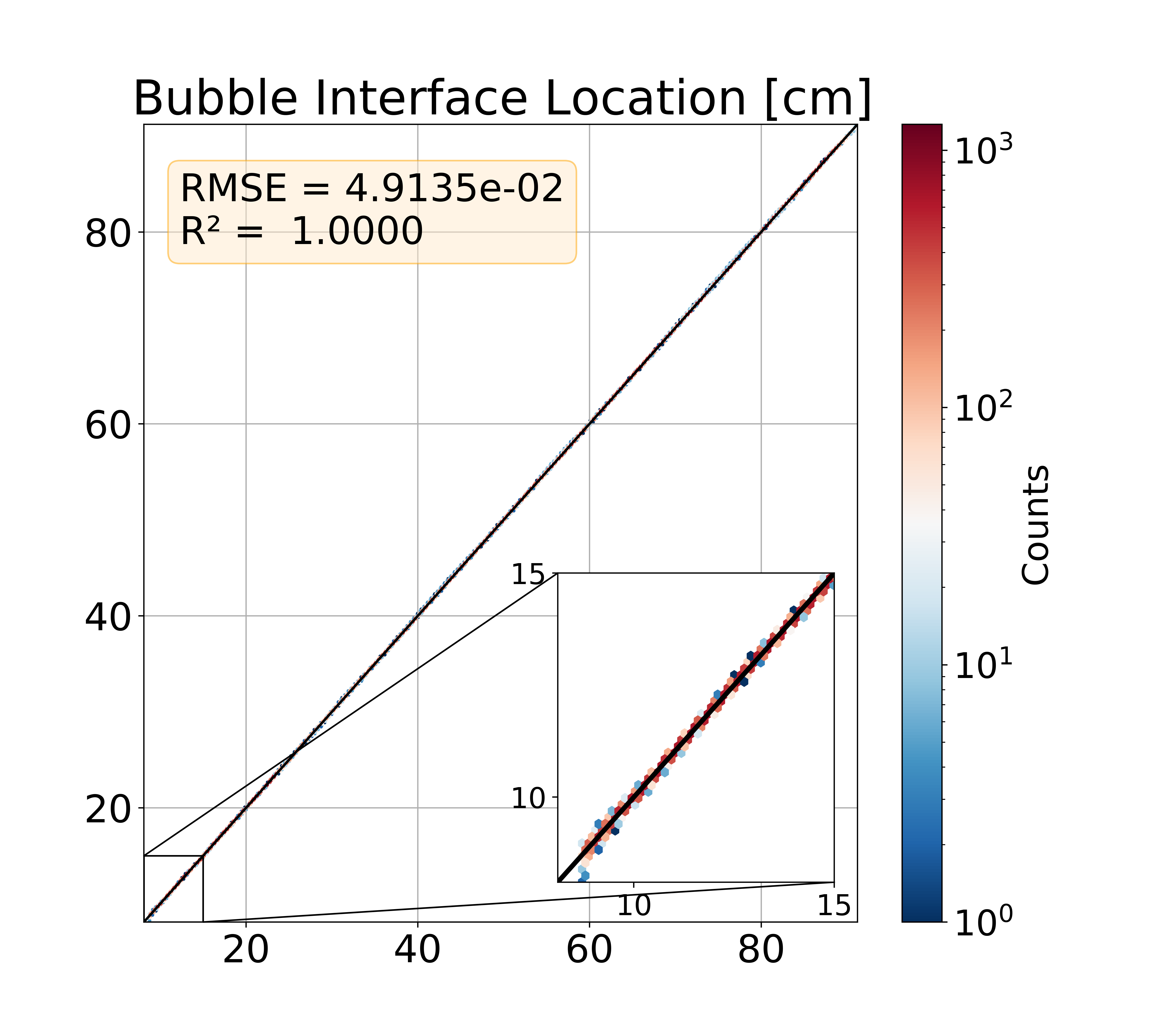}
\includegraphics[width=.3\textwidth]{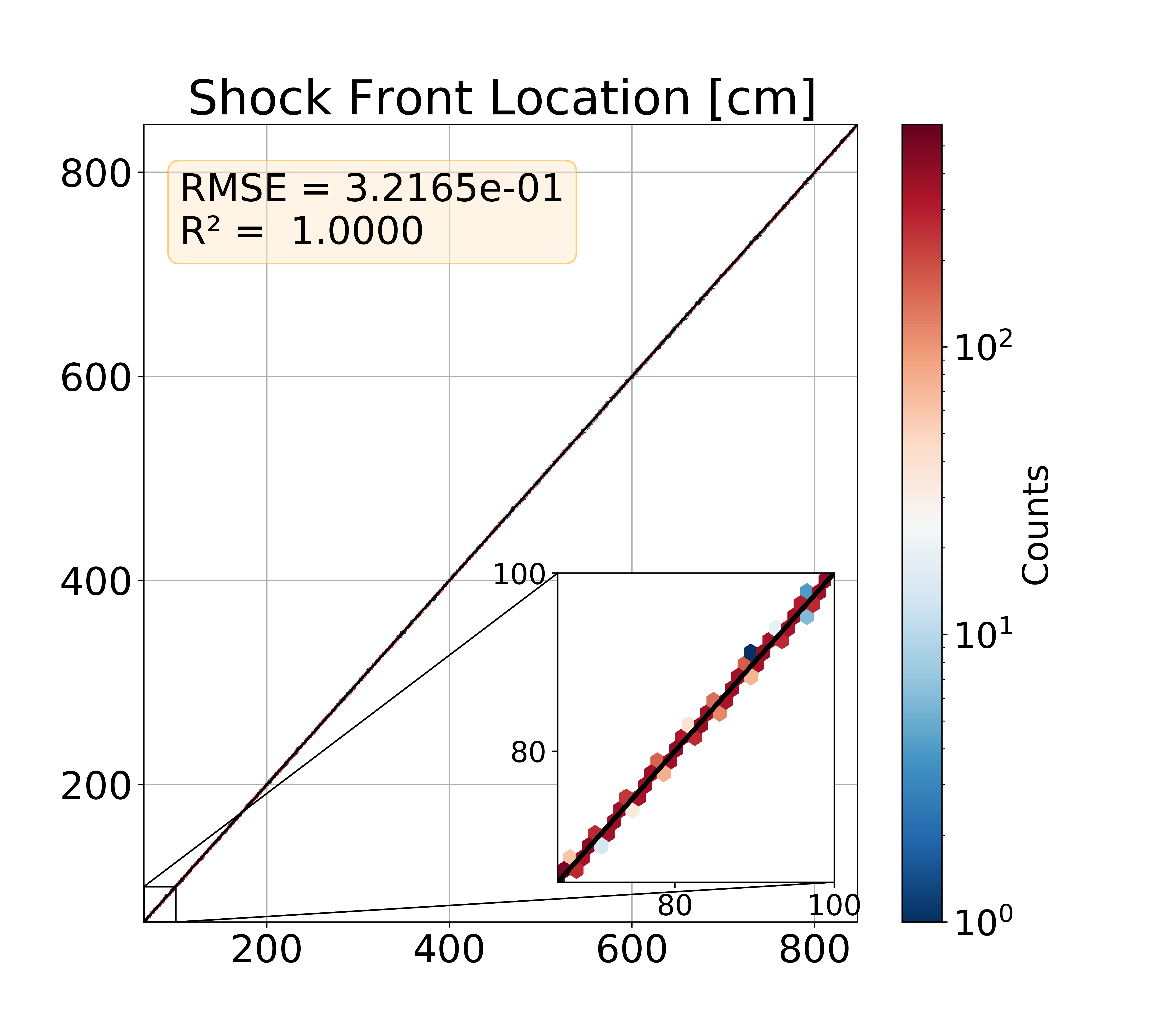}
\includegraphics[width=.3\textwidth]{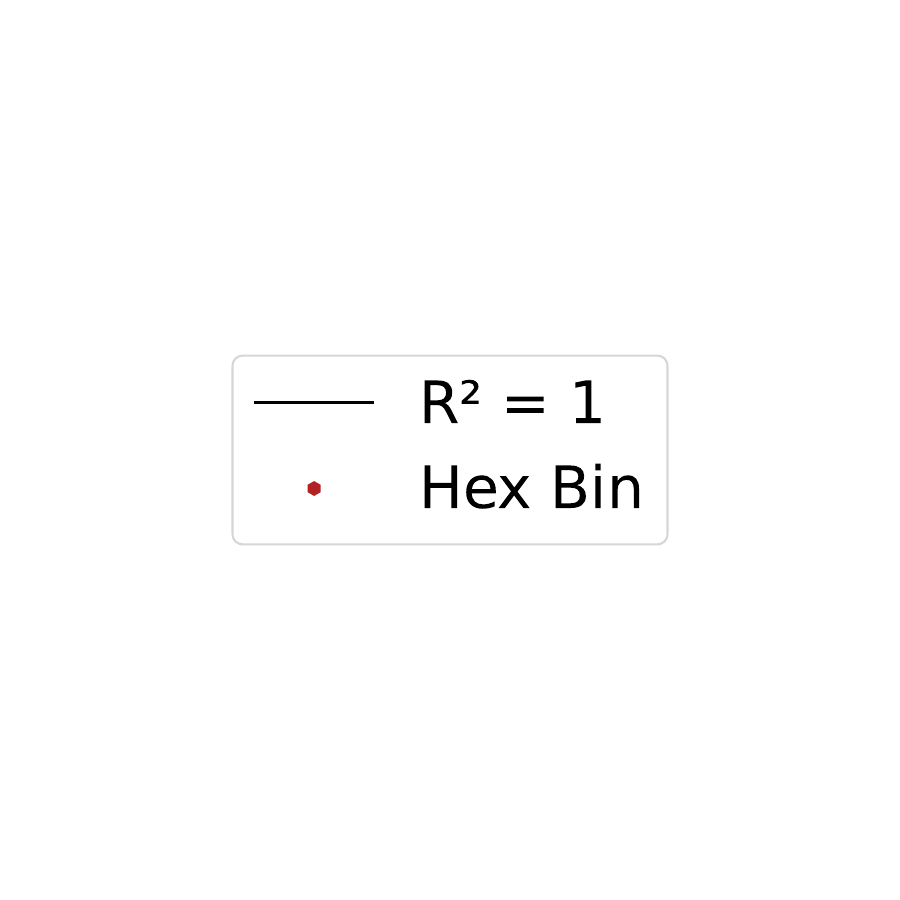}\\
\includegraphics[width=.3\textwidth]{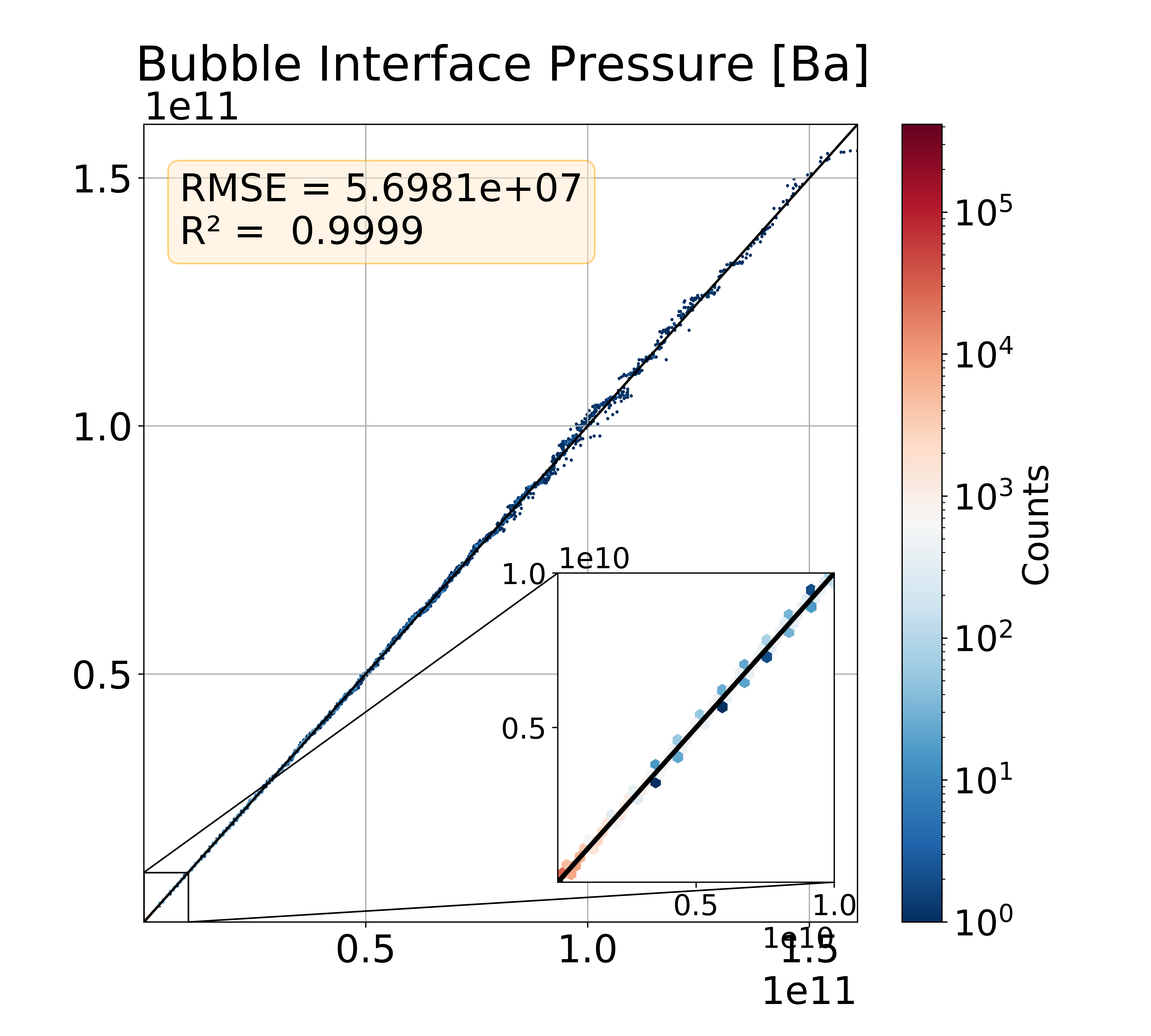}
\includegraphics[width=.3\textwidth]{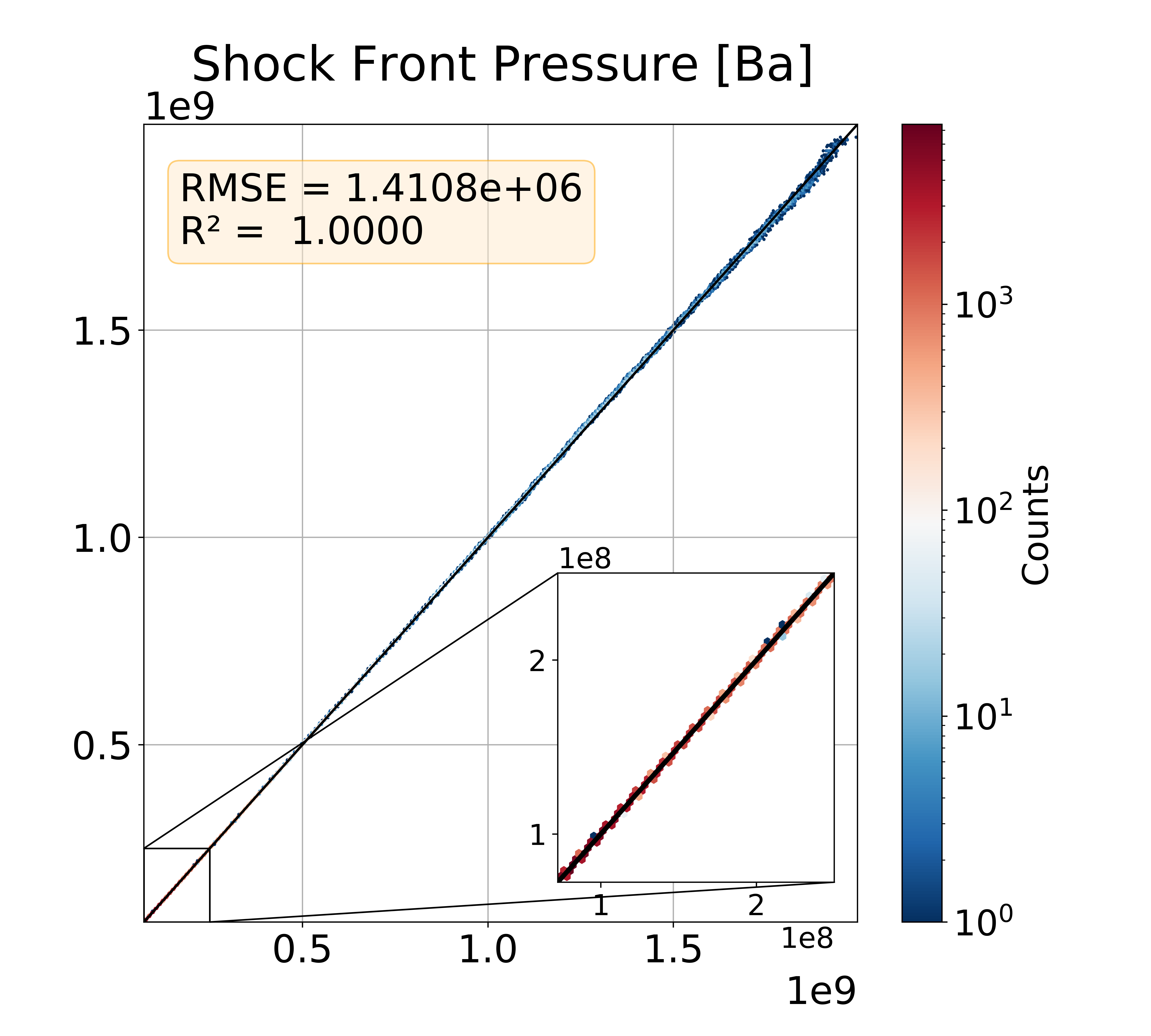}
\includegraphics[width=.3\textwidth]{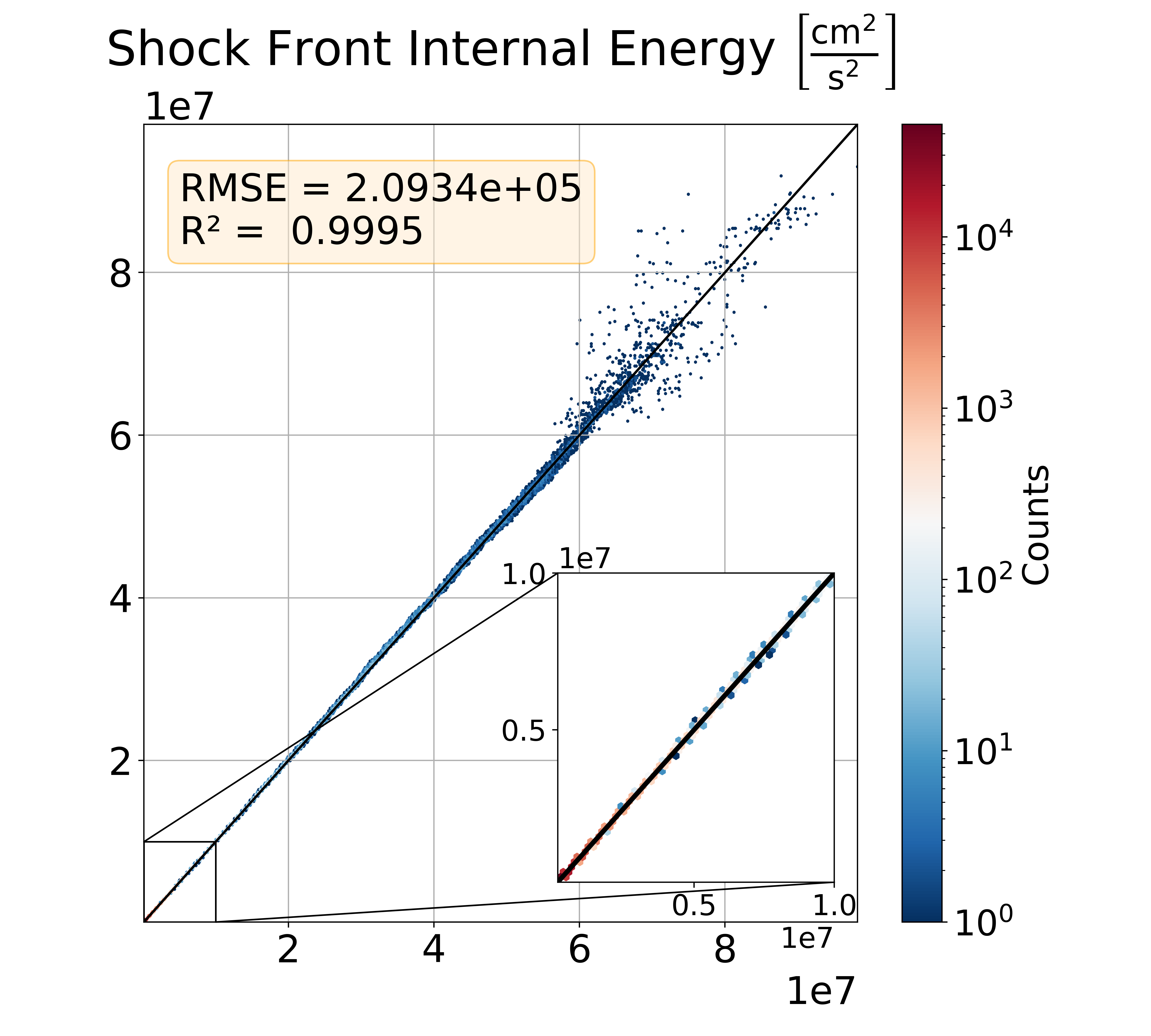}\\
\includegraphics[width=.3\textwidth]{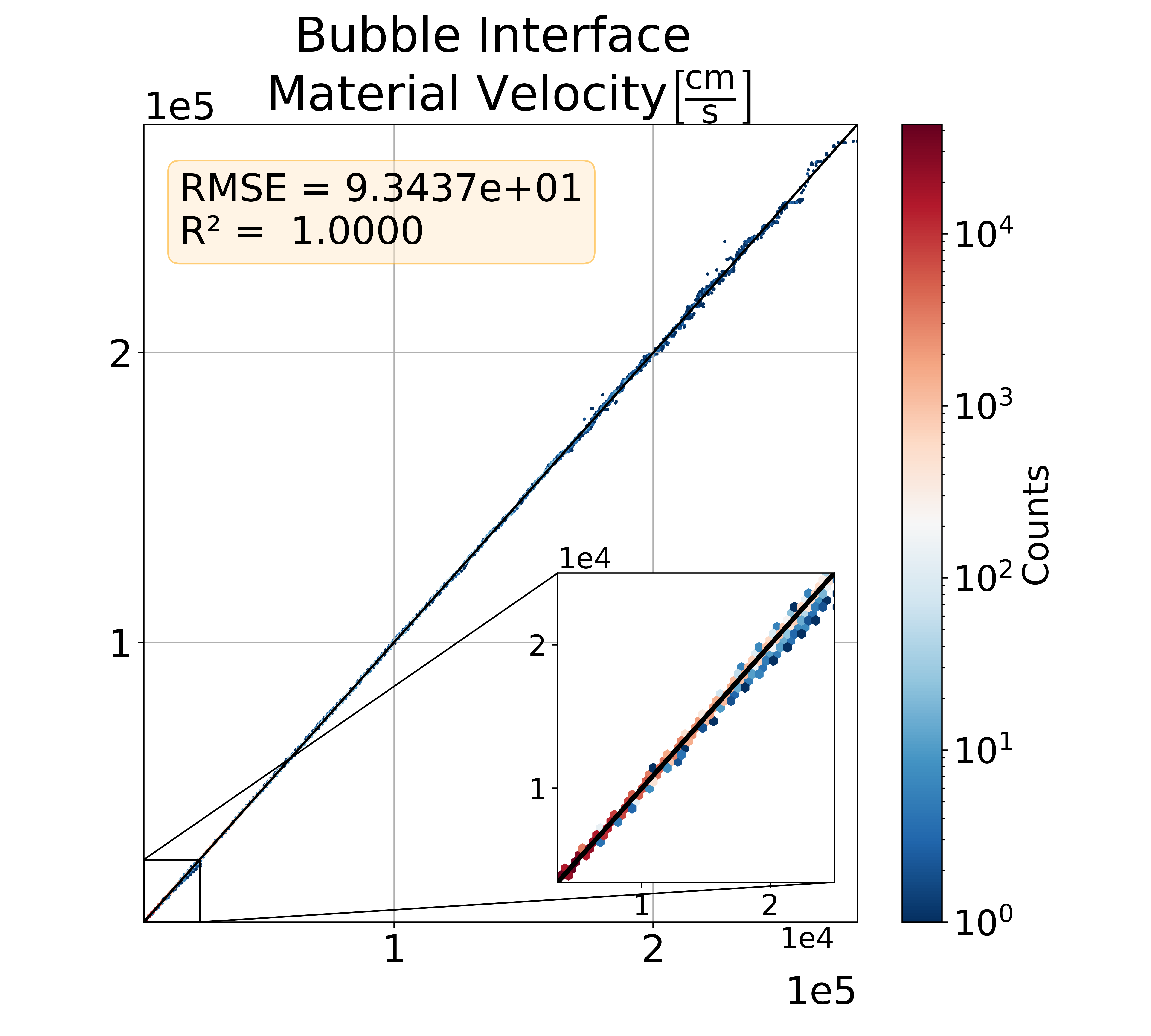}
\includegraphics[width=.3\textwidth]{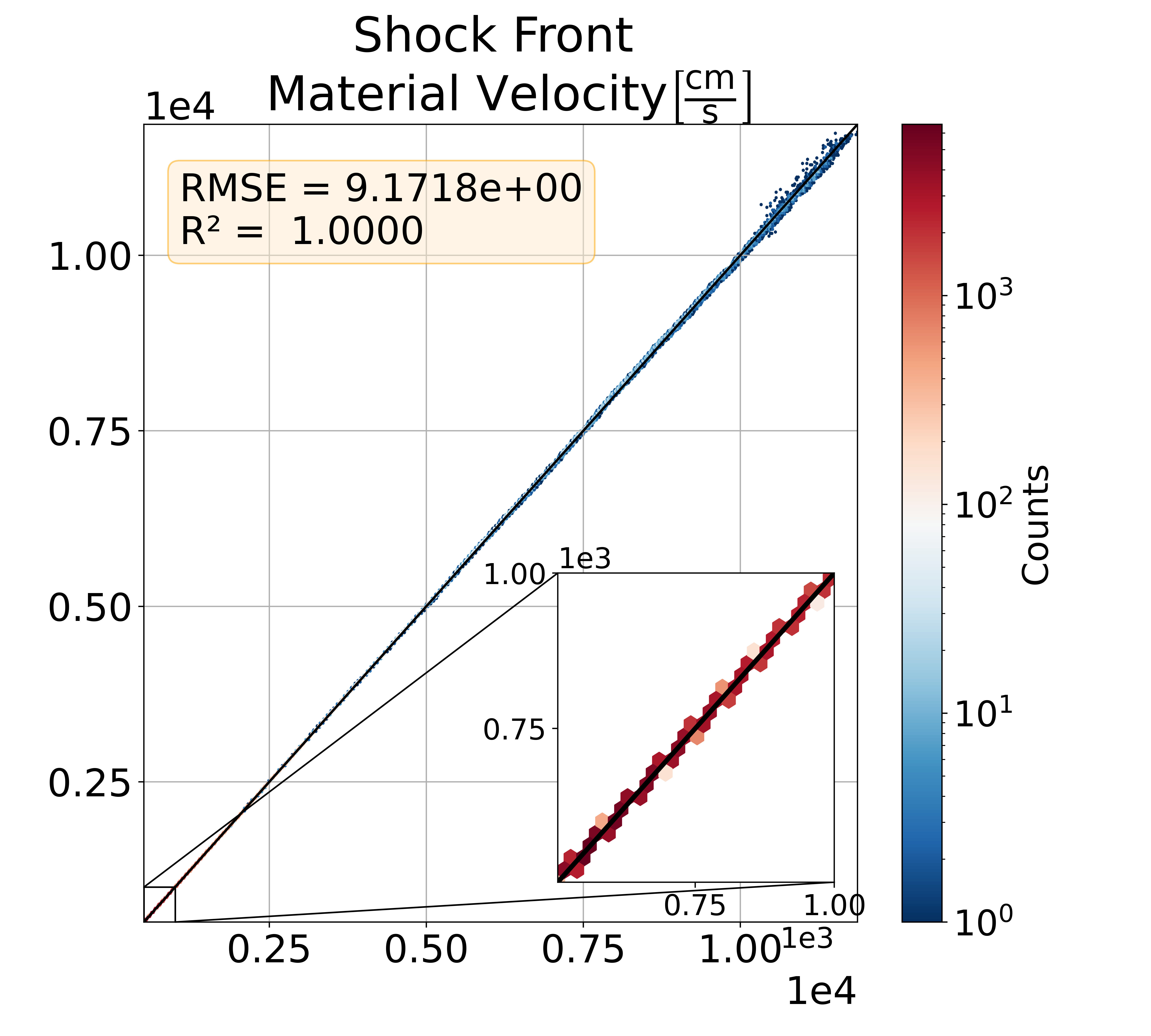}
\includegraphics[width=.3\textwidth]{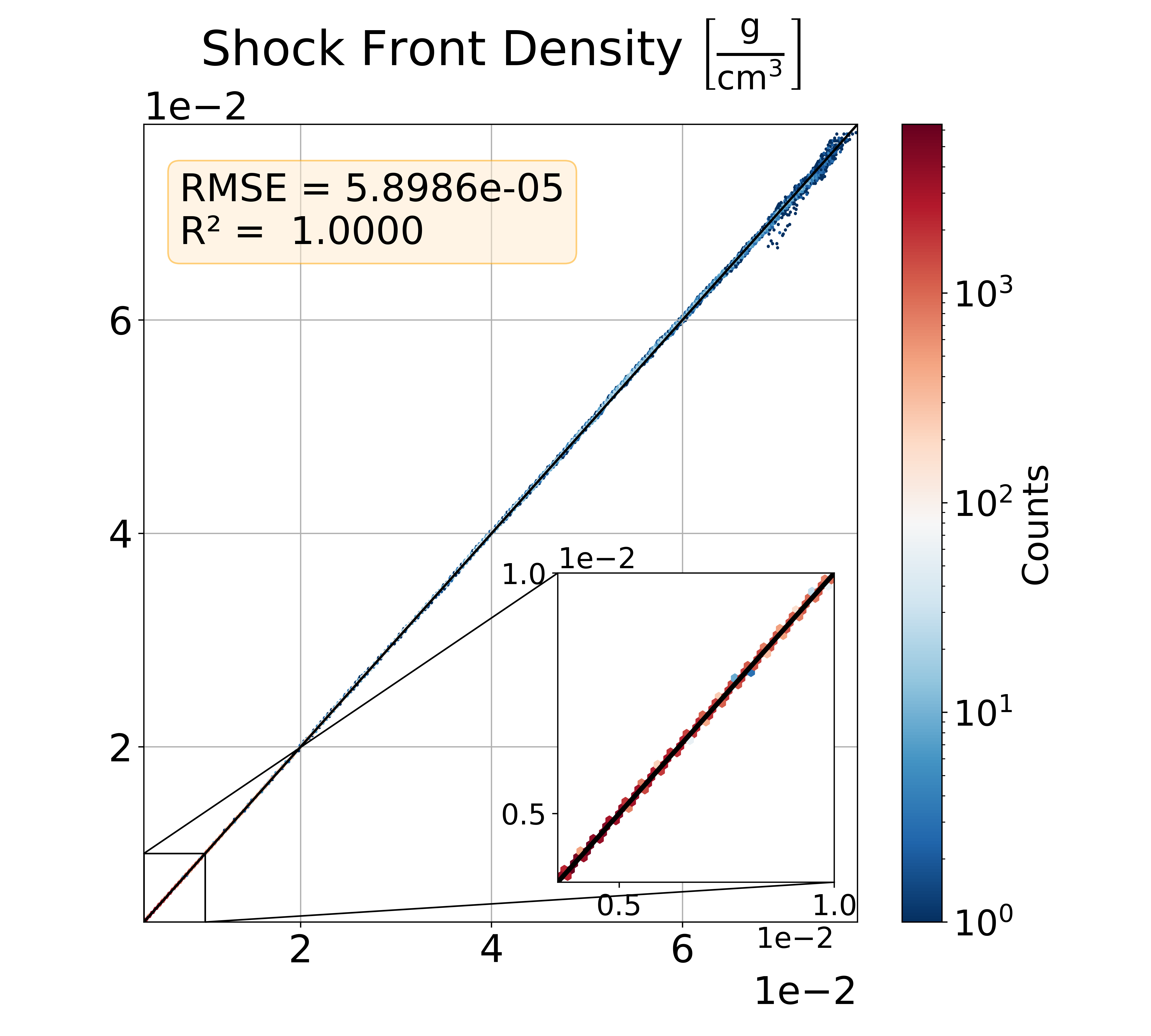}
\caption{\label{fig:parity_plots} \textcolor{black}{Parity plot comparison of test set prediction of the DNN models (y-axis) versus the hydrocode reference data (x-axis). The 45 degree black line on each plot indicates perfect prediction accuracy. Proximal data points are binned together, each bin is colored according to the counts of the data falling in that bin.}}
\end{figure*}

A more rich representation of model accuracy is provided by parity plots comparing model predictions against the ground truth dynamics for all test set data. The parity plots for all shock and interface model outputs are given in Fig. \ref{fig:parity_plots}, where a perfectly accurate model would produce outputs which fall on the 45 degree diagonal. The parity plots indicate that the surrogate models maintain good accuracy in their predictions across the full range of values exhibited by the fluid dynamics state variables. Note that the data distributions are skewed to lower magnitudes due to the rapid expansion and exponential decay nature of an explosive event. To accommodate the highly skewed distribution of the state variables we use a hexbin visualization approach, where data proximal in magnitude is binned together and the bin color represents the counts of data within that bin. We observe that bins containing the largest counts of data are clustered in the low magnitude regimes (highlighted by the plot insets in Fig. \ref{fig:parity_plots}). We note that relatively larger errors in the model predictions are observed at the higher values of the shock energy and the material interface pressure, this observation is corroborated by the associated $R^2$ values which fall slightly below unity. Improved accuracy at higher pressure and energy regimes could be achieved through a loss weighting scheme that more heavily penalizes errors in such regimes. 

\begin{table*}
\renewcommand{\arraystretch}{1.3} 
\caption{\label{tab:model_acc}DNN Accuracy Metrics}
\begin{ruledtabular}
\begin{tabular}{cc|ccc|ccc|ccc}
 & & \multicolumn{3}{c|}{RMSE} & \multicolumn{3}{c|}{$R^2$} & \multicolumn{3}{c}{MAPE} \\
 & & train & val & test & train & val& test & train & val & test \\ \hline
 \multirow{ 3}{*}{Interface} & $r$ & 4.88$\times10^{-2}$ & 4.42$\times10^{-2}$ & 4.91$\times10^{-2}$ & 1.0000 & 1.0000 & 1.0000 & 9.38$\times10^{-4}$ & 9.02$\times10^{-4}$ & 9.53$\times10^{-4}$\\ 
 & $p$ & 5.81$\times10^{+7}$ & 5.33$\times10^{+7}$ & 5.70$\times10^{+7}$ &  0.9999 & 0.9999 & 0.9999 & 4.81$\times10^{-3}$ & 4.83$\times10^{-3}$ & 5.11$\times10^{-3}$\\ 
 & $u$ & 9.08$\times10^{+1}$ & 8.93$\times10^{+1}$ & 9.34$\times10^{+1}$ & 1.0000 & 1.0000 & 1.0000 & 1.16$\times10^{-3}$ & 1.17$\times10^{-3}$ & 1.17$\times10^{-3}$ \\ \hline
 \multirow{ 5}{*}{Shock} & $r$ & 3.24$\times10^{-1}$ & 3.17$\times10^{-1}$ & 3.22$\times10^{-1}$ & 1.0000 & 1.0000 & 1.0000 & 6.69$\times10^{-4}$ & 6.62$\times10^{-4}$ & 6.63$\times10^{-4}$\\ 
 & $p$ & 1.40$\times10^{+6}$ & 1.37$\times10^{+6}$ & 1.41$\times10^{+6}$ & 1.0000 & 1.0000 & 1.0000 & 7.97$\times10^{-4}$ & 7.80$\times10^{-4}$ & 7.94$\times10^{-4}$\\ 
 & $e$ & 2.11$\times10^{+5}$ & 2.04$\times10^{+5}$ & 2.09$\times10^{+5}$ & 0.9995 & 0.9995 & 0.9995 & 2.09$\times10^{-3}$ & 2.07$\times10^{-3}$ & 2.08$\times10^{-3}$ \\ 
 & $\rho$ & 5.87$\times10^{-5}$ & 5.73$\times10^{-5}$ & 5.90$\times10^{-5}$ & 1.0000 & 1.0000 & 1.0000 & 7.90$\times10^{-4}$ & 7.50$\times10^{-4}$ & 7.66$\times10^{-4}$ \\ 
 & $u$ & 9.17$\times10^{+0}$ & 8.98$\times10^{+0}$ & 9.17$\times10^{+0}$ & 1.0000 & 1.0000 & 1.0000 & 7.90$\times10^{-4}$ & 7.80$\times10^{-4}$ & 7.85$\times10^{-4}$ \\
\end{tabular}
\end{ruledtabular}
\end{table*}

\begin{figure*}[ht!]
\includegraphics[width=.3\textwidth]{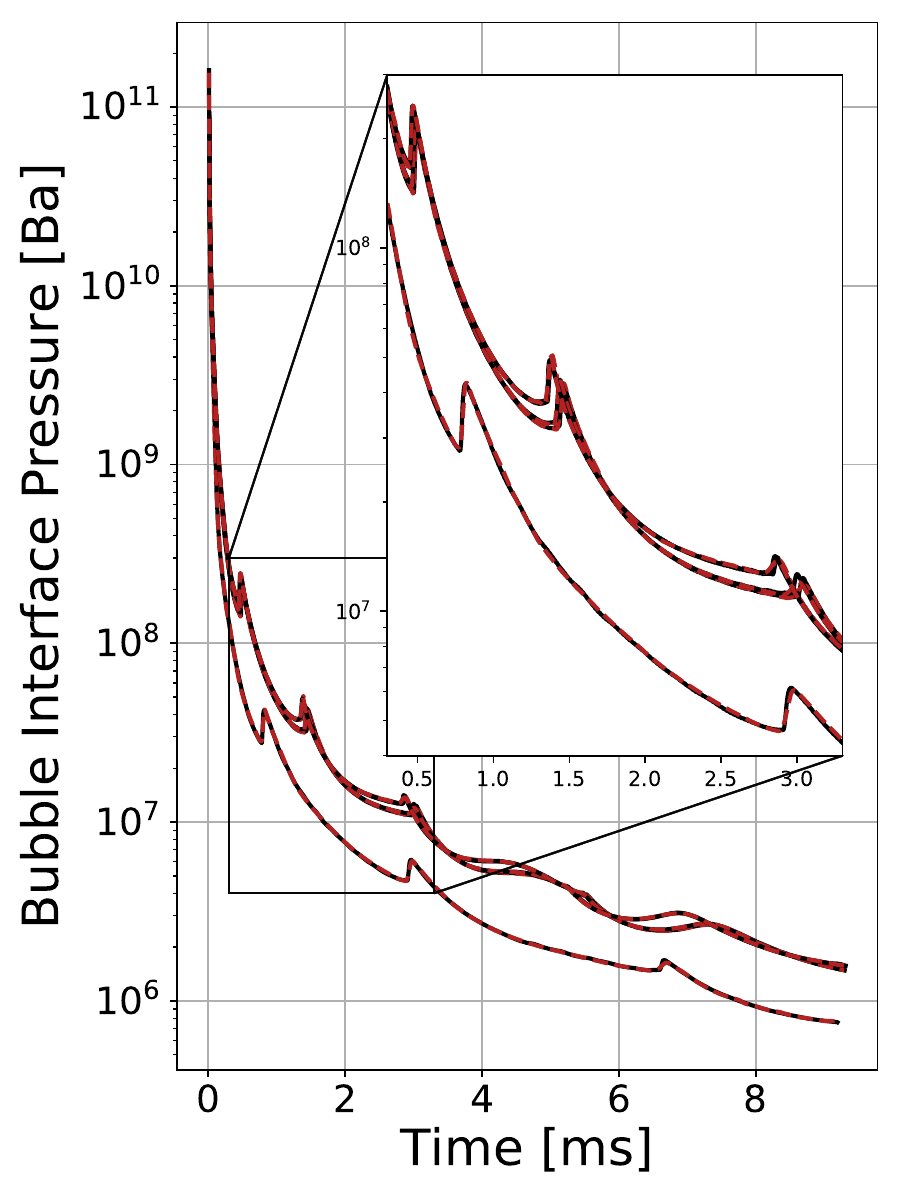}
\includegraphics[width=.3\textwidth]{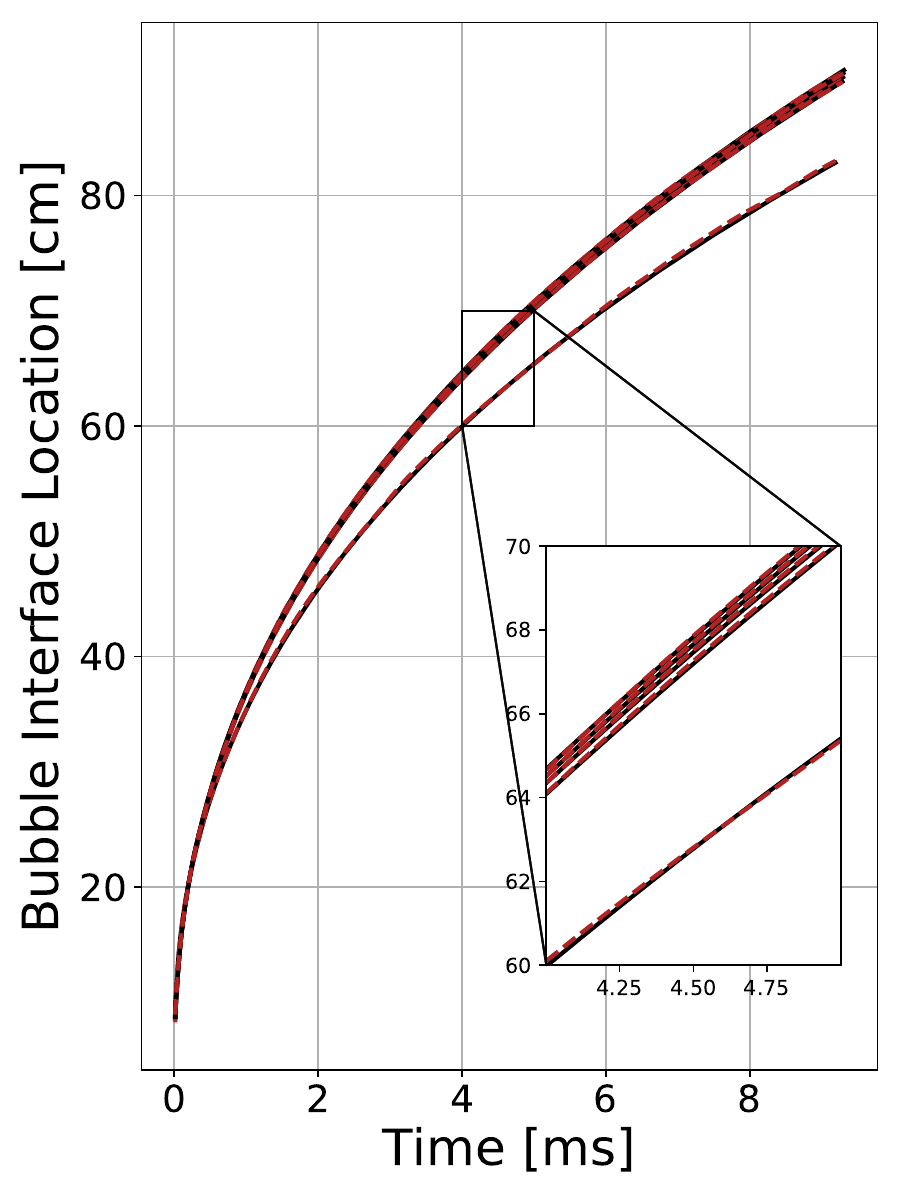}
\includegraphics[width=.3\textwidth]{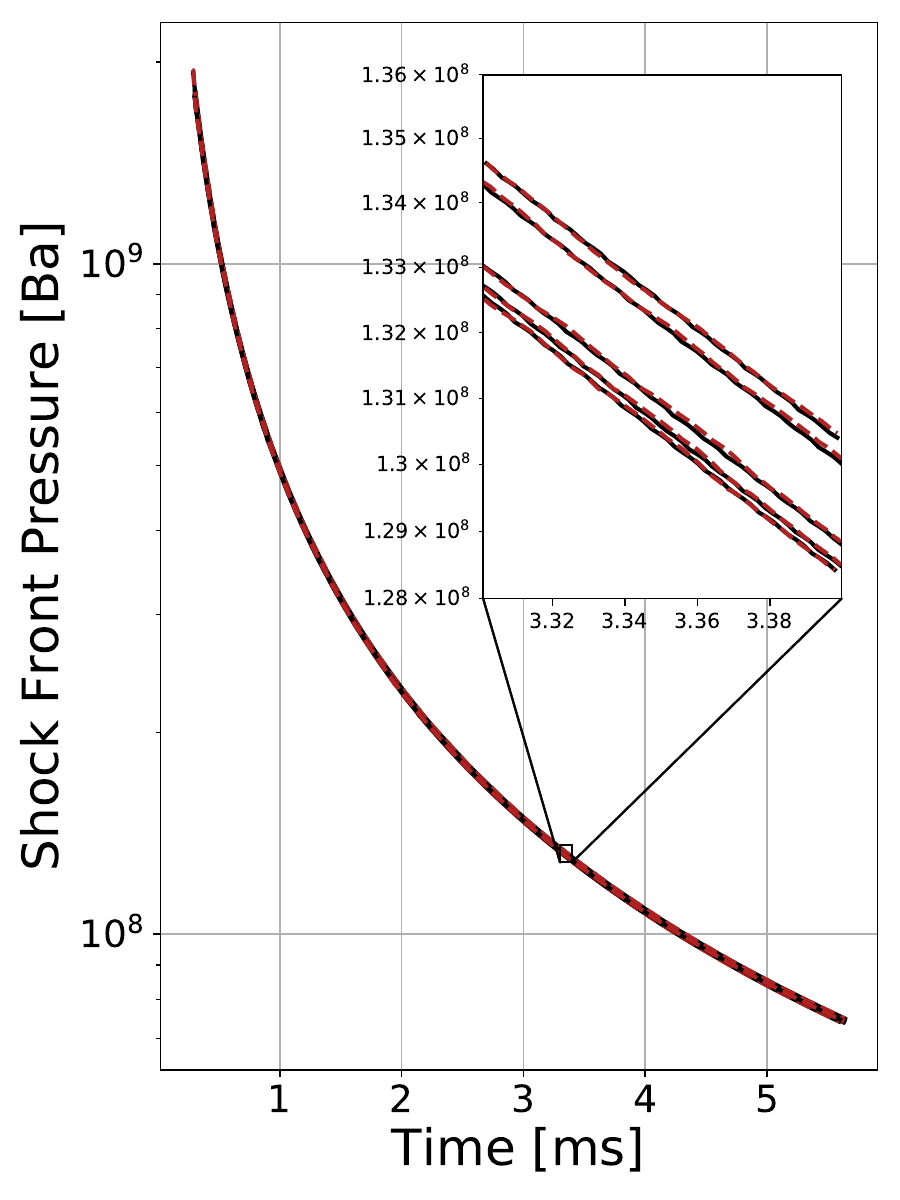}
\includegraphics[width=.33\textwidth]{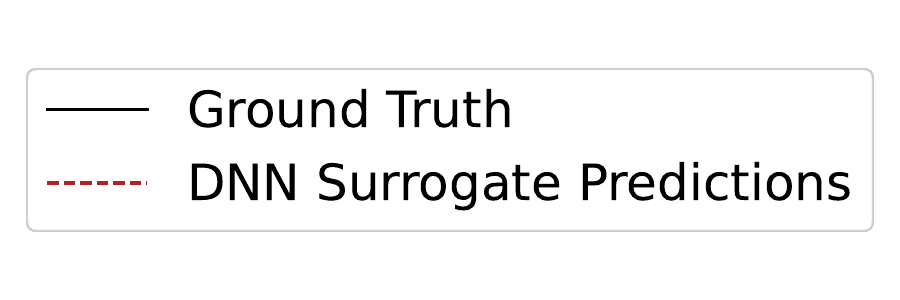}
\caption{\label{fig:time_comparison} \textcolor{black}{Selected fluid variable dynamics at the material interface (left and center) and shock front (right) for five randomly sampled JWL parameter sets. The DNN surrogate predictions are shown as a dashed red line, while the ground-truth data corresponds to the solid black line.}}
\end{figure*}

We now provide a more granular demonstration of the learned mapping between HE material parameters and resulting fluid dynamics. Namely, we consider five JWL parameters sets randomly drawn from the test set, and the predictions of the material interface pressure, material interface location, and shock front pressure dynamics. These predictions are compared against the ground-truth data generated by the hydrodynamics simulators and the visualizations are given in Fig. \ref{fig:time_comparison}. Inspecting the material interface pressure figure, we note that the predicted dynamics are accurate and across the entire time domain for all JWL parameter sets. The model predictions are almost visually indistinguishable from the reference data, successfully capturing the overall pressure decay behavior and transient "spikes" which quickly decay. These short-lived transient structures in the dynamics are a result of wave dynamics internal to the gas bubble, where reverberating shockwaves impart a short-lived increase in pressure upon interacting with the bubble interface. The DNN model exhibits exceptional accuracy in predicting the time instance and magnitude of these pressure spikes, and thus has encoded information about the influence of the internal wave dynamics. The excellent predictive accuracy is also displayed in model predictions of the material interface location, where once again the predicted dynamics are nearly visually indistinguishable from the ground truth data. Several of the sampled HE materials produce front dynamics which are quite similar, however close examination (highlighted by the figure inset) reveals that the DNN model is able to accurately distinguish amongst each material and reproduce the corresponding dynamics. Finally, considering shock front pressure dynamics predictions, the DNN model accurately predicts the behavior for different HE material parameters. The similarity in the dynamics amongst all considered JWL parameter sets was noted in Sec. \ref{sec:data_gen}, yet the inset reveals that our model successfully distinguishes and predicts the pressure behavior at the shock front.

\textcolor{black}{The focus of this section is on evaluating the surrogate model's ability to accurately reproduce hydrocode results rather than direct experimental validation. While future comparisons against experimental data will be valuable, it is important to note that the reference simulator (DYSMAS) used in this study has been extensively validated across a wide range of underwater explosion scenarios. DYSMAS has demonstrated its ability to accurately capture shock and bubble phenomena in small-scale cylindrical explosion tests \cite{wardlaw1998coupled, wardlaw1998spherical} as well as larger-scale shock dynamics \cite{vangessel2022}. Given this established validation history, the surrogate's ability to match DYSMAS results provides a strong foundation for real-world applicability. Future studies will focus on extending this validation framework to include direct experimental comparisons where feasible.}

One objective of developing DNN surrogate models is to leverage their capability for making predictions far more rapidly than their traditional hydrodynamic solver counterparts. Indeed, the time required to run the traditional solver, averaged across 100 randomly sampled JWL parameter sets, is 624 seconds ($\sim$ 10 minutes) on a single AMD 7713 Milan CPU. In contrast, the averaged DNN inference time to predict the dynamics for the same JWL parameter and temporal grids produced by the fluid solver, takes only 0.155 seconds on a single A40 GPU. Therefore, our surrogates make predictions 4,025 times faster than the traditional solver counterpart. Furthermore, these surrogates approximate a temporal operator and are not auto-regressive in nature, providing predictions at any point in time without the need to simulate preceding dynamics.

The results of this section provide clear evidence that the surrogates have learned an accurate mapping from the thermodynamic parameters, characterizing a broad swathe of HE materials, to the fluid dynamics produced by detonating those HE materials. Surprisingly, this accuracy has been achieved using a quite simple fully-connected neural architecture comprised of only 3155 and 3053 trainable parameters respectively. While future efforts will explore the use of more complex neural operator architectures, the results of this work indicate that simple models can encode the rich physics of high pressure multi-material fluid dynamic systems parameterized by a thermodynamic equation of state. In the following two sections, we will highlight the physical insights and design capabilities afforded by the high-accuracy yet fast-running surrogate models.

\section{\label{sec:feature_importance}Feature Importance}

\begin{figure*}[ht!]
\includegraphics[width=.3\textwidth]{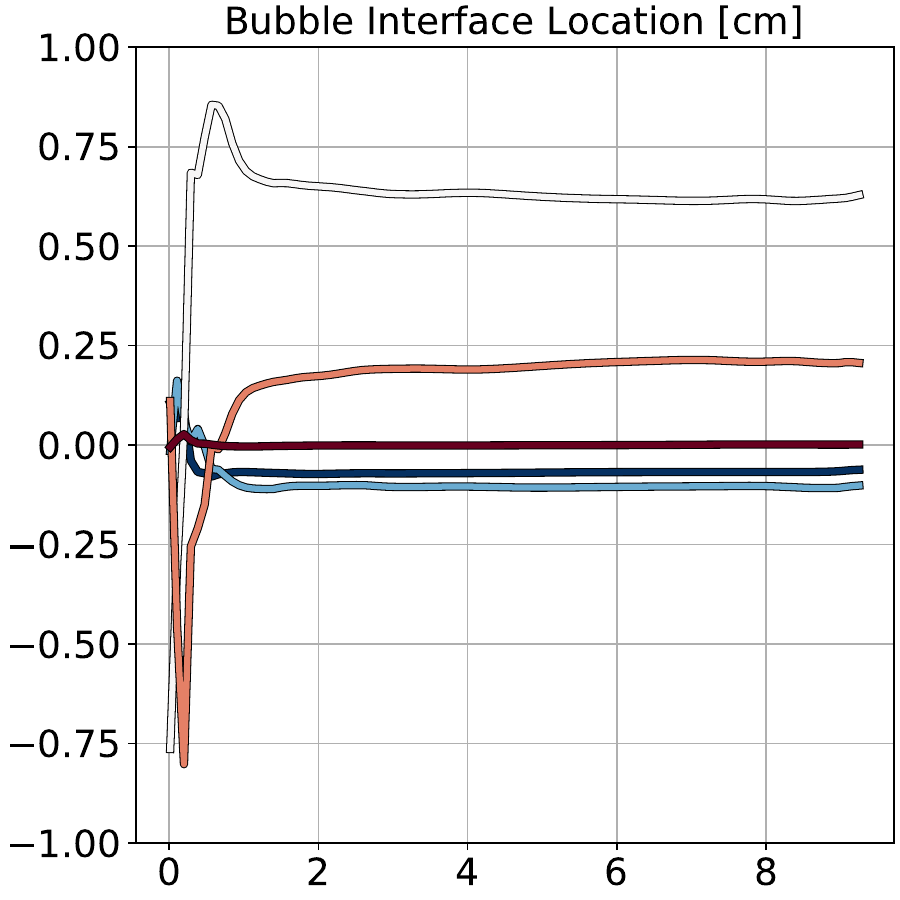}
\includegraphics[width=.3\textwidth]{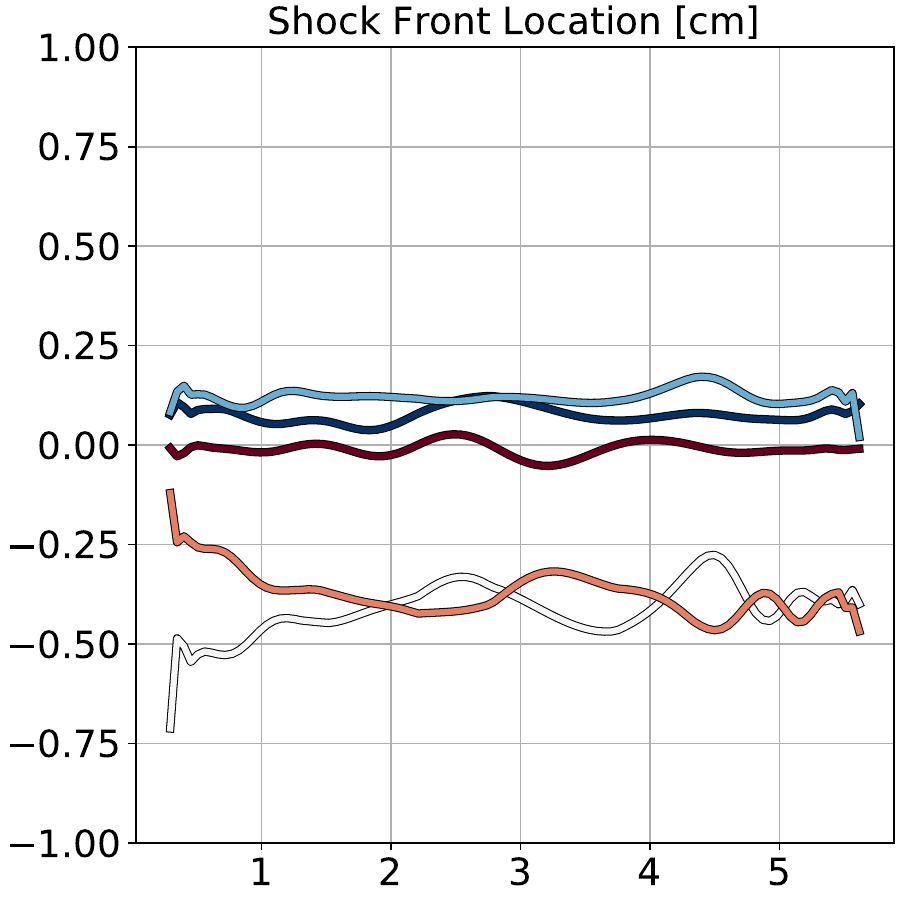}
\includegraphics[width=.3\textwidth]{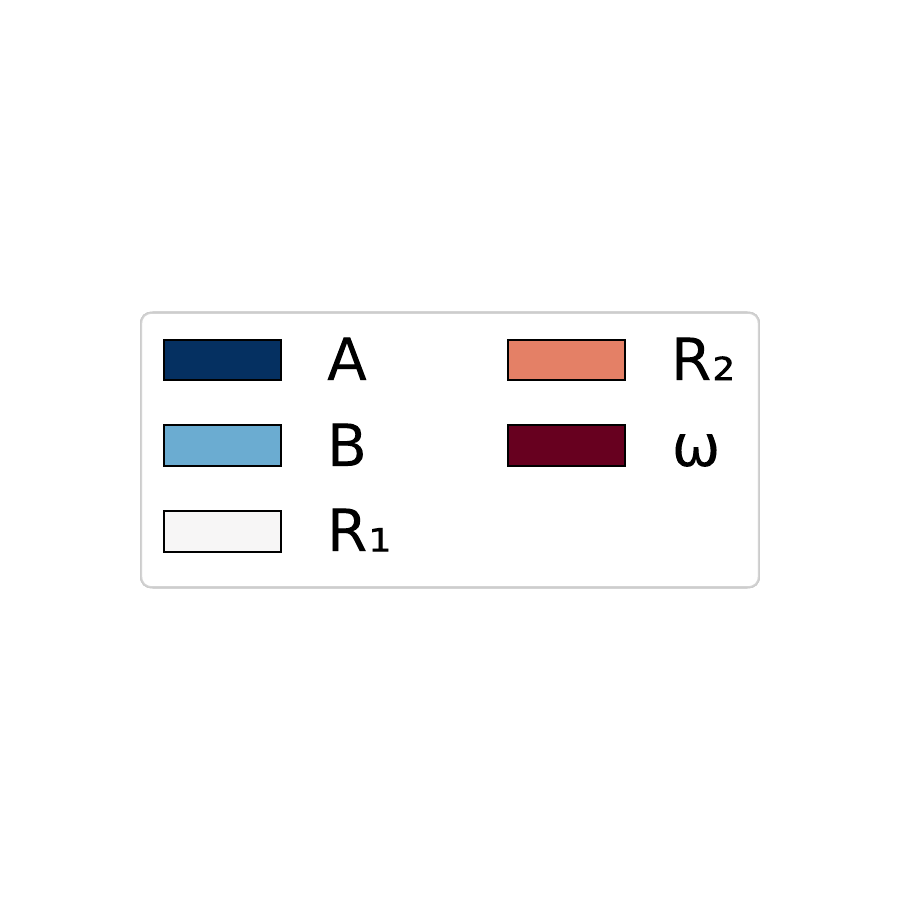}\\
\includegraphics[width=.3\textwidth]{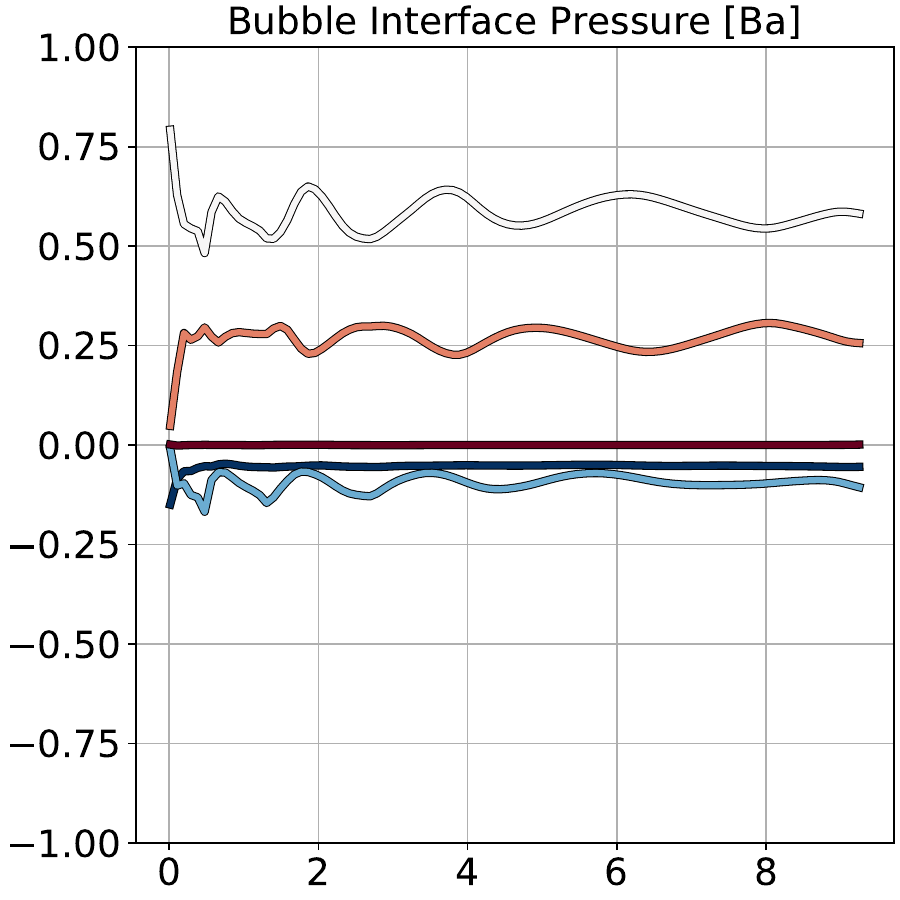}
\includegraphics[width=.3\textwidth]{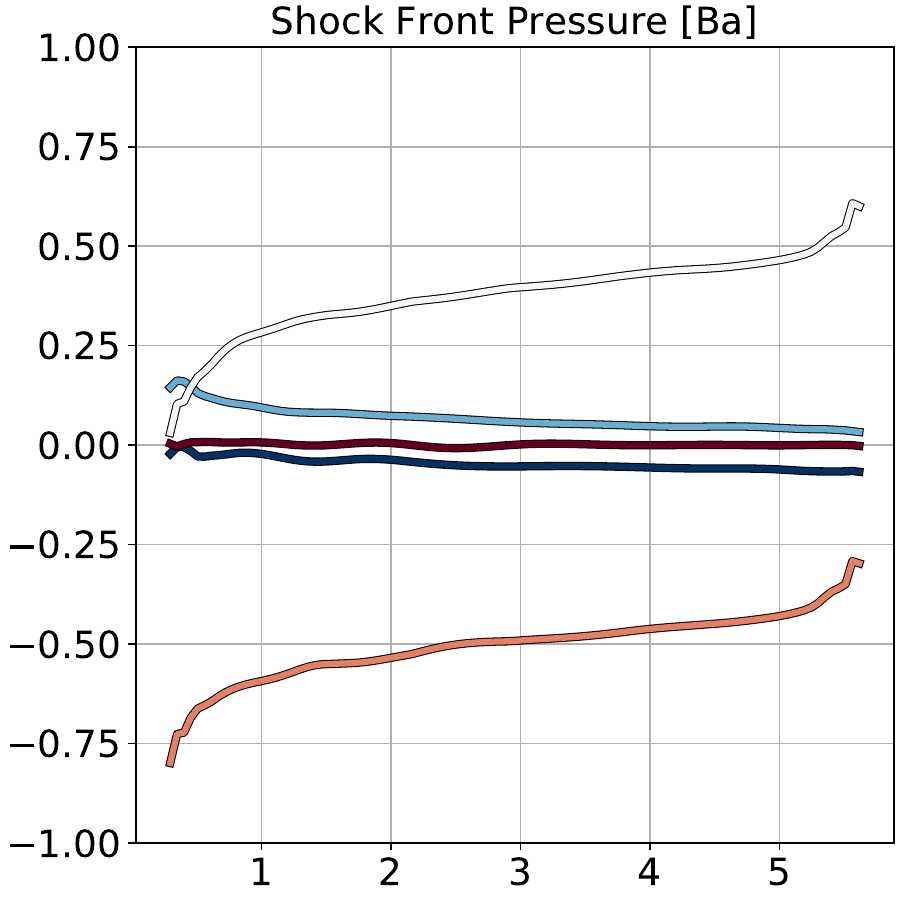}
\includegraphics[width=.3\textwidth]{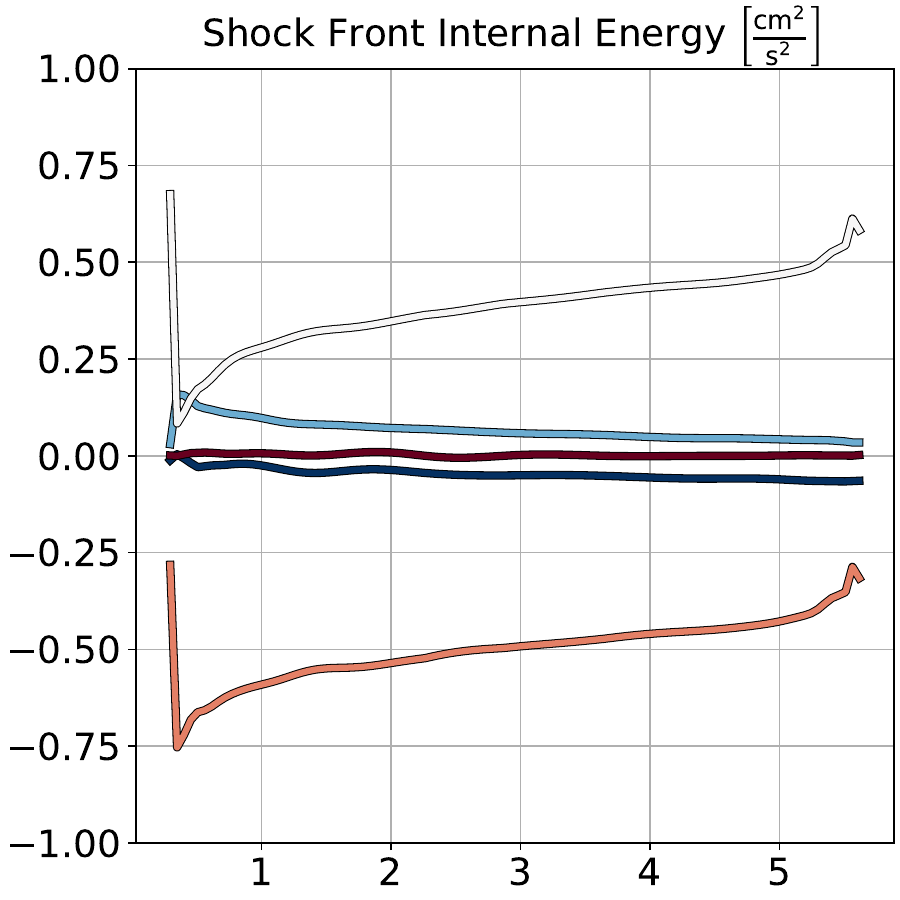}\\
\includegraphics[width=.3\textwidth]{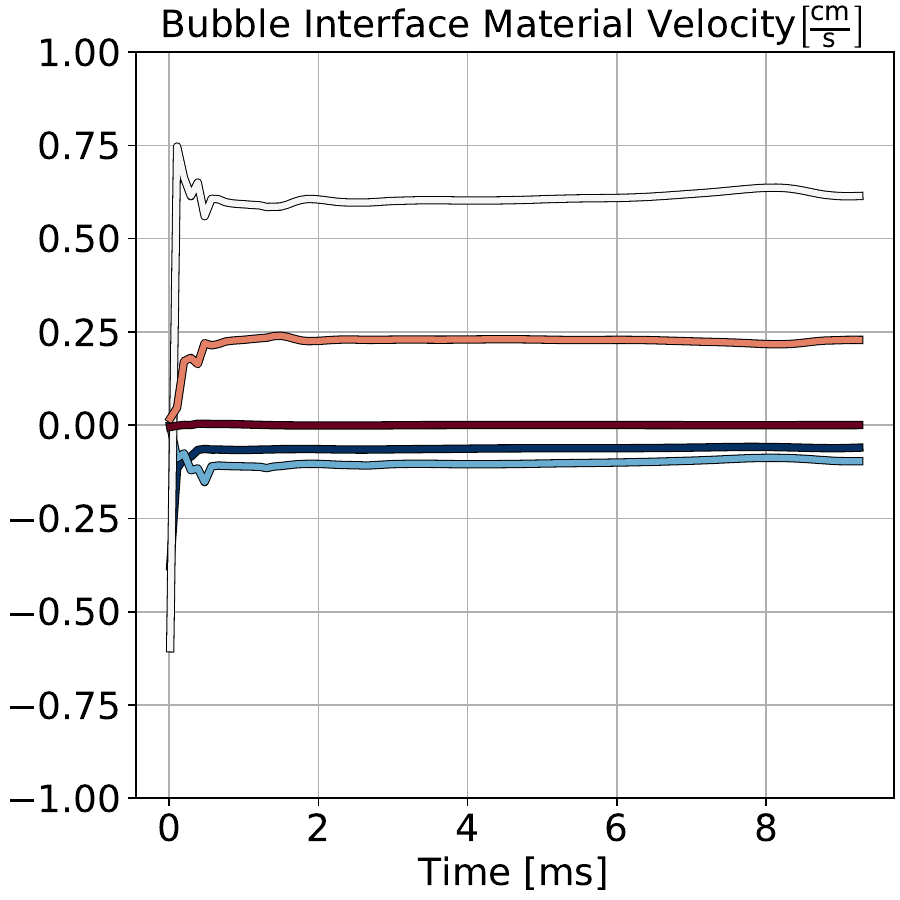}
\includegraphics[width=.3\textwidth]{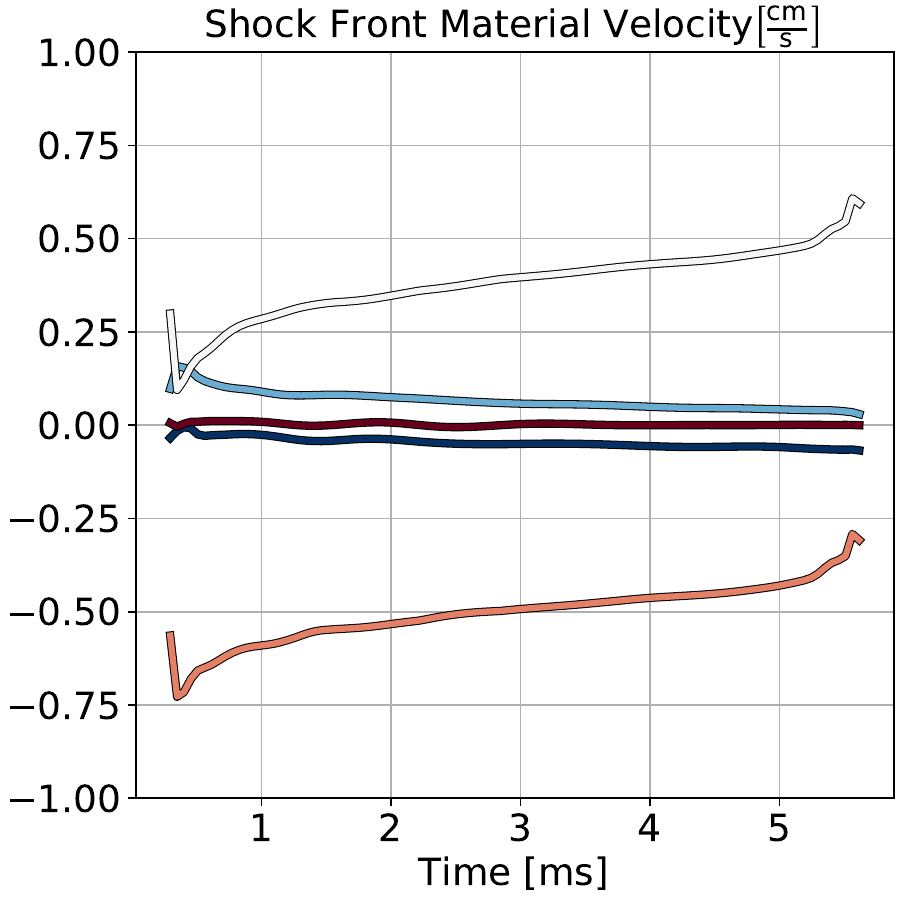}
\includegraphics[width=.3\textwidth]{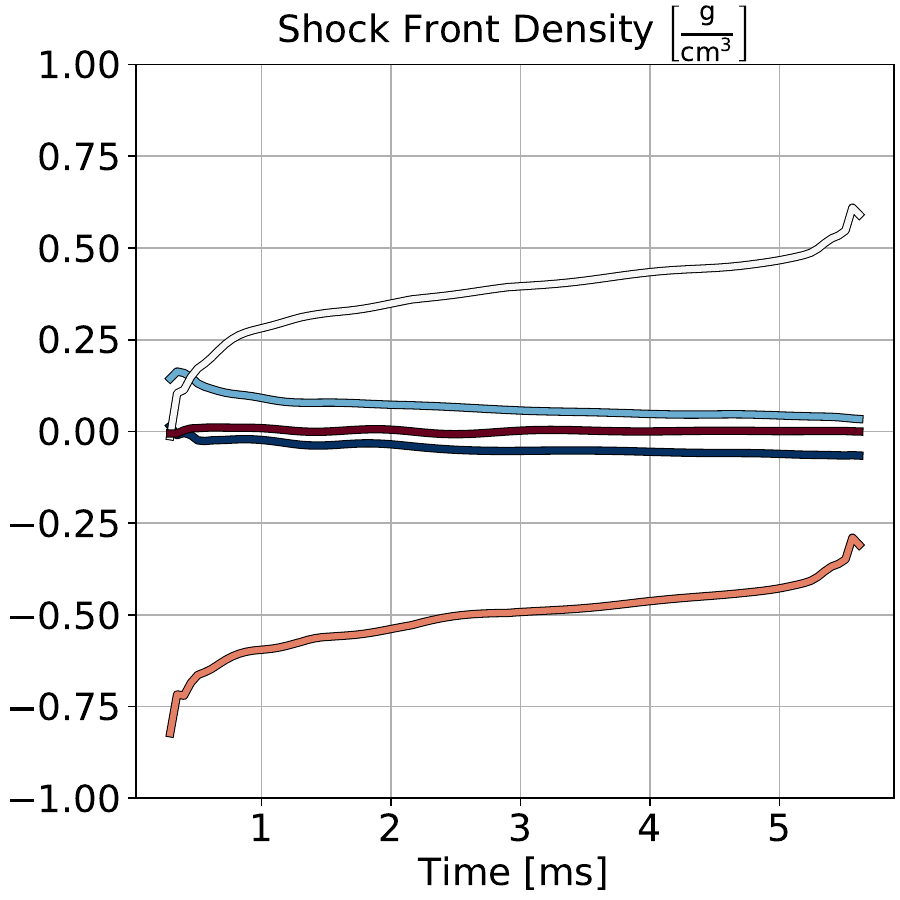}
\caption{\label{fig:fa_plots}Feature ablation values (Eq. \ref{eq:fa}) for each DNN output.}
\end{figure*}

The feature importance technique (see Sec. \ref{subsec:feature_sensitivity}) is used to interpret the degree of influence of an individual JWL thermodynamic parameter on the fluid dynamic response. The feature ablation values (Eq. \ref{eq:fa}), for outputs of both the shock and material interface surrogates, are plotted in Fig. \ref{fig:fa_plots}. The parameters $R_1$ and $R_2$ are shown to be the most influential as they are assigned the largest magnitude ablation values for the majority of output variables across the majority of the time domain. Recalling the doubly exponential functional form of the JWL thermodynamic model, given in Eq. \ref{eq:jwl} and repeated here, 
\begin{equation*}
p(\rho, e) = A\left[1- \frac{\omega \rho}{R_1 \rho_0}\right] e^{-\frac{R_1 \rho_0 }{\rho}} + B\left[1- \frac{\omega \rho}{R_2 \rho_0}\right] e^{-\frac{R_2 \rho_0 }{\rho}} + \omega \rho e
\end{equation*}
$R_1$ and $R_2$ control the rate at which the reacted HE gas pressure decays as the density decreases (i.e. the bubble expands). The parameters $A$ and $B$, which control the total magnitude of the exponential terms, are found to have generally lower ablation values and thus play a secondary role. Finally, the $\omega$ parameter, controlling the asymptotic ideal gas behavior of the system, is assigned an approximately $0$ ablation value and thus plays a trivial role in influencing the dynamics. Note, that at large expansion regimes of the HE bubble, the exponential terms of the EOS may sufficiently decay such that the $\omega$ parameter becomes more influential. For the remainder of this section we will focus on the two most influential features, namely $R_1$ and $R_2$ (referred to jointly as $R_{1,2}$), deriving physical insights of their influence on the shock and material interface dynamics.

\begin{figure*}[ht]
\includegraphics[width=.8\textwidth]{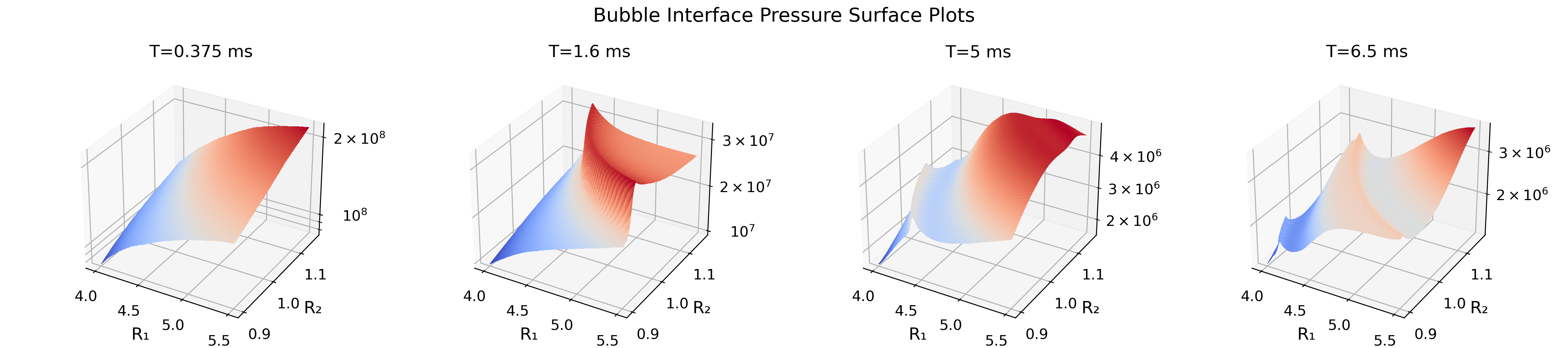}\\
\includegraphics[width=.8\textwidth]{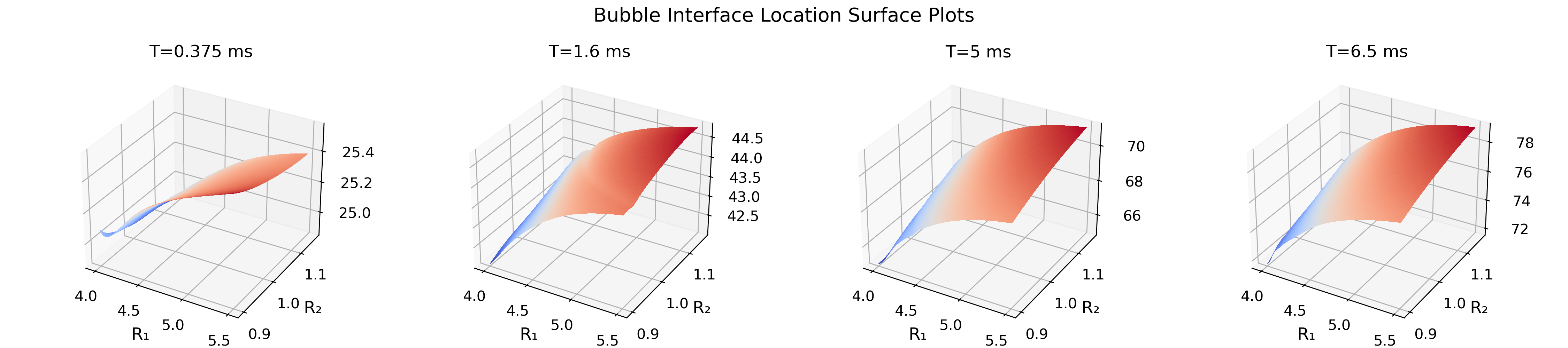}\\
\caption{\label{fig:mat_int_param_plots} \textcolor{black}{Parameter plots for the material interface pressure and location at selected instances in time. The response surfaces are plotted with respect to the JWL parameters, $R_1$ and $R_2$.}}
\end{figure*}

Consider the material interface feature ablation plots in the first column of Fig. \ref{fig:fa_plots}. The parameters $R_{1,2}$ are always positive for times greater than $0.5$ seconds. Positivity in those FA values indicates that location, pressure, and velocity variables at the material interface are positively correlated with $R_{1,2}$. This positive correlation is clearly shown in Fig. \ref{fig:mat_int_param_plots} where the material interface pressure and location response surfaces are plotted with respect to $R_{1,2}$. Ripples in the pressure surfaces indicate internal shockwave interaction with the material interface. Furthermore, at early time instances there is a negative correlation between $R_2$ and interface location. However a prevailing positive trend exists between $R_{1,2}$ and both pressure and location across the temporal range of interest.

The positive correlation of $R_{1,2}$ with interface location and pressure is physically non-intuitive given that HE EOS pressure is a monotonically decreasing function with respect to both of these EOS parameters. We hypothesize that while an increase in $R_{1,2}$ decreases the EOS pressure for an isolated single material HE gas system, it also alters the similarity between the water EOS and the HE EOS. Namely that increasing $R_{1,2}$ decreases the normed distance between the water and gas EOS surfaces for the thermodynamic regime sampled by the multi-material system. Thus the observed positive correlation behavior is a multi-material {\it compatibility} phenomenon, whereby stronger coupling (i.e. thermodynamic function similarity) between the two materials enhances energy transfer across the interface. Greater efficiency in energy transfer across the gas-water interface leads to higher sustained pressure at the interface, allowing the bubble has greater capacity for mechanical work on the surrounding water system leading to larger interface velocities. This intriguing multi-material behavior warrants deeper analysis which is reserved for future work.

This section has demonstrated that feature importance, when applied to surrogates of a high pressure multi-material fluid system, provides physical intuition and a deeper understanding of the system dynamics. The feature ranking capabilities indicate which of the HE thermodynamic parameters exert the highest degree of influence on the shock and material interface dynamics. Furthermore, insights provided by feature importance, coupled with a foundational knowledge of fluid dynamics, provide a richer description of the physical mechanisms governing this system. Namely, non-intuitive positive correlations between the thermodynamic parameters $R_{1,2}$ and the material interface dynamics were discovered. It was hypothesized that multi-material compatibility effects, quantified by the normed distance between EOS functions, control the ability of the HE system to do work on the surrounding water. While traditional design of experiment techniques, such as sensitivity analysis, could in theory provide the same insights and understanding, the rapid inference and accuracy capabilities of our DNN surrogates render fine-grained interrogation of the thermodynamic-fluid dynamic linkage significantly more tractable. One could conceptualize exploiting these newly identified system behaviors through altering the composition of HE materials to induce changes in the thermodynamic parameters, enabling routes to tailored HE system behaviors such as engineering shock and material interface dynamics.

\section{\label{sec:inv_design}Inverse Design for Parameter Discovery}

\begin{figure*}[ht!]
\includegraphics[width=1\textwidth]{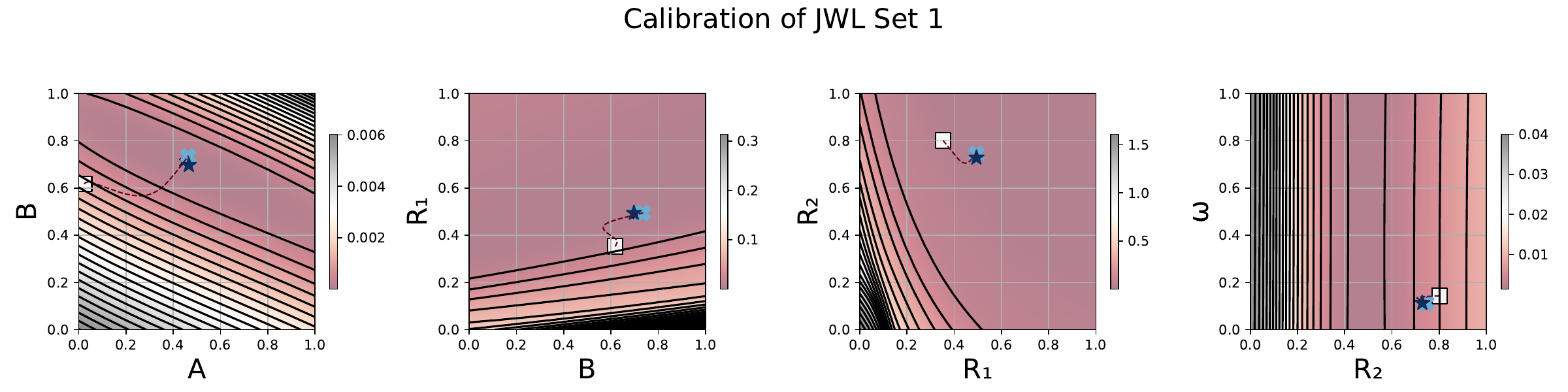}\\
\includegraphics[width=1\textwidth]{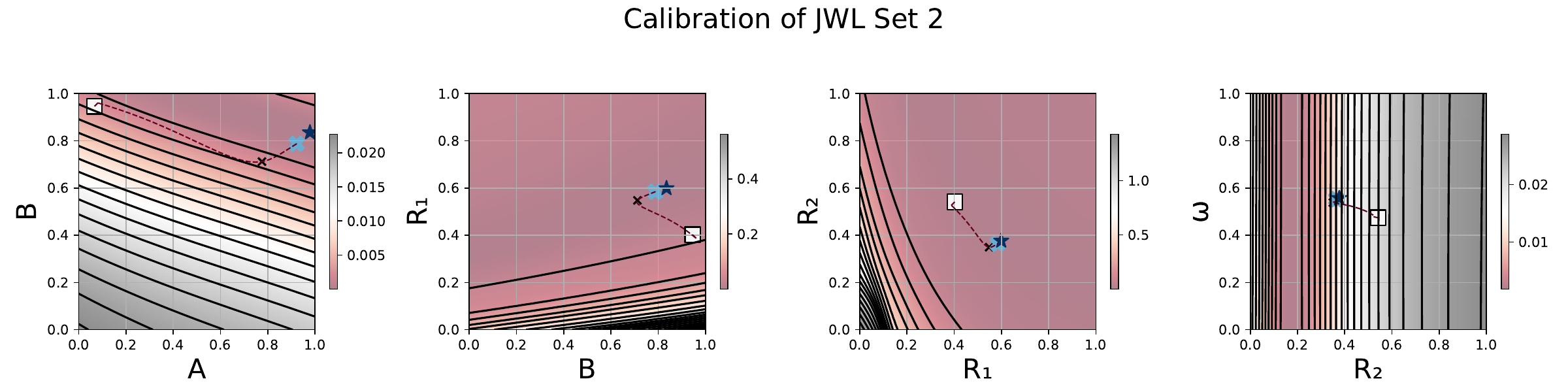}\\
\includegraphics[width=1\textwidth]{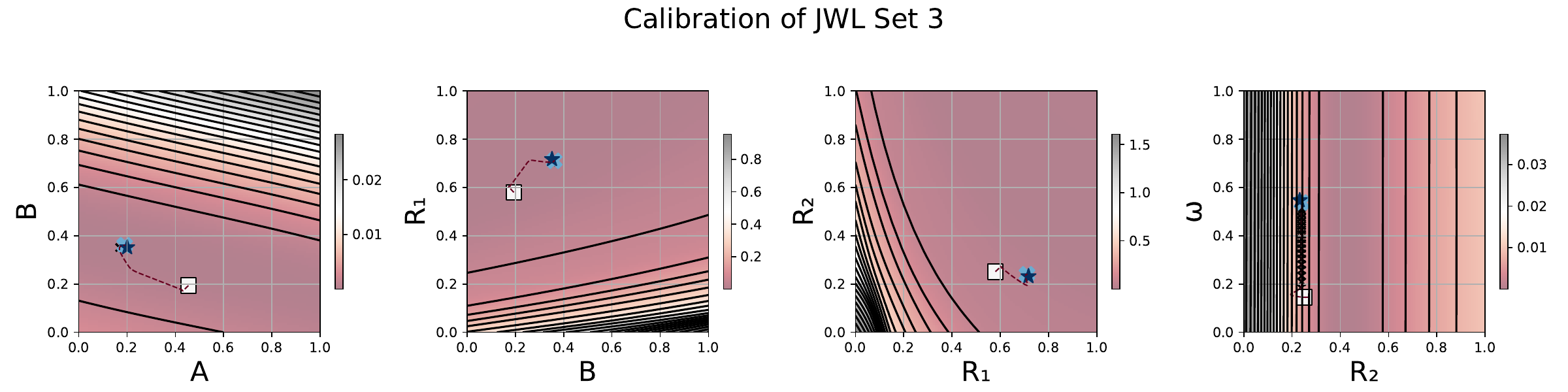}\\
\includegraphics[width=.5\textwidth]{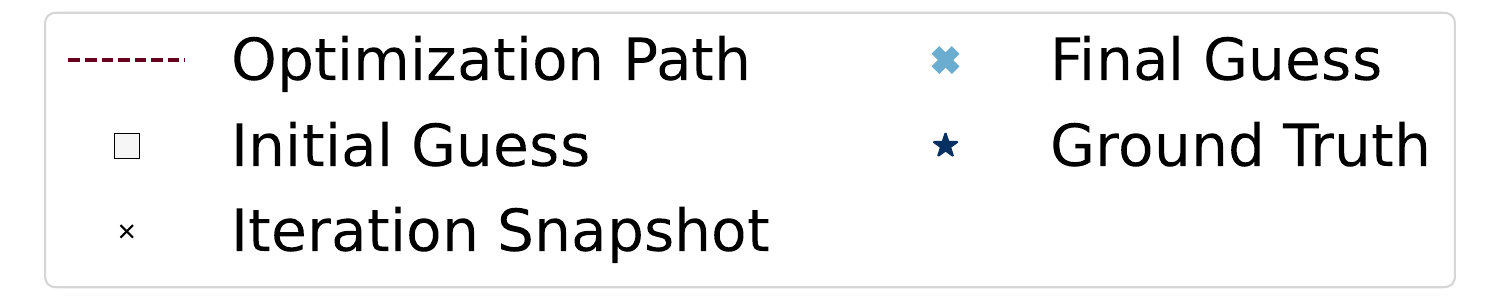}\\
\caption{\label{fig:optimization} \textcolor{black}{Plots of the parameter discovery problem  (each row corresponding to a unique JWL parameter discovery problem). Within each row, each plot is a two dimensional projection of the five dimensional JWL parameter space. The contour shading represents the loss landscape defined in Algorithm \ref{alg:param_discovery}.}}
\end{figure*}

Parameter discovery, specifically identifying the unknown JWL parameters which generated an observed fluid dynamic response, is performed using the derivative-based algorithmic approach outlined in alg.\ref{alg:param_discovery}. The robustness and accuracy of the parameter discovery algorithm, is tested for 100 randomly sampled JWL parameter sets and associated dynamics. For each trial, material interface dynamics are generated by the hydrodynamic simulator for a randomly sampled JWL parameter set, where we ensure each sampled parameter set does not appear in the training or validation dataset. Next, the dynamics are presented to the parameter discovery algorithm while the corresponding JWL parameters are withheld. Three example trials of this discovery process are demonstrated in Fig. \ref{fig:optimization}, where each row corresponds to a unique EOS parameter set, i.e. parameters we are attempting to {\it discover}. For each visualized trial, the algorithm begins with a uniformly sampled initial parameter guess (white square) which may be quite far from the unknown ground truth value. Gradient descent, enabled by differentiating the trained surrogate model with respect to the EOS parameter inputs, is used to iteratively update the parameters (dashed-dotted red line). At each iteration the difference between the dynamics associated with the current parameter values and the observed ground-truth dynamics is reduced in a normed sense. The final guess (light blue cross) is within 0.1\% of the ground truth value for all features except for $\omega$. We attribute the relatively higher error for the $\omega$ parameter to the fact that the dynamics, over the time scales considered, are highly insensitive to this parameter as corroborated by the ablation values discussed in the previous section and plotted in Fig. \ref{fig:fa_plots}. \textcolor{black}{Furthermore, by simultaneously calibrating against both shock {\it and} bubble interface dynamics across a comprehensive thermodynamic state space, we ensure solution uniqueness in the parameter discovery problems in this study. This dual-dynamics approach effectively constrains the high-dimensional JWL parameter space, preventing convergence to local minima that might satisfy single-front dynamics but fail to capture the complete system behavior.}

Two types of accuracy metrics are applied to the parameter discovery experiment with their values are given in Table \ref{tab:inv_metrics}. First, we report the mean absolute error between the ground truth parameters, and the recovered parameters. These values are reported directly in the min-max scaled feature space within which the optimization is performed. The absolute error is less than $4 \times 10^{-2}$ for all parameters except for $\omega$ (for reasons highlighted in the previous paragraph), note the input features range between 0 and 1 within this feature space. We also report the mean absolute percent error in the non-scaled parameter space corresponding to the raw data values of the EOS parameters. Once again excellent parameter recovery accuracy is demonstrated with errors ranging between $.03-.1\%$ for the parameters aside from $\omega$. The accuracy metrics reported here are corroborated in the supplementary materials sec. \ref{app:recovered_dynamics}, where the recovered dynamics are visually indistinguishable from the ground-truth dynamics (see. Fig. \ref{fig:optimization_dynamics}). Thus, the parameter discovery algorithm has shown to be robust and accurate for a wide range of physically meaningful HE EOS parameters and is therefore a feasible approach to characterizing and calibrating new HE materials. Furthermore the differentiable surrogate accelerates the parameter discovery process by {\it at least three orders of magnitude} over traditional hydrodynamic solvers through the combination of accelerated inference capabilities, 4,025 times faster than the traditional solver, and enabling gradient-based optimization. Gradient-based techniques, which are impossible to utilize in conjunction with traditional hydrocode simulators,  reach minima in fewer iterations than gradient-free approaches. In addition to parameter discovery, these benefits of our neural surrogates have the potential to accelerate a wider array of inverse design problems such as EOS design for tailoring HE material fluid dynamics.

\textcolor{black}{While the parameter discovery framework demonstrated in this study shows excellent performance, it is important to note that our current work focuses exclusively on calibration to synthetic data generated by the same simulation framework (DYSMAS). This approach was intentionally chosen to establish the fundamental viability of our differentiable surrogate method. Real-world application would require addressing experimental UNDEX measurements, which present additional challenges due to extreme conditions, non-linear fluid phenomena, and microsecond time-scales \cite{li2022measurement}. Experimental techniques such as piezoelectric-based pressure measurements are susceptible to errors up to 10\% or more \cite{cole1948underwater}. In future work, we plan to extend our framework to accommodate these challenges through two potential approaches: (1) statistical averaging of multiple measurements to provide more robust calibration targets, and (2) incorporating Bayesian model calibration methods similar to \cite{walters2018bayesian}, which can rigorously account for both modeling and experimental uncertainties. These extensions would preserve the computational advantages of our differentiable surrogate while adapting to the inherent noise and variability in experimental data.}

\begin{table*}
\renewcommand{\arraystretch}{1.5}
\caption{\label{tab:inv_metrics} Aggregated accuracy metrics for the inverse design experiment. We present both the absolute error, calculated with respect to the scaled features, and the percent error, calculated with respect to the raw JWL parameter data values.}
\begin{ruledtabular}
\begin{tabular}{l|ccccc}
 Error Measure & $A$ [Ba] & $B$ [Ba] &  $R_1$ [-] &$R_2$ [-] &$\omega$ [-]\\ \hline
 Mean Absolute Error (Scaled) & 3.55$\times10^{-2}$ & 4.03$\times10^{-2}$ & 1.09$\times10^{-2}$ & 1.69$\times10^{-2}$ &  3.13$\times10^{-1}$ \\ 
 Mean Percent Error (Non-scaled) & 1.34$\times10^{-1}$ & 3.49$\times10^{-2}$ & 4.85$\times10^{-2}$ & 7.78$\times10^{-2}$ &  1.30$\times10^{+0}$ \\ 
\end{tabular}
\end{ruledtabular}
\end{table*}

\section{\label{sec:conclusions}Conclusions} 

The shock and material interface dynamics produced by underwater energetic systems are fluid phenomena with critical defense and industrial applications. Furthermore, the phenomenon present within these systems, i.e. multiphase, discontinuous, thermochemical, and highly compressible behaviors, arise in numerous adjacent domains. Therefore, accurate and rapid prediction of the HE-induced fluid response is critical not only for prediction of known and notional HE materials, but holds broader technological relevance as well. We highlighted two prevailing issues inherent in traditional hydrocode modeling of these HE systems, namely significant computational cost and lack of gradient information. To address these shortcomings, we trained two neural surrogate models to reproduce synthetic high-fidelity shock and material interface dynamics data to a high degree of accuracy. The trained models achieved a 4025x calculation speed-up with respect to the traditional hydrocode solver while maintaining high accuracy predictions ($R^2>0.9995$ and MAPE $<5\times 10^{-3}$) for all target state variables. The high-accuracy surrogates were able to capture the effects of complex wave interactions, internal to the bubble, on pressure transients at the bubble interface. Furthermore, despite the similarity in the shock front dynamics among the notional energetic materials, the models can accurately distinguish between the dynamics produced by dissimilar energetic materials.

The accelerated prediction capability of the surrogate provides a computationally efficient route to assessing the feature importance of the system. Feature importance, when applied to our models of the learned dynamics, affords insights into relationships between HE thermodynamic characteristics of the energetic and the dynamics of the fluid structures it produces. Feature ablation calculations demonstrated that the energetic parameters controlling the exponential decay characteristics of the EOS pressure surface were the most influential on the resulting dynamics. Therefore, engineering HE material ingredients in a manner which alters these parameters will provide the largest degree of control over the system dynamics. Furthermore, feature importance revealed non-intuitive system behavior whereby an overall reduction, or softening, in the energetic EOS pressure response led to an increase in the bubble pressure and expansion rate at the interface. We hypothesize that this behavior is a result of multi-material EOS compatibility dictating the efficiency of energy transfer from the energetic product gas to the surrounding water. This behavior, and its potential for engineering HE system dynamics, will be the focus of future research. Thus, scientific insights provided by feature importance analysis of neural surrogates hold potential for scientific insights and exerting control over complex HE-fluid systems.

Finally, we demonstrated the utility of our differentiable models for parameter discovery, and more broadly inverse design, applications. The use of differentiable surrogates overcomes the incompatibility between gradient-based optimization and traditional hydrocode solvers which lack the required derivative information. A gradient-based parameter discovery algorithm was shown to robustly recover the EOS parameters corresponding to unknown (synthetic) dynamics data. The recovered parameters are, on average, within 0.1\% of the ground truth values, yielding time series indistinguishable from the ground truth dynamics. Our algorithmic approach could also be adapted to the design of energetic system thermodynamics properties, producing EOS parameters which achieve shock and bubble dynamics adhering to a desired performance objective. The efficient search of design space enabled by gradient-based optimization, coupled by the three orders of magnitude calculation speed-up of the neural surrogates, enables inverse design and ultimately engineering of energetic systems. 

\textcolor{black}{The computational framework developed in this work can be adapted to similar simulation challenges faced across multiple related domains, offering potential advancements in technologies ranging from hypersonic aerodynamics to medical cavitation therapeutics that share the complex fluid physics characteristic of underwater explosions systems.}

\begin{acknowledgments}
We thank Dr. Peter Chung (University of Maryland) for helpful discussions. Support is gratefully acknowledged for this research from OUSD(R\&E) provided through the JTCG/ME MWIPT and from the Center for Engineering Concepts and Development via the CECD Trailblazers' Internal Seed Grant Program.
\end{acknowledgments}

\section*{Author Contributions}
\textbf{Frank VanGessel:} Conceptualization (lead); Formal analysis (lead); Investigation (equal); Methodology (equal); Data curation (lead); Software (supporting); Validation (equal); Visualization (supporting); Writing - original draft (lead); Writing - review \& editing (lead); Funding acquisition (lead); Project administration (lead).

\textbf{Mitul Pandya:} Conceptualization (supporting); Formal analysis (supporting); Investigation (equal); Methodology (equal); Data curation (supporting); Software (lead); Validation (equal); Visualization (lead); Writing - review \& editing (supporting).

\section*{Data Availability Statement}
The data that support the findings of this study are available on request from the corresponding author. The data and code are not publicly available due to Naval Surface Warfare Center Indian Head Division data sharing policy.

\section*{Release Statement}
DISTRIBUTION A (Log No. 25-022). Approved for Public Release; distribution unlimited.

\bibliography{aipsamp}

\providecommand{\noopsort}[1]{}\providecommand{\singleletter}[1]{#1}%
\begin{thebibliography}{71}%
\makeatletter
\providecommand \@ifxundefined [1]{%
 \@ifx{#1\undefined}
}%
\providecommand \@ifnum [1]{%
 \ifnum #1\expandafter \@firstoftwo
 \else \expandafter \@secondoftwo
 \fi
}%
\providecommand \@ifx [1]{%
 \ifx #1\expandafter \@firstoftwo
 \else \expandafter \@secondoftwo
 \fi
}%
\providecommand \natexlab [1]{#1}%
\providecommand \enquote  [1]{``#1''}%
\providecommand \bibnamefont  [1]{#1}%
\providecommand \bibfnamefont [1]{#1}%
\providecommand \citenamefont [1]{#1}%
\providecommand \href@noop [0]{\@secondoftwo}%
\providecommand \href [0]{\begingroup \@sanitize@url \@href}%
\providecommand \@href[1]{\@@startlink{#1}\@@href}%
\providecommand \@@href[1]{\endgroup#1\@@endlink}%
\providecommand \@sanitize@url [0]{\catcode `\\12\catcode `\$12\catcode `\&12\catcode `\#12\catcode `\^12\catcode `\_12\catcode `\%12\relax}%
\providecommand \@@startlink[1]{}%
\providecommand \@@endlink[0]{}%
\providecommand \url  [0]{\begingroup\@sanitize@url \@url }%
\providecommand \@url [1]{\endgroup\@href {#1}{\urlprefix }}%
\providecommand \urlprefix  [0]{URL }%
\providecommand \Eprint [0]{\href }%
\providecommand \doibase [0]{https://doi.org/}%
\providecommand \selectlanguage [0]{\@gobble}%
\providecommand \bibinfo  [0]{\@secondoftwo}%
\providecommand \bibfield  [0]{\@secondoftwo}%
\providecommand \translation [1]{[#1]}%
\providecommand \BibitemOpen [0]{}%
\providecommand \bibitemStop [0]{}%
\providecommand \bibitemNoStop [0]{.\EOS\space}%
\providecommand \EOS [0]{\spacefactor3000\relax}%
\providecommand \BibitemShut  [1]{\csname bibitem#1\endcsname}%
\let\auto@bib@innerbib\@empty
\bibitem [{\citenamefont {Cole}(1948)}]{cole1948underwater}%
  \BibitemOpen
  \bibfield  {author} {\bibinfo {author} {\bibfnamefont {R.~H.}\ \bibnamefont {Cole}},\ }\bibfield  {title} {\enquote {\bibinfo {title} {Underwater explosions},}\ }\href@noop {} {\bibfield  {journal} {\bibinfo  {journal} {(No Title)}\ } (\bibinfo {year} {1948})}\BibitemShut {NoStop}%
\bibitem [{\citenamefont {Denner}(2024)}]{denner2024kirkwood}%
  \BibitemOpen
  \bibfield  {author} {\bibinfo {author} {\bibfnamefont {F.}~\bibnamefont {Denner}},\ }\bibfield  {title} {\enquote {\bibinfo {title} {The kirkwood--bethe hypothesis for bubble dynamics, cavitation, and underwater explosions},}\ }\href@noop {} {\bibfield  {journal} {\bibinfo  {journal} {Physics of Fluids}\ }\textbf {\bibinfo {volume} {36}} (\bibinfo {year} {2024})}\BibitemShut {NoStop}%
\bibitem [{\citenamefont {Yusvika}\ \emph {et~al.}(2020)\citenamefont {Yusvika}, \citenamefont {Prabowo}, \citenamefont {Baek},\ and\ \citenamefont {Tjahjana}}]{yusvika2020achievements}%
  \BibitemOpen
  \bibfield  {author} {\bibinfo {author} {\bibfnamefont {M.}~\bibnamefont {Yusvika}}, \bibinfo {author} {\bibfnamefont {A.~R.}\ \bibnamefont {Prabowo}}, \bibinfo {author} {\bibfnamefont {S.~J.}\ \bibnamefont {Baek}},\ and\ \bibinfo {author} {\bibfnamefont {D.~D. D.~P.}\ \bibnamefont {Tjahjana}},\ }\bibfield  {title} {\enquote {\bibinfo {title} {Achievements in observation and prediction of cavitation: Effect and damage on the ship propellers},}\ }\href@noop {} {\bibfield  {journal} {\bibinfo  {journal} {Procedia Structural Integrity}\ }\textbf {\bibinfo {volume} {27}},\ \bibinfo {pages} {109--116} (\bibinfo {year} {2020})}\BibitemShut {NoStop}%
\bibitem [{\citenamefont {Bertin}\ and\ \citenamefont {Cummings}(2006)}]{bertin2006critical}%
  \BibitemOpen
  \bibfield  {author} {\bibinfo {author} {\bibfnamefont {J.~J.}\ \bibnamefont {Bertin}}\ and\ \bibinfo {author} {\bibfnamefont {R.~M.}\ \bibnamefont {Cummings}},\ }\bibfield  {title} {\enquote {\bibinfo {title} {Critical hypersonic aerothermodynamic phenomena},}\ }\href@noop {} {\bibfield  {journal} {\bibinfo  {journal} {Annu. Rev. Fluid Mech.}\ }\textbf {\bibinfo {volume} {38}},\ \bibinfo {pages} {129--157} (\bibinfo {year} {2006})}\BibitemShut {NoStop}%
\bibitem [{\citenamefont {Gaffney}\ \emph {et~al.}(2018)\citenamefont {Gaffney}, \citenamefont {Hu}, \citenamefont {Arnault}, \citenamefont {Becker}, \citenamefont {Benedict}, \citenamefont {Boehly}, \citenamefont {Celliers}, \citenamefont {Ceperley}, \citenamefont {{\v{C}}ert{\'\i}k}, \citenamefont {Cl{\'e}rouin} \emph {et~al.}}]{gaffney2018review}%
  \BibitemOpen
  \bibfield  {author} {\bibinfo {author} {\bibfnamefont {J.}~\bibnamefont {Gaffney}}, \bibinfo {author} {\bibfnamefont {S.}~\bibnamefont {Hu}}, \bibinfo {author} {\bibfnamefont {P.}~\bibnamefont {Arnault}}, \bibinfo {author} {\bibfnamefont {A.}~\bibnamefont {Becker}}, \bibinfo {author} {\bibfnamefont {L.}~\bibnamefont {Benedict}}, \bibinfo {author} {\bibfnamefont {T.}~\bibnamefont {Boehly}}, \bibinfo {author} {\bibfnamefont {P.}~\bibnamefont {Celliers}}, \bibinfo {author} {\bibfnamefont {D.}~\bibnamefont {Ceperley}}, \bibinfo {author} {\bibfnamefont {O.}~\bibnamefont {{\v{C}}ert{\'\i}k}}, \bibinfo {author} {\bibfnamefont {J.}~\bibnamefont {Cl{\'e}rouin}}, \emph {et~al.},\ }\bibfield  {title} {\enquote {\bibinfo {title} {A review of equation-of-state models for inertial confinement fusion materials},}\ }\href@noop {} {\bibfield  {journal} {\bibinfo  {journal} {High Energy Density Physics}\ }\textbf {\bibinfo {volume} {28}},\ \bibinfo {pages} {7--24} (\bibinfo {year} {2018})}\BibitemShut {NoStop}%
\bibitem [{\citenamefont {Mendible}\ \emph {et~al.}(2021)\citenamefont {Mendible}, \citenamefont {Koch}, \citenamefont {Lange}, \citenamefont {Brunton},\ and\ \citenamefont {Kutz}}]{mendible2021data}%
  \BibitemOpen
  \bibfield  {author} {\bibinfo {author} {\bibfnamefont {A.}~\bibnamefont {Mendible}}, \bibinfo {author} {\bibfnamefont {J.}~\bibnamefont {Koch}}, \bibinfo {author} {\bibfnamefont {H.}~\bibnamefont {Lange}}, \bibinfo {author} {\bibfnamefont {S.~L.}\ \bibnamefont {Brunton}},\ and\ \bibinfo {author} {\bibfnamefont {J.~N.}\ \bibnamefont {Kutz}},\ }\bibfield  {title} {\enquote {\bibinfo {title} {Data-driven modeling of rotating detonation waves},}\ }\href@noop {} {\bibfield  {journal} {\bibinfo  {journal} {Physical Review Fluids}\ }\textbf {\bibinfo {volume} {6}},\ \bibinfo {pages} {050507} (\bibinfo {year} {2021})}\BibitemShut {NoStop}%
\bibitem [{\citenamefont {Tong}\ \emph {et~al.}(2022)\citenamefont {Tong}, \citenamefont {Wang}, \citenamefont {Yan},\ and\ \citenamefont {Li}}]{tong2022fluid}%
  \BibitemOpen
  \bibfield  {author} {\bibinfo {author} {\bibfnamefont {S.-Y.}\ \bibnamefont {Tong}}, \bibinfo {author} {\bibfnamefont {S.-P.}\ \bibnamefont {Wang}}, \bibinfo {author} {\bibfnamefont {S.}~\bibnamefont {Yan}},\ and\ \bibinfo {author} {\bibfnamefont {S.}~\bibnamefont {Li}},\ }\bibfield  {title} {\enquote {\bibinfo {title} {Fluid--structure interactions between a near-field underwater explosion bubble and a suspended plate},}\ }\href@noop {} {\bibfield  {journal} {\bibinfo  {journal} {AIP Advances}\ }\textbf {\bibinfo {volume} {12}} (\bibinfo {year} {2022})}\BibitemShut {NoStop}%
\bibitem [{\citenamefont {Haskell}\ \emph {et~al.}(2023)\citenamefont {Haskell}, \citenamefont {Lu}, \citenamefont {Stocker}, \citenamefont {Xu},\ and\ \citenamefont {Sukovich}}]{haskell2023monitoring}%
  \BibitemOpen
  \bibfield  {author} {\bibinfo {author} {\bibfnamefont {S.~C.}\ \bibnamefont {Haskell}}, \bibinfo {author} {\bibfnamefont {N.}~\bibnamefont {Lu}}, \bibinfo {author} {\bibfnamefont {G.~E.}\ \bibnamefont {Stocker}}, \bibinfo {author} {\bibfnamefont {Z.}~\bibnamefont {Xu}},\ and\ \bibinfo {author} {\bibfnamefont {J.~R.}\ \bibnamefont {Sukovich}},\ }\bibfield  {title} {\enquote {\bibinfo {title} {Monitoring cavitation dynamics evolution in tissue mimicking hydrogels for repeated exposures via acoustic cavitation emissions},}\ }\href@noop {} {\bibfield  {journal} {\bibinfo  {journal} {The Journal of the Acoustical Society of America}\ }\textbf {\bibinfo {volume} {153}},\ \bibinfo {pages} {237--247} (\bibinfo {year} {2023})}\BibitemShut {NoStop}%
\bibitem [{\citenamefont {Brunton}\ and\ \citenamefont {Kutz}(2022)}]{brunton2022data}%
  \BibitemOpen
  \bibfield  {author} {\bibinfo {author} {\bibfnamefont {S.~L.}\ \bibnamefont {Brunton}}\ and\ \bibinfo {author} {\bibfnamefont {J.~N.}\ \bibnamefont {Kutz}},\ }\href@noop {} {\emph {\bibinfo {title} {Data-driven science and engineering: Machine learning, dynamical systems, and control}}}\ (\bibinfo  {publisher} {Cambridge University Press},\ \bibinfo {year} {2022})\BibitemShut {NoStop}%
\bibitem [{\citenamefont {Chen}, \citenamefont {Chiu},\ and\ \citenamefont {Fuge}(2019)}]{chen2019aerodynamic}%
  \BibitemOpen
  \bibfield  {author} {\bibinfo {author} {\bibfnamefont {W.}~\bibnamefont {Chen}}, \bibinfo {author} {\bibfnamefont {K.}~\bibnamefont {Chiu}},\ and\ \bibinfo {author} {\bibfnamefont {M.}~\bibnamefont {Fuge}},\ }\bibfield  {title} {\enquote {\bibinfo {title} {Aerodynamic design optimization and shape exploration using generative adversarial networks},}\ }in\ \href@noop {} {\emph {\bibinfo {booktitle} {AIAA Scitech 2019 forum}}}\ (\bibinfo {year} {2019})\ p.\ \bibinfo {pages} {2351}\BibitemShut {NoStop}%
\bibitem [{\citenamefont {Chen}\ \emph {et~al.}(2018)\citenamefont {Chen}, \citenamefont {Ni}, \citenamefont {Hu},\ and\ \citenamefont {Xue}}]{chen2018model}%
  \BibitemOpen
  \bibfield  {author} {\bibinfo {author} {\bibfnamefont {H.}~\bibnamefont {Chen}}, \bibinfo {author} {\bibfnamefont {B.}~\bibnamefont {Ni}}, \bibinfo {author} {\bibfnamefont {W.}~\bibnamefont {Hu}},\ and\ \bibinfo {author} {\bibfnamefont {Y.}~\bibnamefont {Xue}},\ }\bibfield  {title} {\enquote {\bibinfo {title} {Model experimental study of damage effects of ship structures under the contact jet loads of bubble in a water tank},}\ }\href@noop {} {\bibfield  {journal} {\bibinfo  {journal} {Shock and Vibration}\ }\textbf {\bibinfo {volume} {2018}},\ \bibinfo {pages} {8456925} (\bibinfo {year} {2018})}\BibitemShut {NoStop}%
\bibitem [{\citenamefont {Brennen}(2015)}]{brennen2015cavitation}%
  \BibitemOpen
  \bibfield  {author} {\bibinfo {author} {\bibfnamefont {C.~E.}\ \bibnamefont {Brennen}},\ }\bibfield  {title} {\enquote {\bibinfo {title} {Cavitation in medicine},}\ }\href@noop {} {\bibfield  {journal} {\bibinfo  {journal} {Interface focus}\ }\textbf {\bibinfo {volume} {5}},\ \bibinfo {pages} {20150022} (\bibinfo {year} {2015})}\BibitemShut {NoStop}%
\bibitem [{\citenamefont {Greif}\ and\ \citenamefont {Urban}(2019)}]{greif2019decay}%
  \BibitemOpen
  \bibfield  {author} {\bibinfo {author} {\bibfnamefont {C.}~\bibnamefont {Greif}}\ and\ \bibinfo {author} {\bibfnamefont {K.}~\bibnamefont {Urban}},\ }\bibfield  {title} {\enquote {\bibinfo {title} {Decay of the kolmogorov n-width for wave problems},}\ }\href@noop {} {\bibfield  {journal} {\bibinfo  {journal} {Applied Mathematics Letters}\ }\textbf {\bibinfo {volume} {96}},\ \bibinfo {pages} {216--222} (\bibinfo {year} {2019})}\BibitemShut {NoStop}%
\bibitem [{\citenamefont {Toro}(2013)}]{toro2013riemann}%
  \BibitemOpen
  \bibfield  {author} {\bibinfo {author} {\bibfnamefont {E.~F.}\ \bibnamefont {Toro}},\ }\href@noop {} {\emph {\bibinfo {title} {Riemann solvers and numerical methods for fluid dynamics: a practical introduction}}}\ (\bibinfo  {publisher} {Springer Science \& Business Media},\ \bibinfo {year} {2013})\BibitemShut {NoStop}%
\bibitem [{\citenamefont {Cao}\ \emph {et~al.}(2017)\citenamefont {Cao}, \citenamefont {Fei}, \citenamefont {Grosshans},\ and\ \citenamefont {Cao}}]{cao2017simulation}%
  \BibitemOpen
  \bibfield  {author} {\bibinfo {author} {\bibfnamefont {L.}~\bibnamefont {Cao}}, \bibinfo {author} {\bibfnamefont {W.}~\bibnamefont {Fei}}, \bibinfo {author} {\bibfnamefont {H.}~\bibnamefont {Grosshans}},\ and\ \bibinfo {author} {\bibfnamefont {N.}~\bibnamefont {Cao}},\ }\bibfield  {title} {\enquote {\bibinfo {title} {Simulation of underwater explosions initiated by high-pressure gas bubbles of various initial shapes},}\ }\href@noop {} {\bibfield  {journal} {\bibinfo  {journal} {Applied Sciences}\ }\textbf {\bibinfo {volume} {7}},\ \bibinfo {pages} {880} (\bibinfo {year} {2017})}\BibitemShut {NoStop}%
\bibitem [{\citenamefont {Kira}, \citenamefont {Fujita},\ and\ \citenamefont {Itoh}(1999)}]{kira1999underwater}%
  \BibitemOpen
  \bibfield  {author} {\bibinfo {author} {\bibfnamefont {A.}~\bibnamefont {Kira}}, \bibinfo {author} {\bibfnamefont {M.}~\bibnamefont {Fujita}},\ and\ \bibinfo {author} {\bibfnamefont {S.}~\bibnamefont {Itoh}},\ }\bibfield  {title} {\enquote {\bibinfo {title} {Underwater explosion of spherical explosives},}\ }\href@noop {} {\bibfield  {journal} {\bibinfo  {journal} {Journal of materials processing technology}\ }\textbf {\bibinfo {volume} {85}},\ \bibinfo {pages} {64--68} (\bibinfo {year} {1999})}\BibitemShut {NoStop}%
\bibitem [{\citenamefont {LeVeque}(2002)}]{leveque2002finite}%
  \BibitemOpen
  \bibfield  {author} {\bibinfo {author} {\bibfnamefont {R.~J.}\ \bibnamefont {LeVeque}},\ }\href@noop {} {\emph {\bibinfo {title} {Finite volume methods for hyperbolic problems}}},\ Vol.~\bibinfo {volume} {31}\ (\bibinfo  {publisher} {Cambridge university press},\ \bibinfo {year} {2002})\BibitemShut {NoStop}%
\bibitem [{\citenamefont {Fikl}\ \emph {et~al.}(2016)\citenamefont {Fikl}, \citenamefont {Le~Chenadec}, \citenamefont {Sayadi},\ and\ \citenamefont {Schmid}}]{fikl2016comprehensive}%
  \BibitemOpen
  \bibfield  {author} {\bibinfo {author} {\bibfnamefont {A.}~\bibnamefont {Fikl}}, \bibinfo {author} {\bibfnamefont {V.}~\bibnamefont {Le~Chenadec}}, \bibinfo {author} {\bibfnamefont {T.}~\bibnamefont {Sayadi}},\ and\ \bibinfo {author} {\bibfnamefont {P.}~\bibnamefont {Schmid}},\ }\bibfield  {title} {\enquote {\bibinfo {title} {A comprehensive study of adjoint-based optimization of non-linear systems with application to burgers’ equation},}\ }in\ \href@noop {} {\emph {\bibinfo {booktitle} {46th AIAA Fluid Dynamics Conference}}}\ (\bibinfo {year} {2016})\ p.\ \bibinfo {pages} {3805}\BibitemShut {NoStop}%
\bibitem [{\citenamefont {Kirkwood}\ and\ \citenamefont {Bethe}(1942)}]{kirkwood1942progress}%
  \BibitemOpen
  \bibfield  {author} {\bibinfo {author} {\bibfnamefont {J.}~\bibnamefont {Kirkwood}}\ and\ \bibinfo {author} {\bibfnamefont {H.}~\bibnamefont {Bethe}},\ }\bibfield  {title} {\enquote {\bibinfo {title} {Progress report on ‘the pressure wave produced by an underwater explosion i’: Technical report no. 588},}\ }\href@noop {} {\bibfield  {journal} {\bibinfo  {journal} {Office of Scientific Research and Development, US Navy}\ }\textbf {\bibinfo {volume} {17}} (\bibinfo {year} {1942})}\BibitemShut {NoStop}%
\bibitem [{\citenamefont {Zhang}\ \emph {et~al.}(2021{\natexlab{a}})\citenamefont {Zhang}, \citenamefont {Wang}, \citenamefont {Jia}, \citenamefont {Gao},\ and\ \citenamefont {Ma}}]{zhang2021engineering}%
  \BibitemOpen
  \bibfield  {author} {\bibinfo {author} {\bibfnamefont {J.}~\bibnamefont {Zhang}}, \bibinfo {author} {\bibfnamefont {S.}~\bibnamefont {Wang}}, \bibinfo {author} {\bibfnamefont {X.}~\bibnamefont {Jia}}, \bibinfo {author} {\bibfnamefont {Y.}~\bibnamefont {Gao}},\ and\ \bibinfo {author} {\bibfnamefont {F.}~\bibnamefont {Ma}},\ }\bibfield  {title} {\enquote {\bibinfo {title} {An engineering application of prosperetti and lezzi equation to solve underwater explosion bubbles},}\ }\href@noop {} {\bibfield  {journal} {\bibinfo  {journal} {Physics of Fluids}\ }\textbf {\bibinfo {volume} {33}} (\bibinfo {year} {2021}{\natexlab{a}})}\BibitemShut {NoStop}%
\bibitem [{\citenamefont {Zhang}\ \emph {et~al.}(2021{\natexlab{b}})\citenamefont {Zhang}, \citenamefont {Wang}, \citenamefont {Jia}, \citenamefont {Gao},\ and\ \citenamefont {Ma}}]{zhang2021improved}%
  \BibitemOpen
  \bibfield  {author} {\bibinfo {author} {\bibfnamefont {J.}~\bibnamefont {Zhang}}, \bibinfo {author} {\bibfnamefont {S.}~\bibnamefont {Wang}}, \bibinfo {author} {\bibfnamefont {X.}~\bibnamefont {Jia}}, \bibinfo {author} {\bibfnamefont {Y.}~\bibnamefont {Gao}},\ and\ \bibinfo {author} {\bibfnamefont {F.}~\bibnamefont {Ma}},\ }\bibfield  {title} {\enquote {\bibinfo {title} {An improved kirkwood--bethe model for calculating near-field shockwave propagation of underwater explosions},}\ }\href@noop {} {\bibfield  {journal} {\bibinfo  {journal} {AIP Advances}\ }\textbf {\bibinfo {volume} {11}} (\bibinfo {year} {2021}{\natexlab{b}})}\BibitemShut {NoStop}%
\bibitem [{\citenamefont {Raissi}, \citenamefont {Perdikaris},\ and\ \citenamefont {Karniadakis}(2019)}]{raissi2019physics}%
  \BibitemOpen
  \bibfield  {author} {\bibinfo {author} {\bibfnamefont {M.}~\bibnamefont {Raissi}}, \bibinfo {author} {\bibfnamefont {P.}~\bibnamefont {Perdikaris}},\ and\ \bibinfo {author} {\bibfnamefont {G.~E.}\ \bibnamefont {Karniadakis}},\ }\bibfield  {title} {\enquote {\bibinfo {title} {Physics-informed neural networks: A deep learning framework for solving forward and inverse problems involving nonlinear partial differential equations},}\ }\href@noop {} {\bibfield  {journal} {\bibinfo  {journal} {Journal of Computational physics}\ }\textbf {\bibinfo {volume} {378}},\ \bibinfo {pages} {686--707} (\bibinfo {year} {2019})}\BibitemShut {NoStop}%
\bibitem [{\citenamefont {Mao}, \citenamefont {Jagtap},\ and\ \citenamefont {Karniadakis}(2020)}]{mao2020physics}%
  \BibitemOpen
  \bibfield  {author} {\bibinfo {author} {\bibfnamefont {Z.}~\bibnamefont {Mao}}, \bibinfo {author} {\bibfnamefont {A.~D.}\ \bibnamefont {Jagtap}},\ and\ \bibinfo {author} {\bibfnamefont {G.~E.}\ \bibnamefont {Karniadakis}},\ }\bibfield  {title} {\enquote {\bibinfo {title} {Physics-informed neural networks for high-speed flows},}\ }\href@noop {} {\bibfield  {journal} {\bibinfo  {journal} {Computer Methods in Applied Mechanics and Engineering}\ }\textbf {\bibinfo {volume} {360}},\ \bibinfo {pages} {112789} (\bibinfo {year} {2020})}\BibitemShut {NoStop}%
\bibitem [{\citenamefont {Taira}\ \emph {et~al.}(2017)\citenamefont {Taira}, \citenamefont {Brunton}, \citenamefont {Dawson}, \citenamefont {Rowley}, \citenamefont {Colonius}, \citenamefont {McKeon}, \citenamefont {Schmidt}, \citenamefont {Gordeyev}, \citenamefont {Theofilis},\ and\ \citenamefont {Ukeiley}}]{taira2017modal}%
  \BibitemOpen
  \bibfield  {author} {\bibinfo {author} {\bibfnamefont {K.}~\bibnamefont {Taira}}, \bibinfo {author} {\bibfnamefont {S.~L.}\ \bibnamefont {Brunton}}, \bibinfo {author} {\bibfnamefont {S.~T.}\ \bibnamefont {Dawson}}, \bibinfo {author} {\bibfnamefont {C.~W.}\ \bibnamefont {Rowley}}, \bibinfo {author} {\bibfnamefont {T.}~\bibnamefont {Colonius}}, \bibinfo {author} {\bibfnamefont {B.~J.}\ \bibnamefont {McKeon}}, \bibinfo {author} {\bibfnamefont {O.~T.}\ \bibnamefont {Schmidt}}, \bibinfo {author} {\bibfnamefont {S.}~\bibnamefont {Gordeyev}}, \bibinfo {author} {\bibfnamefont {V.}~\bibnamefont {Theofilis}},\ and\ \bibinfo {author} {\bibfnamefont {L.~S.}\ \bibnamefont {Ukeiley}},\ }\bibfield  {title} {\enquote {\bibinfo {title} {Modal analysis of fluid flows: An overview},}\ }\href@noop {} {\bibfield  {journal} {\bibinfo  {journal} {Aiaa Journal}\ }\textbf {\bibinfo {volume} {55}},\ \bibinfo {pages} {4013--4041} (\bibinfo {year} {2017})}\BibitemShut {NoStop}%
\bibitem [{\citenamefont {Mendible}\ \emph {et~al.}(2020)\citenamefont {Mendible}, \citenamefont {Brunton}, \citenamefont {Aravkin}, \citenamefont {Lowrie},\ and\ \citenamefont {Kutz}}]{mendible2020dimensionality}%
  \BibitemOpen
  \bibfield  {author} {\bibinfo {author} {\bibfnamefont {A.}~\bibnamefont {Mendible}}, \bibinfo {author} {\bibfnamefont {S.~L.}\ \bibnamefont {Brunton}}, \bibinfo {author} {\bibfnamefont {A.~Y.}\ \bibnamefont {Aravkin}}, \bibinfo {author} {\bibfnamefont {W.}~\bibnamefont {Lowrie}},\ and\ \bibinfo {author} {\bibfnamefont {J.~N.}\ \bibnamefont {Kutz}},\ }\bibfield  {title} {\enquote {\bibinfo {title} {Dimensionality reduction and reduced-order modeling for traveling wave physics},}\ }\href@noop {} {\bibfield  {journal} {\bibinfo  {journal} {Theoretical and Computational Fluid Dynamics}\ }\textbf {\bibinfo {volume} {34}},\ \bibinfo {pages} {385--400} (\bibinfo {year} {2020})}\BibitemShut {NoStop}%
\bibitem [{\citenamefont {Lu}, \citenamefont {Jin},\ and\ \citenamefont {Karniadakis}(2019)}]{lu2019deeponet}%
  \BibitemOpen
  \bibfield  {author} {\bibinfo {author} {\bibfnamefont {L.}~\bibnamefont {Lu}}, \bibinfo {author} {\bibfnamefont {P.}~\bibnamefont {Jin}},\ and\ \bibinfo {author} {\bibfnamefont {G.~E.}\ \bibnamefont {Karniadakis}},\ }\bibfield  {title} {\enquote {\bibinfo {title} {Deeponet: Learning nonlinear operators for identifying differential equations based on the universal approximation theorem of operators},}\ }\href@noop {} {\bibfield  {journal} {\bibinfo  {journal} {arXiv preprint arXiv:1910.03193}\ } (\bibinfo {year} {2019})}\BibitemShut {NoStop}%
\bibitem [{\citenamefont {Kovachki}\ \emph {et~al.}(2023)\citenamefont {Kovachki}, \citenamefont {Li}, \citenamefont {Liu}, \citenamefont {Azizzadenesheli}, \citenamefont {Bhattacharya}, \citenamefont {Stuart},\ and\ \citenamefont {Anandkumar}}]{kovachki2023neural}%
  \BibitemOpen
  \bibfield  {author} {\bibinfo {author} {\bibfnamefont {N.}~\bibnamefont {Kovachki}}, \bibinfo {author} {\bibfnamefont {Z.}~\bibnamefont {Li}}, \bibinfo {author} {\bibfnamefont {B.}~\bibnamefont {Liu}}, \bibinfo {author} {\bibfnamefont {K.}~\bibnamefont {Azizzadenesheli}}, \bibinfo {author} {\bibfnamefont {K.}~\bibnamefont {Bhattacharya}}, \bibinfo {author} {\bibfnamefont {A.}~\bibnamefont {Stuart}},\ and\ \bibinfo {author} {\bibfnamefont {A.}~\bibnamefont {Anandkumar}},\ }\bibfield  {title} {\enquote {\bibinfo {title} {Neural operator: Learning maps between function spaces with applications to pdes},}\ }\href@noop {} {\bibfield  {journal} {\bibinfo  {journal} {Journal of Machine Learning Research}\ }\textbf {\bibinfo {volume} {24}},\ \bibinfo {pages} {1--97} (\bibinfo {year} {2023})}\BibitemShut {NoStop}%
\bibitem [{\citenamefont {Li}\ \emph {et~al.}(2024{\natexlab{a}})\citenamefont {Li}, \citenamefont {Zheng}, \citenamefont {Kovachki}, \citenamefont {Jin}, \citenamefont {Chen}, \citenamefont {Liu}, \citenamefont {Azizzadenesheli},\ and\ \citenamefont {Anandkumar}}]{li2024physics}%
  \BibitemOpen
  \bibfield  {author} {\bibinfo {author} {\bibfnamefont {Z.}~\bibnamefont {Li}}, \bibinfo {author} {\bibfnamefont {H.}~\bibnamefont {Zheng}}, \bibinfo {author} {\bibfnamefont {N.}~\bibnamefont {Kovachki}}, \bibinfo {author} {\bibfnamefont {D.}~\bibnamefont {Jin}}, \bibinfo {author} {\bibfnamefont {H.}~\bibnamefont {Chen}}, \bibinfo {author} {\bibfnamefont {B.}~\bibnamefont {Liu}}, \bibinfo {author} {\bibfnamefont {K.}~\bibnamefont {Azizzadenesheli}},\ and\ \bibinfo {author} {\bibfnamefont {A.}~\bibnamefont {Anandkumar}},\ }\bibfield  {title} {\enquote {\bibinfo {title} {Physics-informed neural operator for learning partial differential equations},}\ }\href@noop {} {\bibfield  {journal} {\bibinfo  {journal} {ACM/JMS Journal of Data Science}\ }\textbf {\bibinfo {volume} {1}},\ \bibinfo {pages} {1--27} (\bibinfo {year} {2024}{\natexlab{a}})}\BibitemShut {NoStop}%
\bibitem [{\citenamefont {Mao}\ \emph {et~al.}(2021)\citenamefont {Mao}, \citenamefont {Lu}, \citenamefont {Marxen}, \citenamefont {Zaki},\ and\ \citenamefont {Karniadakis}}]{mao2021deepm}%
  \BibitemOpen
  \bibfield  {author} {\bibinfo {author} {\bibfnamefont {Z.}~\bibnamefont {Mao}}, \bibinfo {author} {\bibfnamefont {L.}~\bibnamefont {Lu}}, \bibinfo {author} {\bibfnamefont {O.}~\bibnamefont {Marxen}}, \bibinfo {author} {\bibfnamefont {T.~A.}\ \bibnamefont {Zaki}},\ and\ \bibinfo {author} {\bibfnamefont {G.~E.}\ \bibnamefont {Karniadakis}},\ }\bibfield  {title} {\enquote {\bibinfo {title} {Deepm\&mnet for hypersonics: Predicting the coupled flow and finite-rate chemistry behind a normal shock using neural-network approximation of operators},}\ }\href@noop {} {\bibfield  {journal} {\bibinfo  {journal} {Journal of computational physics}\ }\textbf {\bibinfo {volume} {447}},\ \bibinfo {pages} {110698} (\bibinfo {year} {2021})}\BibitemShut {NoStop}%
\bibitem [{\citenamefont {Patel}\ \emph {et~al.}(2022)\citenamefont {Patel}, \citenamefont {Manickam}, \citenamefont {Trask}, \citenamefont {Wood}, \citenamefont {Lee}, \citenamefont {Tomas},\ and\ \citenamefont {Cyr}}]{patel2022thermodynamically}%
  \BibitemOpen
  \bibfield  {author} {\bibinfo {author} {\bibfnamefont {R.~G.}\ \bibnamefont {Patel}}, \bibinfo {author} {\bibfnamefont {I.}~\bibnamefont {Manickam}}, \bibinfo {author} {\bibfnamefont {N.~A.}\ \bibnamefont {Trask}}, \bibinfo {author} {\bibfnamefont {M.~A.}\ \bibnamefont {Wood}}, \bibinfo {author} {\bibfnamefont {M.}~\bibnamefont {Lee}}, \bibinfo {author} {\bibfnamefont {I.}~\bibnamefont {Tomas}},\ and\ \bibinfo {author} {\bibfnamefont {E.~C.}\ \bibnamefont {Cyr}},\ }\bibfield  {title} {\enquote {\bibinfo {title} {Thermodynamically consistent physics-informed neural networks for hyperbolic systems},}\ }\href@noop {} {\bibfield  {journal} {\bibinfo  {journal} {Journal of Computational Physics}\ }\textbf {\bibinfo {volume} {449}},\ \bibinfo {pages} {110754} (\bibinfo {year} {2022})}\BibitemShut {NoStop}%
\bibitem [{\citenamefont {Jagtap}\ \emph {et~al.}(2022)\citenamefont {Jagtap}, \citenamefont {Mao}, \citenamefont {Adams},\ and\ \citenamefont {Karniadakis}}]{jagtap2022physics}%
  \BibitemOpen
  \bibfield  {author} {\bibinfo {author} {\bibfnamefont {A.~D.}\ \bibnamefont {Jagtap}}, \bibinfo {author} {\bibfnamefont {Z.}~\bibnamefont {Mao}}, \bibinfo {author} {\bibfnamefont {N.}~\bibnamefont {Adams}},\ and\ \bibinfo {author} {\bibfnamefont {G.~E.}\ \bibnamefont {Karniadakis}},\ }\bibfield  {title} {\enquote {\bibinfo {title} {Physics-informed neural networks for inverse problems in supersonic flows},}\ }\href@noop {} {\bibfield  {journal} {\bibinfo  {journal} {Journal of Computational Physics}\ }\textbf {\bibinfo {volume} {466}},\ \bibinfo {pages} {111402} (\bibinfo {year} {2022})}\BibitemShut {NoStop}%
\bibitem [{\citenamefont {Jagtap}, \citenamefont {Kharazmi},\ and\ \citenamefont {Karniadakis}(2020)}]{jagtap2020conservative}%
  \BibitemOpen
  \bibfield  {author} {\bibinfo {author} {\bibfnamefont {A.~D.}\ \bibnamefont {Jagtap}}, \bibinfo {author} {\bibfnamefont {E.}~\bibnamefont {Kharazmi}},\ and\ \bibinfo {author} {\bibfnamefont {G.~E.}\ \bibnamefont {Karniadakis}},\ }\bibfield  {title} {\enquote {\bibinfo {title} {Conservative physics-informed neural networks on discrete domains for conservation laws: Applications to forward and inverse problems},}\ }\href@noop {} {\bibfield  {journal} {\bibinfo  {journal} {Computer Methods in Applied Mechanics and Engineering}\ }\textbf {\bibinfo {volume} {365}},\ \bibinfo {pages} {113028} (\bibinfo {year} {2020})}\BibitemShut {NoStop}%
\bibitem [{\citenamefont {Qiu}\ \emph {et~al.}(2022)\citenamefont {Qiu}, \citenamefont {Huang}, \citenamefont {Xiao}, \citenamefont {Wang}, \citenamefont {Zhang}, \citenamefont {Yue}, \citenamefont {Zeng},\ and\ \citenamefont {Wang}}]{qiu2022physics}%
  \BibitemOpen
  \bibfield  {author} {\bibinfo {author} {\bibfnamefont {R.}~\bibnamefont {Qiu}}, \bibinfo {author} {\bibfnamefont {R.}~\bibnamefont {Huang}}, \bibinfo {author} {\bibfnamefont {Y.}~\bibnamefont {Xiao}}, \bibinfo {author} {\bibfnamefont {J.}~\bibnamefont {Wang}}, \bibinfo {author} {\bibfnamefont {Z.}~\bibnamefont {Zhang}}, \bibinfo {author} {\bibfnamefont {J.}~\bibnamefont {Yue}}, \bibinfo {author} {\bibfnamefont {Z.}~\bibnamefont {Zeng}},\ and\ \bibinfo {author} {\bibfnamefont {Y.}~\bibnamefont {Wang}},\ }\bibfield  {title} {\enquote {\bibinfo {title} {Physics-informed neural networks for phase-field method in two-phase flow},}\ }\href@noop {} {\bibfield  {journal} {\bibinfo  {journal} {Physics of Fluids}\ }\textbf {\bibinfo {volume} {34}} (\bibinfo {year} {2022})}\BibitemShut {NoStop}%
\bibitem [{\citenamefont {Wang}\ and\ \citenamefont {Perdikaris}(2021)}]{wang2021deep}%
  \BibitemOpen
  \bibfield  {author} {\bibinfo {author} {\bibfnamefont {S.}~\bibnamefont {Wang}}\ and\ \bibinfo {author} {\bibfnamefont {P.}~\bibnamefont {Perdikaris}},\ }\bibfield  {title} {\enquote {\bibinfo {title} {Deep learning of free boundary and stefan problems},}\ }\href@noop {} {\bibfield  {journal} {\bibinfo  {journal} {Journal of Computational Physics}\ }\textbf {\bibinfo {volume} {428}},\ \bibinfo {pages} {109914} (\bibinfo {year} {2021})}\BibitemShut {NoStop}%
\bibitem [{\citenamefont {Buhendwa}, \citenamefont {Adami},\ and\ \citenamefont {Adams}(2021)}]{buhendwa2021inferring}%
  \BibitemOpen
  \bibfield  {author} {\bibinfo {author} {\bibfnamefont {A.~B.}\ \bibnamefont {Buhendwa}}, \bibinfo {author} {\bibfnamefont {S.}~\bibnamefont {Adami}},\ and\ \bibinfo {author} {\bibfnamefont {N.~A.}\ \bibnamefont {Adams}},\ }\bibfield  {title} {\enquote {\bibinfo {title} {Inferring incompressible two-phase flow fields from the interface motion using physics-informed neural networks},}\ }\href@noop {} {\bibfield  {journal} {\bibinfo  {journal} {Machine Learning with Applications}\ }\textbf {\bibinfo {volume} {4}},\ \bibinfo {pages} {100029} (\bibinfo {year} {2021})}\BibitemShut {NoStop}%
\bibitem [{\citenamefont {Cheng}\ \emph {et~al.}(2024)\citenamefont {Cheng}, \citenamefont {Nguyen}, \citenamefont {Seshadri}, \citenamefont {Verma}, \citenamefont {Gray}, \citenamefont {Beerman}, \citenamefont {Udaykumar},\ and\ \citenamefont {Baek}}]{cheng2024physics}%
  \BibitemOpen
  \bibfield  {author} {\bibinfo {author} {\bibfnamefont {X.}~\bibnamefont {Cheng}}, \bibinfo {author} {\bibfnamefont {P.~C.}\ \bibnamefont {Nguyen}}, \bibinfo {author} {\bibfnamefont {P.~K.}\ \bibnamefont {Seshadri}}, \bibinfo {author} {\bibfnamefont {M.}~\bibnamefont {Verma}}, \bibinfo {author} {\bibfnamefont {Z.~J.}\ \bibnamefont {Gray}}, \bibinfo {author} {\bibfnamefont {J.~T.}\ \bibnamefont {Beerman}}, \bibinfo {author} {\bibfnamefont {H.}~\bibnamefont {Udaykumar}},\ and\ \bibinfo {author} {\bibfnamefont {S.~S.}\ \bibnamefont {Baek}},\ }\bibfield  {title} {\enquote {\bibinfo {title} {Physics-aware recurrent convolutional neural networks for modeling multiphase compressible flows},}\ }\href@noop {} {\bibfield  {journal} {\bibinfo  {journal} {International Journal of Multiphase Flow}\ ,\ \bibinfo {pages} {104877}} (\bibinfo {year} {2024})}\BibitemShut {NoStop}%
\bibitem [{\citenamefont {Nguyen}\ \emph {et~al.}(2023)\citenamefont {Nguyen}, \citenamefont {Nguyen}, \citenamefont {Choi}, \citenamefont {Seshadri}, \citenamefont {Udaykumar},\ and\ \citenamefont {Baek}}]{nguyen2023parc}%
  \BibitemOpen
  \bibfield  {author} {\bibinfo {author} {\bibfnamefont {P.~C.}\ \bibnamefont {Nguyen}}, \bibinfo {author} {\bibfnamefont {Y.-T.}\ \bibnamefont {Nguyen}}, \bibinfo {author} {\bibfnamefont {J.~B.}\ \bibnamefont {Choi}}, \bibinfo {author} {\bibfnamefont {P.~K.}\ \bibnamefont {Seshadri}}, \bibinfo {author} {\bibfnamefont {H.}~\bibnamefont {Udaykumar}},\ and\ \bibinfo {author} {\bibfnamefont {S.~S.}\ \bibnamefont {Baek}},\ }\bibfield  {title} {\enquote {\bibinfo {title} {Parc: Physics-aware recurrent convolutional neural networks to assimilate meso scale reactive mechanics of energetic materials},}\ }\href@noop {} {\bibfield  {journal} {\bibinfo  {journal} {Science advances}\ }\textbf {\bibinfo {volume} {9}},\ \bibinfo {pages} {eadd6868} (\bibinfo {year} {2023})}\BibitemShut {NoStop}%
\bibitem [{\citenamefont {Wardlaw}, \citenamefont {McKeown},\ and\ \citenamefont {Luton}(1998)}]{wardlaw1998coupled}%
  \BibitemOpen
  \bibfield  {author} {\bibinfo {author} {\bibfnamefont {A.}~\bibnamefont {Wardlaw}}, \bibinfo {author} {\bibfnamefont {R.}~\bibnamefont {McKeown}},\ and\ \bibinfo {author} {\bibfnamefont {A.}~\bibnamefont {Luton}},\ }\bibfield  {title} {\enquote {\bibinfo {title} {Coupled hydrocode prediction of underwater explosion damage},}\ }\href@noop {} {\bibfield  {journal} {\bibinfo  {journal} {Indian Head, MD: Naval Surface Warfare Center Indian Head Division (NSWC IHD)}\ } (\bibinfo {year} {1998})}\BibitemShut {NoStop}%
\bibitem [{\citenamefont {Wardlaw~Jr}\ and\ \citenamefont {Luton}(2000)}]{wardlaw2000fluid}%
  \BibitemOpen
  \bibfield  {author} {\bibinfo {author} {\bibfnamefont {A.~B.}\ \bibnamefont {Wardlaw~Jr}}\ and\ \bibinfo {author} {\bibfnamefont {J.~A.}\ \bibnamefont {Luton}},\ }\bibfield  {title} {\enquote {\bibinfo {title} {Fluid-structure interaction mechanisms for close-in explosions},}\ }\href@noop {} {\bibfield  {journal} {\bibinfo  {journal} {Shock and Vibration}\ }\textbf {\bibinfo {volume} {7}},\ \bibinfo {pages} {265--275} (\bibinfo {year} {2000})}\BibitemShut {NoStop}%
\bibitem [{\citenamefont {Liu}\ \emph {et~al.}(2018)\citenamefont {Liu}, \citenamefont {Ming}, \citenamefont {Zhang}, \citenamefont {Miao},\ and\ \citenamefont {Liu}}]{liu2018continuous}%
  \BibitemOpen
  \bibfield  {author} {\bibinfo {author} {\bibfnamefont {W.}~\bibnamefont {Liu}}, \bibinfo {author} {\bibfnamefont {F.}~\bibnamefont {Ming}}, \bibinfo {author} {\bibfnamefont {A.}~\bibnamefont {Zhang}}, \bibinfo {author} {\bibfnamefont {X.}~\bibnamefont {Miao}},\ and\ \bibinfo {author} {\bibfnamefont {Y.}~\bibnamefont {Liu}},\ }\bibfield  {title} {\enquote {\bibinfo {title} {Continuous simulation of the whole process of underwater explosion based on eulerian finite element approach},}\ }\href@noop {} {\bibfield  {journal} {\bibinfo  {journal} {Applied Ocean Research}\ }\textbf {\bibinfo {volume} {80}},\ \bibinfo {pages} {125--135} (\bibinfo {year} {2018})}\BibitemShut {NoStop}%
\bibitem [{\citenamefont {Wardlaw~Jr}\ and\ \citenamefont {Mair}(1998)}]{wardlaw1998spherical}%
  \BibitemOpen
  \bibfield  {author} {\bibinfo {author} {\bibfnamefont {A.~B.}\ \bibnamefont {Wardlaw~Jr}}\ and\ \bibinfo {author} {\bibfnamefont {H.~U.}\ \bibnamefont {Mair}},\ }\bibfield  {title} {\enquote {\bibinfo {title} {Spherical solutions of an underwater explosion bubble},}\ }\href@noop {} {\bibfield  {journal} {\bibinfo  {journal} {Shock and Vibration}\ }\textbf {\bibinfo {volume} {5}},\ \bibinfo {pages} {89--102} (\bibinfo {year} {1998})}\BibitemShut {NoStop}%
\bibitem [{\citenamefont {Lee}, \citenamefont {Hornig},\ and\ \citenamefont {Kury}(1968)}]{lee1968adiabatic}%
  \BibitemOpen
  \bibfield  {author} {\bibinfo {author} {\bibfnamefont {E.}~\bibnamefont {Lee}}, \bibinfo {author} {\bibfnamefont {H.}~\bibnamefont {Hornig}},\ and\ \bibinfo {author} {\bibfnamefont {J.}~\bibnamefont {Kury}},\ }\href@noop {} {\enquote {\bibinfo {title} {Adiabatic expansion of high explosive detonation products},}\ }\bibinfo {type} {Tech. Rep.}\ (\bibinfo  {institution} {Univ. of California Radiation Lab. at Livermore, Livermore, CA (United States)},\ \bibinfo {year} {1968})\BibitemShut {NoStop}%
\bibitem [{\citenamefont {Dobratz}(1981)}]{osti_6530310}%
  \BibitemOpen
  \bibfield  {author} {\bibinfo {author} {\bibfnamefont {B.~M.}\ \bibnamefont {Dobratz}},\ }\bibfield  {title} {\enquote {\bibinfo {title} {Llnl explosives handbook: properties of chemical explosives and explosives and explosive simulants},}\ }\href {https://doi.org/10.2172/6530310} {\  (\bibinfo {year} {1981}),\ 10.2172/6530310}\BibitemShut {NoStop}%
\bibitem [{\citenamefont {Baker}\ \emph {et~al.}(2010)\citenamefont {Baker}, \citenamefont {Murphy}, \citenamefont {Stiel},\ and\ \citenamefont {Wrobel}}]{baker2010theory}%
  \BibitemOpen
  \bibfield  {author} {\bibinfo {author} {\bibfnamefont {E.}~\bibnamefont {Baker}}, \bibinfo {author} {\bibfnamefont {D.}~\bibnamefont {Murphy}}, \bibinfo {author} {\bibfnamefont {L.}~\bibnamefont {Stiel}},\ and\ \bibinfo {author} {\bibfnamefont {E.}~\bibnamefont {Wrobel}},\ }\bibfield  {title} {\enquote {\bibinfo {title} {Theory and calibration of jwl and jwlb thermodynamic equations of state},}\ }\href@noop {} {\bibfield  {journal} {\bibinfo  {journal} {WIT Transactions on The Built Environment}\ }\textbf {\bibinfo {volume} {113}},\ \bibinfo {pages} {147--158} (\bibinfo {year} {2010})}\BibitemShut {NoStop}%
\bibitem [{\citenamefont {Giam}, \citenamefont {Toh},\ and\ \citenamefont {Tan}(2020)}]{giam2020numerical}%
  \BibitemOpen
  \bibfield  {author} {\bibinfo {author} {\bibfnamefont {A.}~\bibnamefont {Giam}}, \bibinfo {author} {\bibfnamefont {W.}~\bibnamefont {Toh}},\ and\ \bibinfo {author} {\bibfnamefont {V.~B.~C.}\ \bibnamefont {Tan}},\ }\bibfield  {title} {\enquote {\bibinfo {title} {Numerical review of jones--wilkins--lee parameters for trinitrotoluene explosive in free-air blast},}\ }\href@noop {} {\bibfield  {journal} {\bibinfo  {journal} {Journal of Applied Mechanics}\ }\textbf {\bibinfo {volume} {87}},\ \bibinfo {pages} {051008} (\bibinfo {year} {2020})}\BibitemShut {NoStop}%
\bibitem [{\citenamefont {Jones}(1998)}]{jones1998numerical}%
  \BibitemOpen
  \bibfield  {author} {\bibinfo {author} {\bibfnamefont {D.}~\bibnamefont {Jones}},\ }\bibfield  {title} {\enquote {\bibinfo {title} {Numerical simulation of detonation in condensed phase explosives},}\ }\href@noop {} {\bibfield  {journal} {\bibinfo  {journal} {NASA}\ } (\bibinfo {year} {1998})}\BibitemShut {NoStop}%
\bibitem [{\citenamefont {Pedregosa}\ \emph {et~al.}(2011)\citenamefont {Pedregosa}, \citenamefont {Varoquaux}, \citenamefont {Gramfort}, \citenamefont {Michel}, \citenamefont {Thirion}, \citenamefont {Grisel}, \citenamefont {Blondel}, \citenamefont {Prettenhofer}, \citenamefont {Weiss}, \citenamefont {Dubourg} \emph {et~al.}}]{pedregosa2011scikit}%
  \BibitemOpen
  \bibfield  {author} {\bibinfo {author} {\bibfnamefont {F.}~\bibnamefont {Pedregosa}}, \bibinfo {author} {\bibfnamefont {G.}~\bibnamefont {Varoquaux}}, \bibinfo {author} {\bibfnamefont {A.}~\bibnamefont {Gramfort}}, \bibinfo {author} {\bibfnamefont {V.}~\bibnamefont {Michel}}, \bibinfo {author} {\bibfnamefont {B.}~\bibnamefont {Thirion}}, \bibinfo {author} {\bibfnamefont {O.}~\bibnamefont {Grisel}}, \bibinfo {author} {\bibfnamefont {M.}~\bibnamefont {Blondel}}, \bibinfo {author} {\bibfnamefont {P.}~\bibnamefont {Prettenhofer}}, \bibinfo {author} {\bibfnamefont {R.}~\bibnamefont {Weiss}}, \bibinfo {author} {\bibfnamefont {V.}~\bibnamefont {Dubourg}}, \emph {et~al.},\ }\bibfield  {title} {\enquote {\bibinfo {title} {Scikit-learn: Machine learning in python},}\ }\href@noop {} {\bibfield  {journal} {\bibinfo  {journal} {Journal of machine learning research}\ }\textbf {\bibinfo {volume} {12}},\ \bibinfo {pages} {2825--2830} (\bibinfo {year} {2011})}\BibitemShut {NoStop}%
\bibitem [{\citenamefont {Paszke}\ \emph {et~al.}(2019)\citenamefont {Paszke}, \citenamefont {Gross}, \citenamefont {Massa}, \citenamefont {Lerer}, \citenamefont {Bradbury}, \citenamefont {Chanan}, \citenamefont {Killeen}, \citenamefont {Lin}, \citenamefont {Gimelshein}, \citenamefont {Antiga} \emph {et~al.}}]{paszke2019pytorch}%
  \BibitemOpen
  \bibfield  {author} {\bibinfo {author} {\bibfnamefont {A.}~\bibnamefont {Paszke}}, \bibinfo {author} {\bibfnamefont {S.}~\bibnamefont {Gross}}, \bibinfo {author} {\bibfnamefont {F.}~\bibnamefont {Massa}}, \bibinfo {author} {\bibfnamefont {A.}~\bibnamefont {Lerer}}, \bibinfo {author} {\bibfnamefont {J.}~\bibnamefont {Bradbury}}, \bibinfo {author} {\bibfnamefont {G.}~\bibnamefont {Chanan}}, \bibinfo {author} {\bibfnamefont {T.}~\bibnamefont {Killeen}}, \bibinfo {author} {\bibfnamefont {Z.}~\bibnamefont {Lin}}, \bibinfo {author} {\bibfnamefont {N.}~\bibnamefont {Gimelshein}}, \bibinfo {author} {\bibfnamefont {L.}~\bibnamefont {Antiga}}, \emph {et~al.},\ }\bibfield  {title} {\enquote {\bibinfo {title} {Pytorch: An imperative style, high-performance deep learning library},}\ }\href@noop {} {\bibfield  {journal} {\bibinfo  {journal} {Advances in neural information processing systems}\ }\textbf {\bibinfo {volume} {32}} (\bibinfo {year} {2019})}\BibitemShut {NoStop}%
\bibitem [{\citenamefont {Kokhlikyan}\ \emph {et~al.}(2020)\citenamefont {Kokhlikyan}, \citenamefont {Miglani}, \citenamefont {Martin}, \citenamefont {Wang}, \citenamefont {Alsallakh}, \citenamefont {Reynolds}, \citenamefont {Melnikov}, \citenamefont {Kliushkina}, \citenamefont {Araya}, \citenamefont {Yan} \emph {et~al.}}]{kokhlikyan2020captum}%
  \BibitemOpen
  \bibfield  {author} {\bibinfo {author} {\bibfnamefont {N.}~\bibnamefont {Kokhlikyan}}, \bibinfo {author} {\bibfnamefont {V.}~\bibnamefont {Miglani}}, \bibinfo {author} {\bibfnamefont {M.}~\bibnamefont {Martin}}, \bibinfo {author} {\bibfnamefont {E.}~\bibnamefont {Wang}}, \bibinfo {author} {\bibfnamefont {B.}~\bibnamefont {Alsallakh}}, \bibinfo {author} {\bibfnamefont {J.}~\bibnamefont {Reynolds}}, \bibinfo {author} {\bibfnamefont {A.}~\bibnamefont {Melnikov}}, \bibinfo {author} {\bibfnamefont {N.}~\bibnamefont {Kliushkina}}, \bibinfo {author} {\bibfnamefont {C.}~\bibnamefont {Araya}}, \bibinfo {author} {\bibfnamefont {S.}~\bibnamefont {Yan}}, \emph {et~al.},\ }\bibfield  {title} {\enquote {\bibinfo {title} {Captum: A unified and generic model interpretability library for pytorch},}\ }\href@noop {} {\bibfield  {journal} {\bibinfo  {journal} {arXiv preprint arXiv:2009.07896}\ } (\bibinfo {year} {2020})}\BibitemShut {NoStop}%
\bibitem [{\citenamefont {Li}\ \emph {et~al.}(2024{\natexlab{b}})\citenamefont {Li}, \citenamefont {Chen}, \citenamefont {Huang}, \citenamefont {Yang}, \citenamefont {Song}, \citenamefont {Cao}, \citenamefont {Lu}, \citenamefont {Tian},\ and\ \citenamefont {Hua}}]{li2024automatic}%
  \BibitemOpen
  \bibfield  {author} {\bibinfo {author} {\bibfnamefont {X.-L.}\ \bibnamefont {Li}}, \bibinfo {author} {\bibfnamefont {K.-Q.}\ \bibnamefont {Chen}}, \bibinfo {author} {\bibfnamefont {H.-J.}\ \bibnamefont {Huang}}, \bibinfo {author} {\bibfnamefont {S.}~\bibnamefont {Yang}}, \bibinfo {author} {\bibfnamefont {Q.-G.}\ \bibnamefont {Song}}, \bibinfo {author} {\bibfnamefont {W.}~\bibnamefont {Cao}}, \bibinfo {author} {\bibfnamefont {Z.-H.}\ \bibnamefont {Lu}}, \bibinfo {author} {\bibfnamefont {C.}~\bibnamefont {Tian}},\ and\ \bibinfo {author} {\bibfnamefont {C.}~\bibnamefont {Hua}},\ }\bibfield  {title} {\enquote {\bibinfo {title} {Automatic optimization of jwl-miller parameters of hmx-based aluminized explosive based on genetic algorithm},}\ }\href@noop {} {\bibfield  {journal} {\bibinfo  {journal} {Propellants, Explosives, Pyrotechnics}\ ,\ \bibinfo {pages} {e202300195}} (\bibinfo {year} {2024}{\natexlab{b}})}\BibitemShut {NoStop}%
\bibitem [{\citenamefont {Walters}\ \emph {et~al.}(2018)\citenamefont {Walters}, \citenamefont {Biswas}, \citenamefont {Lawrence}, \citenamefont {Francom}, \citenamefont {Luscher}, \citenamefont {Fredenburg}, \citenamefont {Moran}, \citenamefont {Sweeney}, \citenamefont {Sandberg}, \citenamefont {Ahrens} \emph {et~al.}}]{walters2018bayesian}%
  \BibitemOpen
  \bibfield  {author} {\bibinfo {author} {\bibfnamefont {D.~J.}\ \bibnamefont {Walters}}, \bibinfo {author} {\bibfnamefont {A.}~\bibnamefont {Biswas}}, \bibinfo {author} {\bibfnamefont {E.~C.}\ \bibnamefont {Lawrence}}, \bibinfo {author} {\bibfnamefont {D.~C.}\ \bibnamefont {Francom}}, \bibinfo {author} {\bibfnamefont {D.~J.}\ \bibnamefont {Luscher}}, \bibinfo {author} {\bibfnamefont {D.~A.}\ \bibnamefont {Fredenburg}}, \bibinfo {author} {\bibfnamefont {K.~R.}\ \bibnamefont {Moran}}, \bibinfo {author} {\bibfnamefont {C.~M.}\ \bibnamefont {Sweeney}}, \bibinfo {author} {\bibfnamefont {R.~L.}\ \bibnamefont {Sandberg}}, \bibinfo {author} {\bibfnamefont {J.~P.}\ \bibnamefont {Ahrens}}, \emph {et~al.},\ }\bibfield  {title} {\enquote {\bibinfo {title} {Bayesian calibration of strength parameters using hydrocode simulations of symmetric impact shock experiments of al-5083},}\ }\href@noop {} {\bibfield  {journal} {\bibinfo  {journal} {Journal of Applied Physics}\ }\textbf {\bibinfo {volume} {124}} (\bibinfo {year}
  {2018})}\BibitemShut {NoStop}%
\bibitem [{\citenamefont {Mortensen}\ and\ \citenamefont {Souers}(2017)}]{mortensen2017optimizing}%
  \BibitemOpen
  \bibfield  {author} {\bibinfo {author} {\bibfnamefont {C.}~\bibnamefont {Mortensen}}\ and\ \bibinfo {author} {\bibfnamefont {P.~C.}\ \bibnamefont {Souers}},\ }\bibfield  {title} {\enquote {\bibinfo {title} {Optimizing code calibration of the jwl explosive equation-of-state to the cylinder test},}\ }\href@noop {} {\bibfield  {journal} {\bibinfo  {journal} {Propellants, Explosives, Pyrotechnics}\ }\textbf {\bibinfo {volume} {42}},\ \bibinfo {pages} {616--622} (\bibinfo {year} {2017})}\BibitemShut {NoStop}%
\bibitem [{\citenamefont {Glorot}\ and\ \citenamefont {Bengio}(2010)}]{glorot2010understanding}%
  \BibitemOpen
  \bibfield  {author} {\bibinfo {author} {\bibfnamefont {X.}~\bibnamefont {Glorot}}\ and\ \bibinfo {author} {\bibfnamefont {Y.}~\bibnamefont {Bengio}},\ }\bibfield  {title} {\enquote {\bibinfo {title} {Understanding the difficulty of training deep feedforward neural networks},}\ }in\ \href@noop {} {\emph {\bibinfo {booktitle} {Proceedings of the thirteenth international conference on artificial intelligence and statistics}}}\ (\bibinfo {organization} {JMLR Workshop and Conference Proceedings},\ \bibinfo {year} {2010})\ pp.\ \bibinfo {pages} {249--256}\BibitemShut {NoStop}%
\bibitem [{\citenamefont {VanGessel}\ and\ \citenamefont {McGrath}(2022)}]{vangessel2022}%
  \BibitemOpen
  \bibfield  {author} {\bibinfo {author} {\bibfnamefont {F.~G.}\ \bibnamefont {VanGessel}}\ and\ \bibinfo {author} {\bibfnamefont {T.~P.}\ \bibnamefont {McGrath}},\ }\bibfield  {title} {\enquote {\bibinfo {title} {An improved water equation of state for underwater explosion simulations},}\ }\href@noop {} {\bibfield  {journal} {\bibinfo  {journal} {Journal of DoD Research and Engineering}\ }\textbf {\bibinfo {volume} {4}},\ \bibinfo {pages} {16--27} (\bibinfo {year} {2022})}\BibitemShut {NoStop}%
\bibitem [{\citenamefont {Li}\ \emph {et~al.}(2022)\citenamefont {Li}, \citenamefont {Shi}, \citenamefont {Wang},\ and\ \citenamefont {Zhao}}]{li2022measurement}%
  \BibitemOpen
  \bibfield  {author} {\bibinfo {author} {\bibfnamefont {G.}~\bibnamefont {Li}}, \bibinfo {author} {\bibfnamefont {D.}~\bibnamefont {Shi}}, \bibinfo {author} {\bibfnamefont {L.}~\bibnamefont {Wang}},\ and\ \bibinfo {author} {\bibfnamefont {K.}~\bibnamefont {Zhao}},\ }\bibfield  {title} {\enquote {\bibinfo {title} {Measurement technology of underwater explosion load: A review},}\ }\href@noop {} {\bibfield  {journal} {\bibinfo  {journal} {Ocean Engineering}\ }\textbf {\bibinfo {volume} {254}},\ \bibinfo {pages} {111383} (\bibinfo {year} {2022})}\BibitemShut {NoStop}%
\bibitem [{\citenamefont {Tillotson}(1962)}]{tillotson1962metallic}%
  \BibitemOpen
  \bibfield  {author} {\bibinfo {author} {\bibfnamefont {J.~H.}\ \bibnamefont {Tillotson}},\ }\href@noop {} {\emph {\bibinfo {title} {Metallic equations of state for hypervelocity impact}}}\ (\bibinfo  {publisher} {General Dynamics Falls Church, VA},\ \bibinfo {year} {1962})\BibitemShut {NoStop}%
\bibitem [{\citenamefont {Brundage}(2013)}]{brundage2013implementation}%
  \BibitemOpen
  \bibfield  {author} {\bibinfo {author} {\bibfnamefont {A.~L.}\ \bibnamefont {Brundage}},\ }\bibfield  {title} {\enquote {\bibinfo {title} {Implementation of tillotson equation of state for hypervelocity impact of metals, geologic materials, and liquids},}\ }\href@noop {} {\bibfield  {journal} {\bibinfo  {journal} {Procedia Engineering}\ }\textbf {\bibinfo {volume} {58}},\ \bibinfo {pages} {461--470} (\bibinfo {year} {2013})}\BibitemShut {NoStop}%
\bibitem [{\citenamefont {Plesset}\ and\ \citenamefont {Prosperetti}(1977)}]{plesset1977bubble}%
  \BibitemOpen
  \bibfield  {author} {\bibinfo {author} {\bibfnamefont {M.~S.}\ \bibnamefont {Plesset}}\ and\ \bibinfo {author} {\bibfnamefont {A.}~\bibnamefont {Prosperetti}},\ }\bibfield  {title} {\enquote {\bibinfo {title} {Bubble dynamics and cavitation},}\ }\href@noop {} {\bibfield  {journal} {\bibinfo  {journal} {Annual review of fluid mechanics}\ }\textbf {\bibinfo {volume} {9}},\ \bibinfo {pages} {145--185} (\bibinfo {year} {1977})}\BibitemShut {NoStop}%
\bibitem [{\citenamefont {Sambasivan}\ and\ \citenamefont {UdayKumar}(2010)}]{sambasivan2010sharp}%
  \BibitemOpen
  \bibfield  {author} {\bibinfo {author} {\bibfnamefont {S.~K.}\ \bibnamefont {Sambasivan}}\ and\ \bibinfo {author} {\bibfnamefont {H.}~\bibnamefont {UdayKumar}},\ }\bibfield  {title} {\enquote {\bibinfo {title} {Sharp interface simulations with local mesh refinement for multi-material dynamics in strongly shocked flows},}\ }\href@noop {} {\bibfield  {journal} {\bibinfo  {journal} {Computers \& Fluids}\ }\textbf {\bibinfo {volume} {39}},\ \bibinfo {pages} {1456--1479} (\bibinfo {year} {2010})}\BibitemShut {NoStop}%
\bibitem [{\citenamefont {Kapahi}, \citenamefont {Sambasivan},\ and\ \citenamefont {Udaykumar}(2013)}]{kapahi2013three}%
  \BibitemOpen
  \bibfield  {author} {\bibinfo {author} {\bibfnamefont {A.}~\bibnamefont {Kapahi}}, \bibinfo {author} {\bibfnamefont {S.}~\bibnamefont {Sambasivan}},\ and\ \bibinfo {author} {\bibfnamefont {H.}~\bibnamefont {Udaykumar}},\ }\bibfield  {title} {\enquote {\bibinfo {title} {A three-dimensional sharp interface cartesian grid method for solving high speed multi-material impact, penetration and fragmentation problems},}\ }\href@noop {} {\bibfield  {journal} {\bibinfo  {journal} {Journal of Computational Physics}\ }\textbf {\bibinfo {volume} {241}},\ \bibinfo {pages} {308--332} (\bibinfo {year} {2013})}\BibitemShut {NoStop}%
\bibitem [{\citenamefont {Li}\ \emph {et~al.}(2014)\citenamefont {Li}, \citenamefont {Zhang}, \citenamefont {Wang},\ and\ \citenamefont {Hu}}]{li2014numerical}%
  \BibitemOpen
  \bibfield  {author} {\bibinfo {author} {\bibfnamefont {X.}~\bibnamefont {Li}}, \bibinfo {author} {\bibfnamefont {C.}~\bibnamefont {Zhang}}, \bibinfo {author} {\bibfnamefont {X.}~\bibnamefont {Wang}},\ and\ \bibinfo {author} {\bibfnamefont {X.}~\bibnamefont {Hu}},\ }\bibfield  {title} {\enquote {\bibinfo {title} {Numerical study of underwater shock wave by a modified method of characteristics},}\ }\href@noop {} {\bibfield  {journal} {\bibinfo  {journal} {Journal of Applied Physics}\ }\textbf {\bibinfo {volume} {115}} (\bibinfo {year} {2014})}\BibitemShut {NoStop}%
\bibitem [{\citenamefont {Plesset}(1949)}]{plesset1949dynamics}%
  \BibitemOpen
  \bibfield  {author} {\bibinfo {author} {\bibfnamefont {M.~S.}\ \bibnamefont {Plesset}},\ }\bibfield  {title} {\enquote {\bibinfo {title} {The dynamics of cavitation bubbles},}\ }\href@noop {} {\  (\bibinfo {year} {1949})}\BibitemShut {NoStop}%
\bibitem [{\citenamefont {Prosperetti}\ and\ \citenamefont {Lezzi}(1986)}]{prosperetti1986bubble}%
  \BibitemOpen
  \bibfield  {author} {\bibinfo {author} {\bibfnamefont {A.}~\bibnamefont {Prosperetti}}\ and\ \bibinfo {author} {\bibfnamefont {A.}~\bibnamefont {Lezzi}},\ }\bibfield  {title} {\enquote {\bibinfo {title} {Bubble dynamics in a compressible liquid. part 1. first-order theory},}\ }\href@noop {} {\bibfield  {journal} {\bibinfo  {journal} {Journal of Fluid Mechanics}\ }\textbf {\bibinfo {volume} {168}},\ \bibinfo {pages} {457--478} (\bibinfo {year} {1986})}\BibitemShut {NoStop}%
\bibitem [{\citenamefont {Greenhall}\ \emph {et~al.}(2024)\citenamefont {Greenhall}, \citenamefont {Zerkle}, \citenamefont {Davis}, \citenamefont {Broilo},\ and\ \citenamefont {Pantea}}]{greenhall2024measuring}%
  \BibitemOpen
  \bibfield  {author} {\bibinfo {author} {\bibfnamefont {J.}~\bibnamefont {Greenhall}}, \bibinfo {author} {\bibfnamefont {D.~K.}\ \bibnamefont {Zerkle}}, \bibinfo {author} {\bibfnamefont {E.~S.}\ \bibnamefont {Davis}}, \bibinfo {author} {\bibfnamefont {R.}~\bibnamefont {Broilo}},\ and\ \bibinfo {author} {\bibfnamefont {C.}~\bibnamefont {Pantea}},\ }\bibfield  {title} {\enquote {\bibinfo {title} {Measuring thermal profiles in high explosives using neural networks},}\ }\href@noop {} {\bibfield  {journal} {\bibinfo  {journal} {APL Machine Learning}\ }\textbf {\bibinfo {volume} {2}} (\bibinfo {year} {2024})}\BibitemShut {NoStop}%
\bibitem [{\citenamefont {Sharma}\ \emph {et~al.}(2024)\citenamefont {Sharma}, \citenamefont {Gaffney}, \citenamefont {Tsapetis},\ and\ \citenamefont {Shields}}]{sharma2024learning}%
  \BibitemOpen
  \bibfield  {author} {\bibinfo {author} {\bibfnamefont {H.}~\bibnamefont {Sharma}}, \bibinfo {author} {\bibfnamefont {J.~A.}\ \bibnamefont {Gaffney}}, \bibinfo {author} {\bibfnamefont {D.}~\bibnamefont {Tsapetis}},\ and\ \bibinfo {author} {\bibfnamefont {M.~D.}\ \bibnamefont {Shields}},\ }\bibfield  {title} {\enquote {\bibinfo {title} {Learning thermodynamically constrained equations of state with uncertainty},}\ }\href@noop {} {\bibfield  {journal} {\bibinfo  {journal} {APL Machine Learning}\ }\textbf {\bibinfo {volume} {2}} (\bibinfo {year} {2024})}\BibitemShut {NoStop}%
\bibitem [{\citenamefont {Hinz}\ \emph {et~al.}(2024)\citenamefont {Hinz}, \citenamefont {Yu}, \citenamefont {Pandey}, \citenamefont {Sapkota}, \citenamefont {Yu}, \citenamefont {Mihaylov}, \citenamefont {Karasiev},\ and\ \citenamefont {Hu}}]{hinz2024development}%
  \BibitemOpen
  \bibfield  {author} {\bibinfo {author} {\bibfnamefont {J.}~\bibnamefont {Hinz}}, \bibinfo {author} {\bibfnamefont {D.}~\bibnamefont {Yu}}, \bibinfo {author} {\bibfnamefont {D.~S.}\ \bibnamefont {Pandey}}, \bibinfo {author} {\bibfnamefont {H.}~\bibnamefont {Sapkota}}, \bibinfo {author} {\bibfnamefont {Q.}~\bibnamefont {Yu}}, \bibinfo {author} {\bibfnamefont {D.}~\bibnamefont {Mihaylov}}, \bibinfo {author} {\bibfnamefont {V.}~\bibnamefont {Karasiev}},\ and\ \bibinfo {author} {\bibfnamefont {S.}~\bibnamefont {Hu}},\ }\bibfield  {title} {\enquote {\bibinfo {title} {The development of thermodynamically consistent and physics-informed equation-of-state model through machine learning},}\ }\href@noop {} {\bibfield  {journal} {\bibinfo  {journal} {APL Machine Learning}\ }\textbf {\bibinfo {volume} {2}} (\bibinfo {year} {2024})}\BibitemShut {NoStop}%
\bibitem [{\citenamefont {Cox}\ and\ \citenamefont {Christie}(2015)}]{cox2015fitting}%
  \BibitemOpen
  \bibfield  {author} {\bibinfo {author} {\bibfnamefont {G.}~\bibnamefont {Cox}}\ and\ \bibinfo {author} {\bibfnamefont {M.~A.}\ \bibnamefont {Christie}},\ }\bibfield  {title} {\enquote {\bibinfo {title} {Fitting of a multiphase equation of state with swarm intelligence},}\ }\href@noop {} {\bibfield  {journal} {\bibinfo  {journal} {Journal of Physics: Condensed Matter}\ }\textbf {\bibinfo {volume} {27}},\ \bibinfo {pages} {405201} (\bibinfo {year} {2015})}\BibitemShut {NoStop}%
\bibitem [{\citenamefont {Forte}\ \emph {et~al.}(2018)\citenamefont {Forte}, \citenamefont {Burger}, \citenamefont {Langenbach}, \citenamefont {Hasse},\ and\ \citenamefont {Bortz}}]{forte2018multi}%
  \BibitemOpen
  \bibfield  {author} {\bibinfo {author} {\bibfnamefont {E.}~\bibnamefont {Forte}}, \bibinfo {author} {\bibfnamefont {J.}~\bibnamefont {Burger}}, \bibinfo {author} {\bibfnamefont {K.}~\bibnamefont {Langenbach}}, \bibinfo {author} {\bibfnamefont {H.}~\bibnamefont {Hasse}},\ and\ \bibinfo {author} {\bibfnamefont {M.}~\bibnamefont {Bortz}},\ }\bibfield  {title} {\enquote {\bibinfo {title} {Multi-criteria optimization for parameterization of saft-type equations of state for water},}\ }\href@noop {} {\bibfield  {journal} {\bibinfo  {journal} {AIChE Journal}\ }\textbf {\bibinfo {volume} {64}},\ \bibinfo {pages} {226--237} (\bibinfo {year} {2018})}\BibitemShut {NoStop}%
\bibitem [{\citenamefont {Bergh}, \citenamefont {Wedberg},\ and\ \citenamefont {Lundgren}(2018)}]{bergh2018optimization}%
  \BibitemOpen
  \bibfield  {author} {\bibinfo {author} {\bibfnamefont {M.}~\bibnamefont {Bergh}}, \bibinfo {author} {\bibfnamefont {R.}~\bibnamefont {Wedberg}},\ and\ \bibinfo {author} {\bibfnamefont {J.}~\bibnamefont {Lundgren}},\ }\bibfield  {title} {\enquote {\bibinfo {title} {Optimization of equation of state and burn model parameters for explosives},}\ }in\ \href@noop {} {\emph {\bibinfo {booktitle} {AIP Conference Proceedings}}},\ Vol.\ \bibinfo {volume} {1979}\ (\bibinfo {organization} {AIP Publishing},\ \bibinfo {year} {2018})\BibitemShut {NoStop}%
\bibitem [{\citenamefont {Wang}\ \emph {et~al.}(2016)\citenamefont {Wang}, \citenamefont {Wang}, \citenamefont {Lu}, \citenamefont {Zhou}, \citenamefont {Chen},\ and\ \citenamefont {Yan}}]{wang2016determination}%
  \BibitemOpen
  \bibfield  {author} {\bibinfo {author} {\bibfnamefont {G.}~\bibnamefont {Wang}}, \bibinfo {author} {\bibfnamefont {Y.}~\bibnamefont {Wang}}, \bibinfo {author} {\bibfnamefont {W.}~\bibnamefont {Lu}}, \bibinfo {author} {\bibfnamefont {W.}~\bibnamefont {Zhou}}, \bibinfo {author} {\bibfnamefont {M.}~\bibnamefont {Chen}},\ and\ \bibinfo {author} {\bibfnamefont {P.}~\bibnamefont {Yan}},\ }\bibfield  {title} {\enquote {\bibinfo {title} {On the determination of the mesh size for numerical simulations of shock wave propagation in near field underwater explosion},}\ }\href@noop {} {\bibfield  {journal} {\bibinfo  {journal} {Applied Ocean Research}\ }\textbf {\bibinfo {volume} {59}},\ \bibinfo {pages} {1--9} (\bibinfo {year} {2016})}\BibitemShut {NoStop}%
\bibitem [{\citenamefont {Schittke}\ \emph {et~al.}(1989)\citenamefont {Schittke}, \citenamefont {Mohr}, \citenamefont {Luetje}, \citenamefont {Pfrang}, \citenamefont {Freercks},\ and\ \citenamefont {Niessen}}]{schittke1989program}%
  \BibitemOpen
  \bibfield  {author} {\bibinfo {author} {\bibfnamefont {H.}~\bibnamefont {Schittke}}, \bibinfo {author} {\bibfnamefont {W.}~\bibnamefont {Mohr}}, \bibinfo {author} {\bibfnamefont {H.}~\bibnamefont {Luetje}}, \bibinfo {author} {\bibfnamefont {W.}~\bibnamefont {Pfrang}}, \bibinfo {author} {\bibfnamefont {J.}~\bibnamefont {Freercks}},\ and\ \bibinfo {author} {\bibfnamefont {E.}~\bibnamefont {Niessen}},\ }\bibfield  {title} {\enquote {\bibinfo {title} {The program dysmas/elc and its application on underwater shock loading of vessels},}\ }in\ \href@noop {} {\emph {\bibinfo {booktitle} {60th Shock and Vibration Symposium}}},\ Vol.~\bibinfo {volume} {1}\ (\bibinfo {year} {1989})\ pp.\ \bibinfo {pages} {55--78}\BibitemShut {NoStop}%
\end{thebibliography}%

\clearpage  
\appendix 
\section{\label{app:syn_data_gen}Data Generation}
GEMINI is the fluid solver portion of the DYSMAS hydrocode, a fluid-structure computational simulator that has been validated for a wide array of underwater energetic scenarios \cite{wardlaw2000fluid}. The scenario studied presently involves a 10 pound spherical charge composed of an arbitrary ideal energetic material surrounded by an infinite water medium. This work considers only ideal explosives with fixed mass density and heat of detonation. Under these assumptions, any notional or known explosive is completely represented by the thermodynamic equation of state (EOS) properties of the energetic material. We next define the system geometry, initial conditions, and boundary conditions, all of which are required to simulate the detonation-induced dynamics for any chosen energetic material,.

As we are concerned with the free-field response the system geometry is modeled as spherically symmetric with the charge located at the origin. The system domain is specified to be large enough that boundary reflections of the shockwave do not influence the spatiotemporal regime of interest. The far boundary is assumed to be perfectly reflecting, with the spatial mesh chosen to achieve convergence of all field variables. The initial state of the system is assumed quiescent with the water pressure corresponding to the hydrostatic pressure achieved at a depth of fifty feet. The system dynamics are initialized with a programmed burn which simulates an idealized \textit{Chapman-Jouguet} detonation wave that releases the stored chemical energy of the charge\cite{jones1998numerical}. The resulting system dynamics are thus fully fixed once the water and energetic material thermodynamic models are chosen. A detailed description of the spatial mesh is given in Table \ref{tab:grid_setup}. A two-stage simulation procedure is used to capture both the initial energy release as well as the subsequent shock propagation phase. A remap condition is used to determine when to terminate each stage of the simulation.
\begin{table*}
\caption{\label{tab:grid_setup} Hydrocode Grid Setup}
\begin{ruledtabular}
\begin{tabular}{cccccc}
 Region & Grid Refinement & Size of First Cell [cm] & Cell Ratio & Grid Width [cm] & Remap Condition \\ \hline
 Precalc & Fine & 0.043 & 1 & 65.15 & 7.5*[Charge Radius] \\  \hline
 \multirow{ 2}{*}{Rezone} & Coarse (No stretch) & 0.709 & 1 & 850.58 & - \\ 
 & Coarse (Stretch) & 0.709 & 1.025 & 1222.82 & ~10.5$\tau$ decay\\ 
\end{tabular}
\end{ruledtabular}
\end{table*}

The water equation of state is chosen to be the Tillotson EOS form \cite{tillotson1962metallic} with the fixed parametrization given in Table \ref{tab:grid_setup}. The water EOS used in the fluid solver simulation is a Tillotson functional \cite{tillotson1962metallic} form, given by:
\begin{equation}
\begin{split}
P(\rho, e) =& \ P_0 + \omega \rho (e-e_0) + A \mu + B \mu^2 + C \mu^3 \\ \mu =& \frac{\rho}{\rho_0} - 1
\end{split}
\end{equation}
Originally developed for application to the hypervelocity impact of metals, the Tillotson EOS has been re-calibrated to accurately capture the thermodynamic properties of water \cite{brundage2013implementation}. The Tillotson EOS parameters are expressed in Table \ref{tab:water_params}.

\begin{table*}
\caption{\label{tab:water_params} Tillotson EOS for Water}
\begin{ruledtabular}
\begin{tabular}{cccccccc}
 Material & $e_0$ [erg] & $\rho_0 $ [g/cc] & $A$ [Ba] & $B$ [Ba] &  $C$ [Ba] &$P_0$ [Ba] &$\omega$ [-]\\ \hline
 Water & 3.542E+9 & 1.0 & 2.04815E+10 & 4.8888E+10 &  1.457E+11 & 1E+6 & 0.28 \\ 
\end{tabular}
\end{ruledtabular}
\end{table*}

\section{\label{app:ml_comparision}ML Model Comparison and Hyperparameter Values}
Listed here are the error metric comparisons for the neural network, kernel ridge regression, and gaussian process regression ML methods (Tables \ref{tab:ml_method_comp_bubble} and \ref{tab:ml_method_comp_shock}). It can be seen that the neural network performed significantly better on the more complicated state variable decay, namely pressure and material velocity at the bubble interface and the pressure, density, and material velocity at the shock front. However, note that the other ML methods performed better at predicting the spatial locations of each of the front. However, exhaustive hyperparameter tuning was not part of the model selection procedure and the chosen hyperparameters (Table \ref{tab:ml_hyperparameters}) shown here do not necessarily reflect fully optimized models.
\begin{table*}
\caption{\label{tab:ml_hyperparameters} ML Method Hyperparameters}
\begin{ruledtabular}
\begin{tabular}{cccccc}
 ML Method & Kernel & $\alpha$ & Degree & Zero Coeff. &  $\gamma$ \\ \hline
 KRR & Poly & 1E-3 & 3 & 10 &  100 \\ 
 GPR & Matern ($\nu$=1.5) & 1E-10 & - & - & - \\ 
\end{tabular}
\end{ruledtabular}
\end{table*}

\begin{table*}
\caption{\label{tab:ml_method_comp_bubble} ML Method Comparison: Bubble Interface}
\begin{ruledtabular}
\begin{tabular}{cccccccc}
 & ML Method & RMSE (val) & RMSE (test) & $R^2$ (val)& $R^2$ (test)& MAPE (val) & MAPE (test) \\ \hline
 \multirow{ 3}{*}{$r$} & NN & 4.42E-2 & 4.91E-2 & 1.0000 & 1.0000 & 9.02E-4 & 9.53E-4\\ 
 & KRR & 5.30E-3 & \textbf{4.75E-3} & 1.0000 & 1.0000 & 5.86E-5 & \textbf{6.04E-5}\\ 
 & GPR & 6.49E-2 & 5.15E-2 & 1.0000 & 1.0000 & 5.21E-4 & 5.10E-4 \\ \hline
 \multirow{ 3}{*}{$p$} & NN & 5.33E+7 & \textbf{5.70E+7} & 0.9999 & 0.9999 & 4.83E-3 & \textbf{5.11E-3}\\ 
 & KRR & 7.66E+7 & 8.05E+7 & 0.9999 & 0.9999 & 7.22E-3 & 8.79E-3\\ 
 & GPR & 9.98E+7 & 1.07E+8 & 0.9998 & 0.9998 & 2.70E-2 & 3.48E-2\\ \hline
 \multirow{ 3}{*}{$u$} & NN & 8.93E+1 & \textbf{9.34E+1} & 1.0000 & 1.0000 & 1.17E-3 & \textbf{1.17E-3}\\ 
 & KRR & 9.94E+1 & 1.04E+2 & 1.0000 & 1.0000 & 1.38E-3 & 1.41E-3\\
 & GPR & 1.20E+2 & 1.30E+2 & 1.0000 & 1.0000 & 1.45E-3 & 1.51E-3\\
\end{tabular}
\end{ruledtabular}
\end{table*}

\begin{table*}
\caption{\label{tab:ml_method_comp_shock} ML Method Comparison: Shock Front}
\begin{ruledtabular}
\begin{tabular}{cccccccc}
 & ML Method & RMSE (val) & RMSE (test) & $R^2$ (val)& $R^2$ (test)& MAPE (val) & MAPE (test) \\ \hline
 \multirow{ 3}{*}{$r$} & NN & 3.17E-1 & 3.22E-1 & 1.0000 & 1.0000 & 6.62E-4 & 6.63E-4\\
 & KRR & 1.11E-1 & \textbf{1.11E-1} & 1.0000 & 1.0000 & 2.02E-4 & \textbf{2.08E-4}\\
 & GPR & 2.36E-1 & 2.34E-1 & 1.0000 & 1.0000 & 5.69E-4 & 5.64E-4\\ \hline
 \multirow{ 3}{*}{$p$} & NN & 1.37E+6 & \textbf{1.41E+6} & 1.0000 & 1.0000 & 7.80E-4 & 7.94E-4\\
 & KRR & 1.71E+6 & 1.54E+6 & 1.0000 & 1.0000 & 5.69E-4 & \textbf{5.71E-4}\\ 
 & GPR & 1.58E+6 & 1.60E+6 & 1.0000 & 1.0000 & 8.03E-4 & 7.94E-4\\ \hline
  \multirow{ 3}{*}{$e$} & NN & 2.04E+5 & 2.09E+5 & 0.9995 & 0.9995 & 2.07E-3 & \textbf{2.08E-3} \\
 & KRR & 2.08E+5 & 1.99E+5 & 0.9995 & 0.9995 & 2.45E-3 & 2.51E-3 \\
 & GPR & 1.97E+5 & \textbf{1.97E+5} & 0.9995 & 0.9995 & 2.52E-3 & 2.43E-3 \\ \hline
  \multirow{ 3}{*}{$\rho$} & NN & 5.73E-5 & \textbf{5.90E-5} & 1.0000 & 1.0000 & 7.50E-4 & 7.66E-4 \\ 
 & KRR & 6.90E-5 & 6.20E-5 & 1.0000 & 1.0000 & 5.57E-4 & \textbf{5.60E-4} \\
 & GPR & 6.38E-5 & 6.20E-5 & 1.0000 & 1.0000 & 7.72E-4 & 7.55E-4 \\ \hline
 \multirow{ 3}{*}{$u$} & NN & 8.98E+0 & \textbf{9.17E+0} & 1.0000 & 1.0000 & 7.80E-4 & 7.85E-4 \\
 & KRR & 1.09E+1 & 1.10E+1 & 1.0000 & 1.0000 & 5.76E-4 & \textbf{5.82E-4} \\
 & GPR & 1.03E+1 & 1.01E+1 & 1.0000 & 1.0000 & 7.93E-4 & 7.81E-4 \\
\end{tabular}
\end{ruledtabular}
\end{table*}

\section{\label{app:nn_training}DNN Architecture and Neural Network Training Approach}
\begin{figure*}
\includegraphics[width=.4\textwidth]{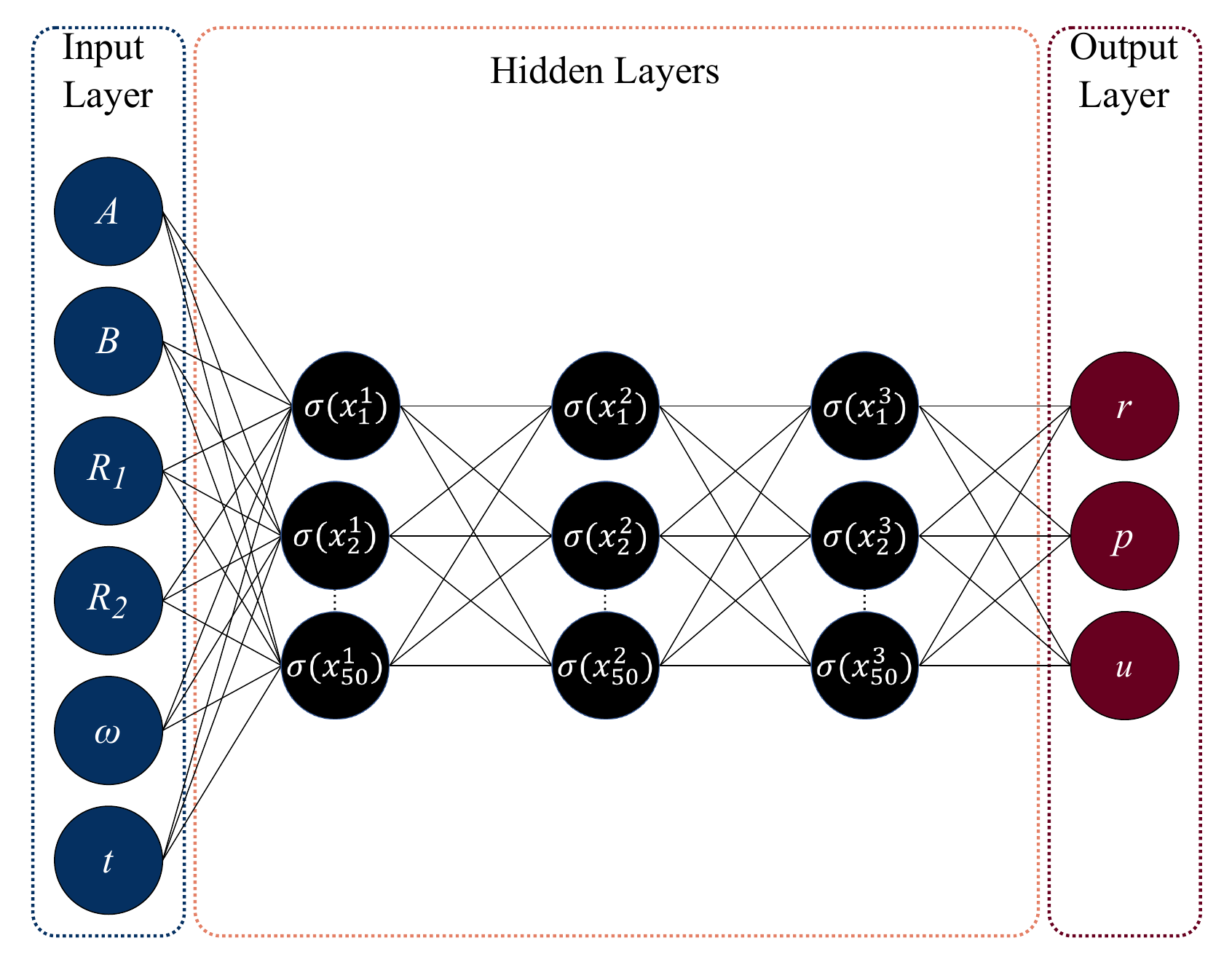}
\includegraphics[width=.4\textwidth]{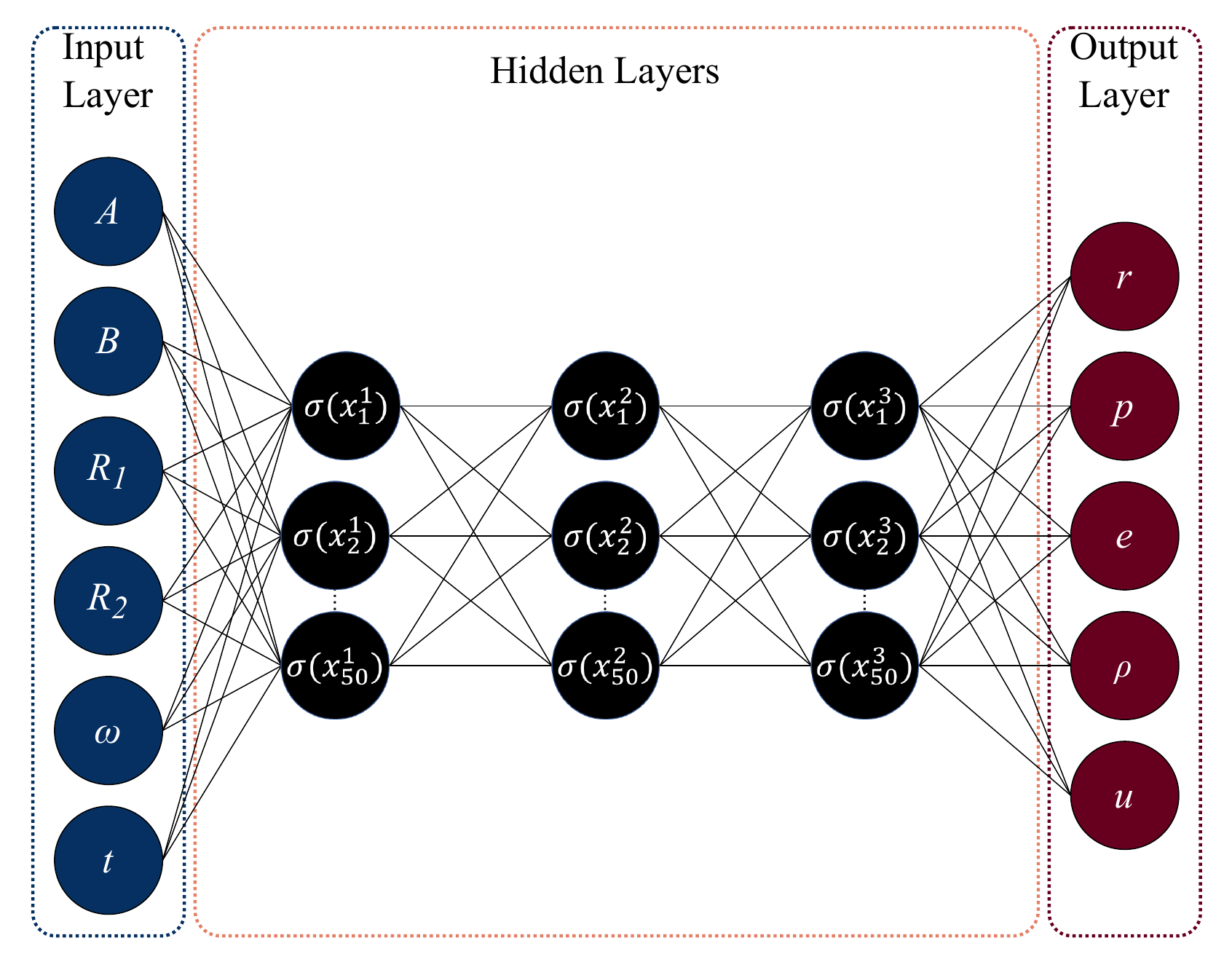}
\caption{\label{fig:DNN} {DNN of the Bubble Interface (left) and Shock Front (right).}}
\end{figure*}

The  DNN (see Fig. \ref{fig:DNN}) is a fully connected architecture using hyperbolic tangent activation functions between linear transformation layers with the final layer containing either 3 or 5 output neurons depending on the front of interest. The loss function is a composite RMSE with equal weight assigned to each of the output variables. Due to the large size of data (approx. 4 million inputs for each model) training was performed using mini-batch gradient descent and a learning rate scheduler was implemented for better training. See Table \ref{tab:nn_hyperparameters} for further information.
\begin{table*}
\caption{\label{tab:nn_hyperparameters} DNN Hyperparameters}
\begin{ruledtabular}
\begin{tabular}{cc}
 Epochs & 10000 \\ \hline
 Batch Size & 4096 \\  \hline
 Weight Initialization & Xavier \\ \hline
 Activation Function (AV) & Tanh \\ \hline
 Architecture (Fully connected) & 6 x 50 x 50 x 3 (bubble interface)\\
& 6 x 50 x 50 x 5 (shock front) \\ \hline
Scaling & Input: MinMaxScaler \\
& Output: Quantile Transformer \\ \hline
Optimizer & Adam \\ \hline
Loss & Composite RMSE \\ \hline
Learning Rate & 1E-03 (LR Scheduler once plateau reached)\\ 
\end{tabular}
\end{ruledtabular}
\end{table*}

\nocite{*}
The process to evaluate the model selection criteria (MSC) is as follows. First, evaluate the relative error between the DNN surrogate model and the ground truth, and evaluate how frequently  this value is less than or equal to $\pm .25\%$. The total number of instances for which this condition holds is averaged across all 100 JWL parameter validation datasets. The functional expression for the model selection criterion is given in Eq. \ref{eq:msc}. The final DNN models are chosen to correspond to the training epoch which maximizes MSC.

\begin{equation}
    \label{eq:msc}
    \begin{split}
   MSC = \frac{1}{N_{val}} \sum_{i=1}^{N_{val}} \frac{1}{N_t} \sum_{j=1}^{N_t} &\mathbf {1} _{[0, 2.5 \times 10^{-3}]} \left(\left |\frac{(f_{\mathbf{\theta}}(t_j; \Theta_i)-f_{GT}(t_j; \Theta_i))}
    {f_{GT}(t_j; \Theta_i)}\right |\right) \\
    &\mathbf {1} _{A}(x) = \begin{cases}1~&{\text{ if }}~x\in A~,\\0~&{\text{ if }}~x\notin A~.\end{cases}
    \end{split}
\end{equation}

\section{\label{app:dnn_response_parameter_plots}DNN Surrogate Response Surface Parameter Plots}
\begin{figure*}
\includegraphics[width=.8\textwidth]{mitul_plots/Pressure_bubble_R_surface_plots.png}\\
\includegraphics[width=.8\textwidth]{mitul_plots/Location_bubble_R_surface_plots.png}\\
\includegraphics[width=.8\textwidth]{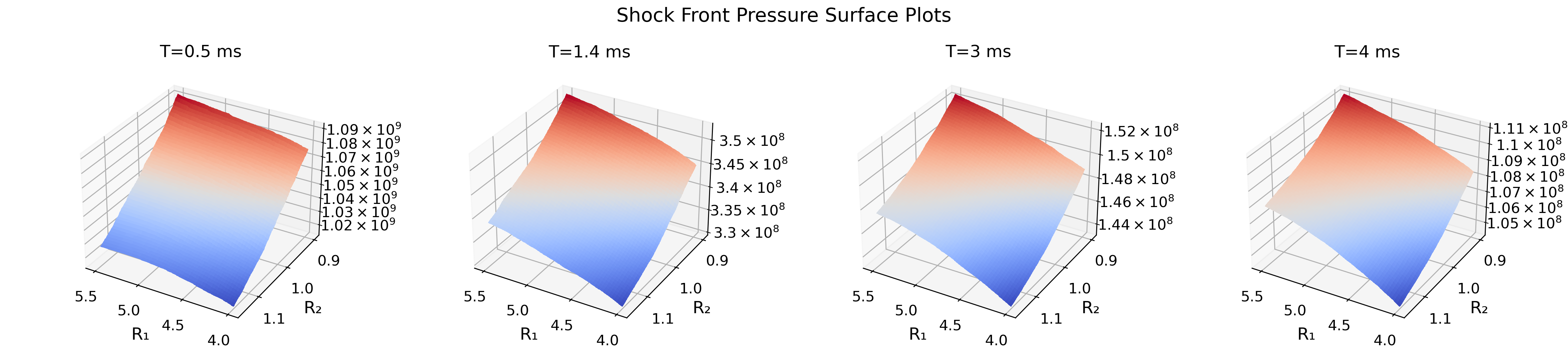}\\
\caption{\label{fig:surface_plots} Parameter Plots for selected state variable dynamics at the bubble (top and center) and shock (bottom) fronts. The principal JWL parameters of interest are $R_1$ and $R_2$.}
\end{figure*}

Due to the rapid prediction time of the trained DNN model, we are able to input thousands of different JWL parameter combinations (i.e., different notional energetic systems) and visualize parameter surface plots which aid in understanding the relationships between input parameters and the resulting state variable dynamics. The plots in Fig. \ref{fig:surface_plots} corroborate with the feature importance results and demonstrate how the state variables are influenced by $R_1$ and $R_2$ at four different snapshots in time. 
\label{app:param_plots}

\section{\label{app:recovered_dynamics}Recovered Dynamics}
\begin{figure*}
\includegraphics[width=.8\textwidth]{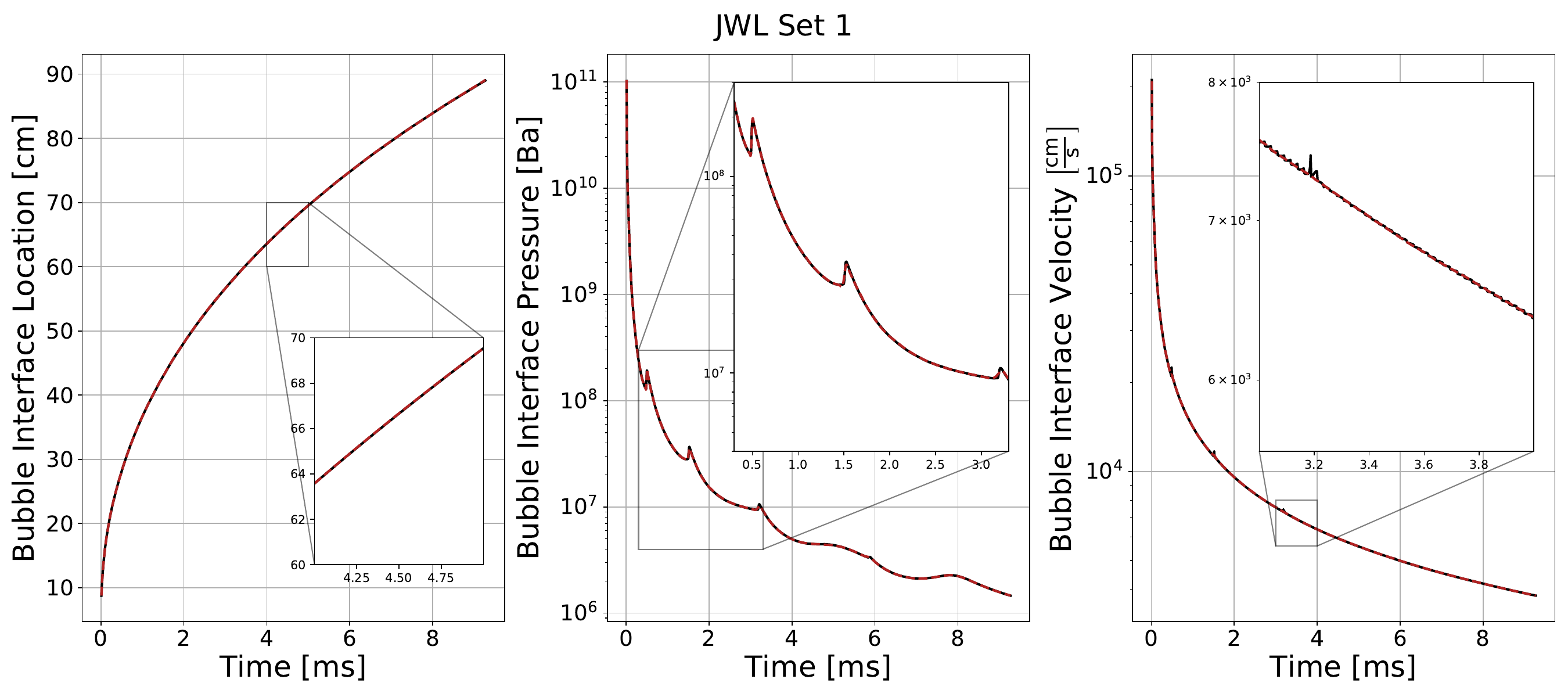}\\
\includegraphics[width=.8\textwidth]{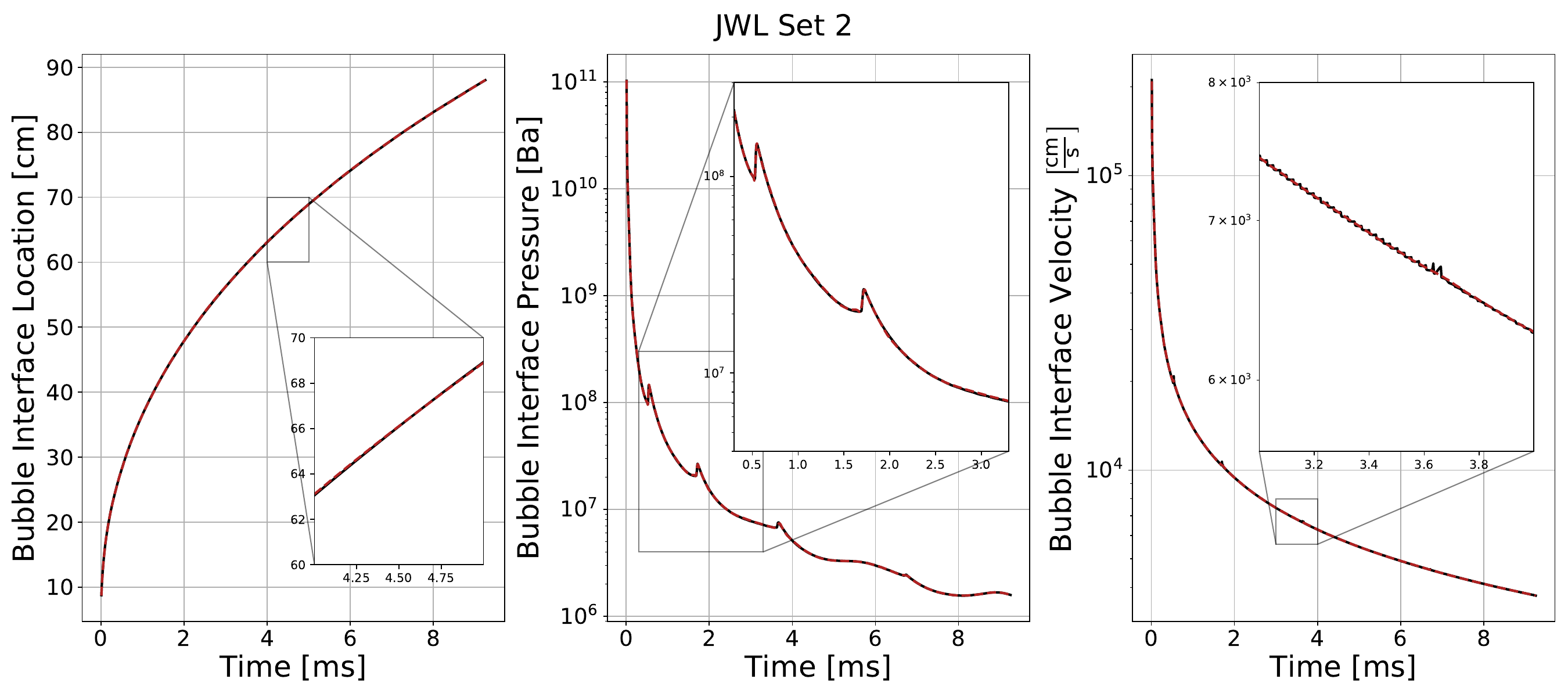}\\
\includegraphics[width=.8\textwidth]{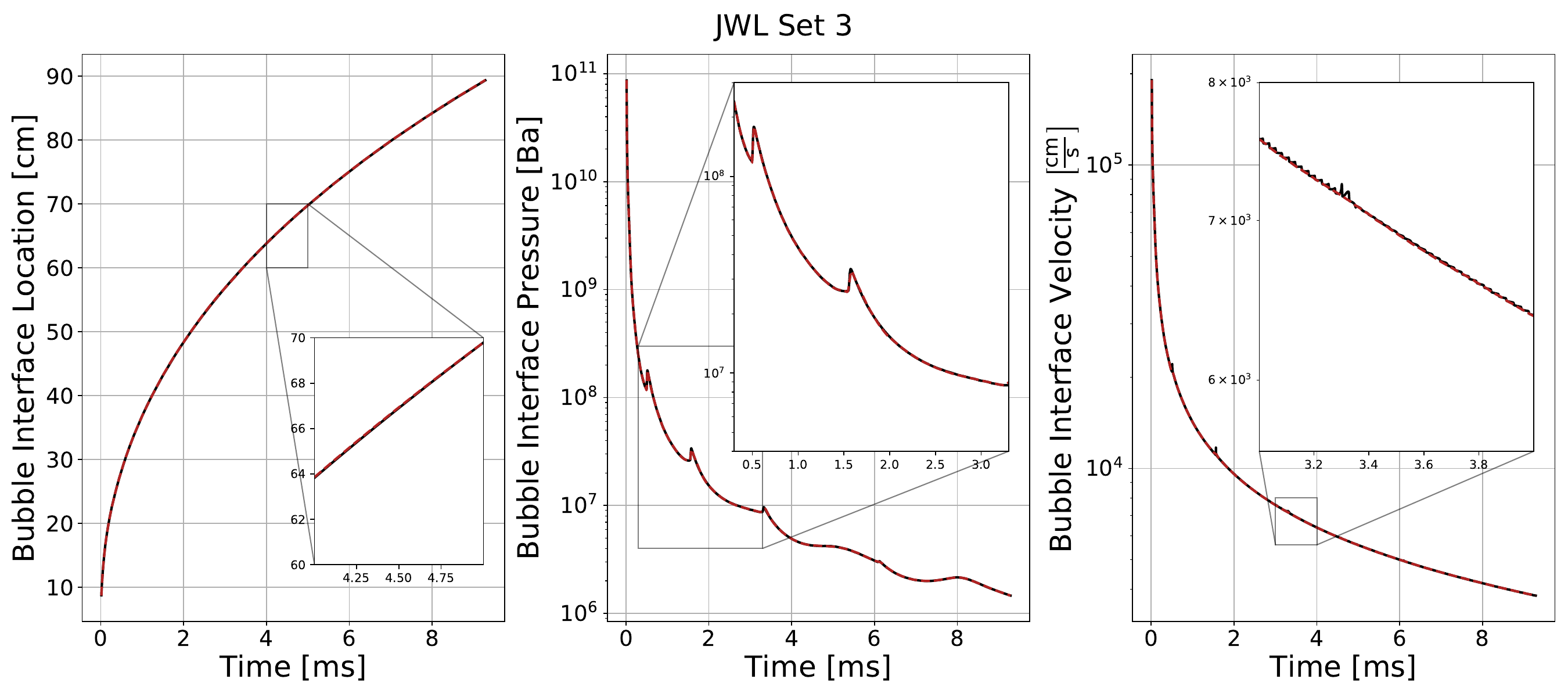}\\
\includegraphics[width=.3\textwidth]{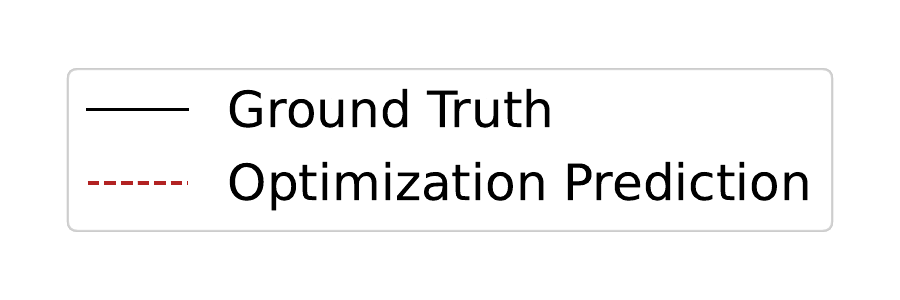}
\caption{\label{fig:optimization_dynamics} Comparison of the state variable dynamics between the ground-truth data and the predictions from the calibrated JWL EOS. The predictions are shown as a dashed dark red line, while the ground-truth data corresponds to the solid black line.}
\end{figure*}

The dynamics recovered from the inverse design algorithm are shown in Fig. \ref{fig:optimization_dynamics}. These recovered time series curves are visually indistinguishable from one another, further corroborating that the parameter discovery algorithm is robust and accurate.

\end{document}